\newif\ifdraft
\draftfalse

\ifdraft
	\documentclass[12pt, journal, a4paper, comsoc, draftclsnofoot, onecolumn]{IEEEtran}
\else
\documentclass[journal, a4paper, comsoc]{IEEEtran}
\fi

\ifdraft
\def\subparagraph{}
\usepackage[compact]{titlesec}
\fi

\usepackage[utf8]{luainputenc}

\usepackage[noadjust]{cite}

\usepackage{graphicx}
\graphicspath{{./illustrationer/}}

\usepackage[cmex10]{amsmath}
\usepackage{array}

\usepackage{booktabs}

\ifCLASSOPTIONcompsoc
  \usepackage[caption=false,font=normalsize,labelfont=sf,textfont=sf]{subfig}
\else
  \usepackage[caption=false,font=footnotesize]{subfig}
\fi

\usepackage{url}

\usepackage{microtype}
\usepackage{amssymb,mathtools}

\DeclarePairedDelimiter\abs{\lvert}{\rvert}
\DeclarePairedDelimiter\bracket{\lbrack}{\rbrack}

\usepackage{minamakron}
\usepackage{siunitx}
\usepackage{pgfplots}
\usetikzlibrary{fadings}

\ifdraft
\pgfplotsset{
	small,
	legend style={font=\footnotesize},
	label style={font=\footnotesize},
	width=252pt,
	height=24ex,
}
\fi

\usepackage{tikz}

\newcommand{\sigmaxm}{\ensuremath\sigma_{\kern-.25emx_m}}

\newcommand*\circled[1]{\tikz[baseline=(char.base)]
	{
		\node[shape=circle,draw,inner sep=0.05pt] (char) {#1};}}

\newsavebox{\elementwiseconv}
\savebox{\elementwiseconv}{\circled{$\scriptstyle\star$}}

\newcommand{\circledstar}{\mathrel{\raisebox{.12pt}{\usebox{\elementwiseconv}}}}

\usepgfplotslibrary{external}
{distortion_from_large_arrays}
\tikzsetexternalprefix{ext_figures/}
\tikzset{external/force remake=false}

\begin{document}
	\title{%
		\ifdraft
		\rule{0pt}{0pt}\clap{Spatial Characteristics of Distortion Radiated from}\rule{0pt}{0pt}\\\rule{0pt}{0pt}\clap{Antenna Arrays with Transceiver Nonlinearities}\rule{0pt}{0pt}%
		\else%
		Spatial Characteristics of Distortion Radiated from Antenna Arrays with Transceiver Nonlinearities%
		\fi%
		}
	
	\author{Christopher~Mollén, Ulf~Gustavsson, Thomas~Eriksson, Erik~G.~Larsson%
			\thanks{C. Mollén och E. Larsson are with the Department
			of Electrical Engineering, Linköping University, 581~83 Linköping, Sweden e-mail: christopher.mollen@liu.se, erik.g.larsson@liu.se.}
		\thanks{U. Gustavsson is with Ericsson Research, Ericsson AB, Lindholmspiren 11, 417~56 Gothenburg, Sweden e-mail: ulf.gustavsson@ericsson.com.}
		\thanks{T.~Eriksson is with the Department of Signals and Systems, Chalmers University of Technology, 412~96 Gothenburg, Sweden e-mail: thomase@chalmers.se}%
		\thanks{Research leading to these results has received funding from Vetenskapsrådet (The Swedish Research Council) and ELLIIT.  Parts of the work have been done at the GHz Centre, which is sponsored by VINNOVA.}%
		\thanks{Part of this work was presented at IEEE International Conference on Communications in 2016 \cite{mollen2016OOB}.}}
	
	\maketitle
	\ifdraft
	\vspace{-8ex}
	\fi
	\begin{abstract}
		The distortion from massive \MIMO (multiple-input--multiple-output) base stations with nonlinear amplifiers is studied and its radiation pattern is derived.  The distortion is analyzed both in-band and out-of-band.  By using an orthogonal Hermite representation of the amplified signal, the spatial cross-correlation matrix of the nonlinear distortion is obtained.  It shows that, if the input signal to the amplifiers has a dominant beam, the distortion is beamformed in the same way as that beam.  When there are multiple beams without any one being dominant, it is shown that the distortion is practically isotropic.  The derived theory is useful to predict how the nonlinear distortion will behave, to analyze the out-of-band radiation, to do reciprocity calibration, and to schedule users in the frequency plane to minimize the effect of in-band distortion.
	\end{abstract}
	
	
	\IEEEpeerreviewmaketitle
		
	\section{Introduction}

\IEEEPARstart{N}{onlinear} hardware
causes signal distortion that degrades both the performance of the own
        transmission, so called \emph{in-band distortion}, and the
        performance of systems using adjacent frequency channels, so
        called \emph{out-of-band radiation}.  Often the power
        amplifier is the main cause of nonlinear distortion.  Other
        nonlinear components are digital-to-analog converters and
        mixers.  While nonlinear distortion from single-antenna
        transmitters is a well-investigated phenomenon, it has been
        far less studied in the context of large arrays.
	
	The radiation pattern of the nonlinear distortion from large
	arrays has recently attracted  attention because of its
	potential impact on the performance of massive \MIMO
	systems \cite{6690}.  Papers such
	as \cite{bjornson2013massive,blandino2017analysis} have
	suggested that the distortion combines non-coherently at the
	served users and vanishes with an increasing number of
	transmit antennas.  These results were also corroborated to some extent (for in-band distortion)
        via
	simulations in \cite{UGUSGC14}.  However, contrary to these
	results, \cite{zou2015impact} showed that the amplifier
	distortion can combine coherently and degrade the performance
	significantly.  Just like \cite{bjornson2013massive}, however,
	the results in \cite{zou2015impact} rely on an over-simplified
	symbol-sampled system model and on frequency-flat fading.  The
	aim of this paper is to give a rigorous description of the
	distortion created by nonlinear hardware in multi-antenna
	transmitters, and to quantify to which degree the distortion
	combines coherently.
	
	The contribution of this paper is to give a rigorous
	continuous-time system model of a multi-antenna transmitter
	for both single-carrier and \OFDM (orthogonal
	frequency-division multiplexing) transmission that uses
	digital precoding to beamform to multiple users.  Orthogonal
	polynomials are used to partition the amplified transmit
	signal into a desired signal---the linearly amplified
	signal---and a distortion term that is uncorrelated from the
	desired signal in order to analyze both the in-band and
	out-of-band distortion separately from the desired linear
	signal.  The orthogonal representation also allows for a
	straightforward derivation of the radiation pattern of the
	distortion and its spatial characteristics.  If $K$ is the
	number of served users and $L$ is the number of significant
	channel taps, it is shown that the number of directions that
	the distortion is beamformed in scales as $\ordo(K^3 L^2)$.
	If all users are served with the same power, the distortion is isotropic when this number is greater than the number of antennas, and it is beamformed otherwise.  The beamforming gain of the distortion, however, is not larger
	than that of the desired signal.
	
	The analysis is based on the assumptions that the signals follow a
	Gaussian distribution because then the Itô-Hermite polynomials
	form an orthogonal basis, in which the nonlinearities can be
	described.  Other distributions may require other polynomial
	bases.  However, many massive \MIMO signals closely follow
	a Gaussian distribution after modulation and (linear) precoding.  It is
	also assumed that the amplifier nonlinearities can be
	described by memory polynomials, which is a commonly used
	model for amplifiers whose memory effects can be captured by
	one-dimensional kernels \cite{pedro2005comparative,
	morgan2006generalized}.  For a general nonlinearity, it might
	be possible, albeit tedious, to use the method
	in \cite{barrett1980formula} to generalize our results.
	
	\subsection*{Other Related Work}

Orthogonal polynomials have been used before to analyze power
amplifiers \cite{gard1999characterization, raich2004orthogonal,
raich2004orthogonal2}.  Previous work, however, is limited to
single-antenna transmitters and cannot be directly generalized to analyze the
radiation pattern from a transmitter with multiple antennas.  Only
some special cases of arrays have been considered before in the case
of line-of-sight propagation.  For example,
in \cite{sandrin1973spatial}, the directivity of the distortion in
phased arrays for satellite communication is studied and,
in \cite{hemmi2002pattern}, a phased array with two beams is studied.
	
	Previously, we have addressed the topic of out-of-band
	radiation in \cite{mollen2016OOB}
	and \cite{mollen2016outofbandArXiv}.  A preliminary study was
	conducted in \cite{mollen2016OOB}, where a polynomial model of
	degree three was considered and the cross-correlations were
	derived without the help of orthogonal polynomials.
	In \cite{mollen2016outofbandArXiv}, the spatial behavior of
	the out-of-band radiation was explained without giving any
	mathematical details.  In this paper, we present a deeper
	analysis of the distortion from large arrays, not only the
	out-of-band radiation, but also the in-band distortion.  In
	doing so, we use the theory about Hermite expansions that we
	have derived and presented in \cite{mollen2017nonlinearInThesis} for a
	general nonlinearity.
	
	\section{System Model}

\begin{figure*}
		\centering
		\ifdraft
		\footnotesize
		\def\svgwidth{.9\linewidth}
		\fi
		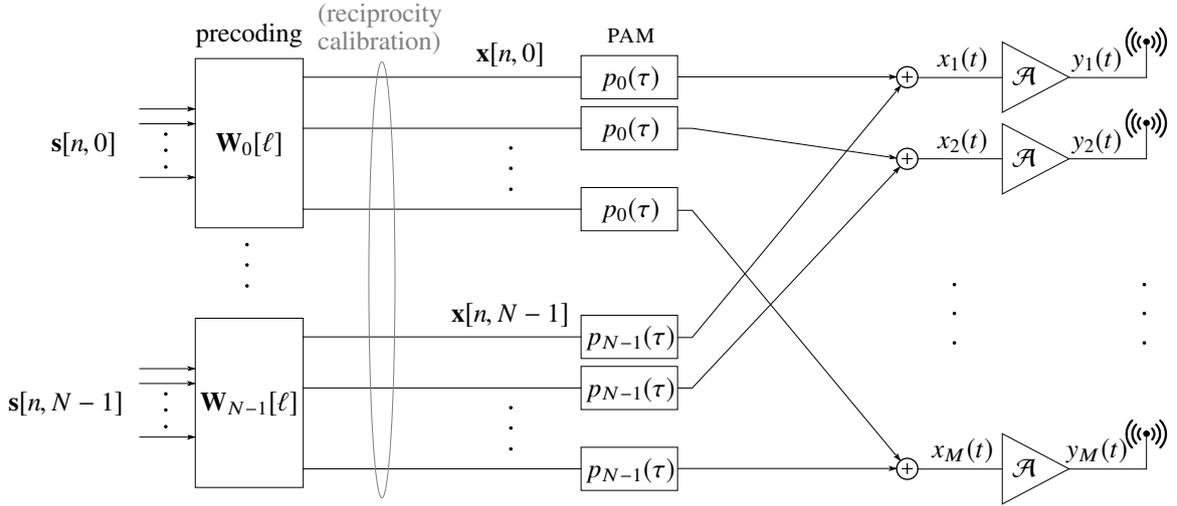
		\caption{A block diagram of the multi-carrier transmitter.  In a single-carrier transmitter, there is only one branch $N = 1$.  The position of the reciprocity filter discussed in Section~\ref{sec:reciprocity_filter1923883902} is marked in grey.}
		\label{fig:system_model9109823491}
	\end{figure*}

The transmission from an array with $M$ antennas is studied.  The
block diagram in Figure~\ref{fig:system_model9109823491} shows the
transmitter that will be explained in this section.  The input and
output signals to the amplifier at antenna $m$ is denoted by $x_m(t)$
and $y_m(t)$ respectively.  By denoting the operation of the amplifier
$\symcal{A}$, and the amplified transmit signal is then: \begin{align}
y_m(t) = \symcal{A}\left(x_m(t)\right).  \end{align} For later use,
the following vector notation is
introduced: \begin{align} \symbf{x}(t)
&\triangleq \bigl(x_1(t), \ldots, x_M(t)\bigr)^\tr,\\ \symbf{y}(t)
&\triangleq \bigl(y_1(t), \ldots, y_M(t)\bigr)^\tr.  \end{align}
	
	The signal received at location $\symfrak{x}$ is given by
	\begin{align}\label{eq:rx_signal1929292}
		r_\symfrak{x}(t) = \sqrt{\beta_\symfrak{x}} \int_{-\infty}^{\infty} \symbf{h}^\tr_\symfrak{x}(\tau) \symbf{y}(t - \tau) \symrm{d}\tau,
	\end{align}
	where the $M$\mbox{-}dimensional impulse response $\symbf{h}_\symfrak{x}(\tau)$ models the small-scale fading from the array to location $\symfrak{x}$ and $\beta_\symfrak{x} \in \symbb{R}^+$ models the large-scale fading, i.e.\ the slowly changing signal attenuation due to distance and shadowing.  In a real system, the received signal will be corrupted by thermal noise, which commonly is modeled as an additional noise term in \eqref{eq:rx_signal1929292}.  The noise term is neglected as it has no impact on the distortion.
	
	\subsection{Multi-Carrier Transmission}\label{sec:OFDM_transmission18923y64}
	It is assumed that the analog transmit signal $x_m(t)$ is generated from pulse amplitude modulation.  In multi-carrier transmission, $N$ pulses $p_\nu(\tau)$ are used to modulate the digital signals $x_m[n, \nu]$, where $n$ is a time index and $\nu \in \{0,\ldots,N-1\}$ the index of the pulse.  The complex baseband representation of the analog transmit signal is given by:
	\begin{align}\label{eq:10981238844}
		x_m(t) = \frac{1}{\sqrt{N}} \sum_{\nu=0}^{N-1} \sum_{n=-\infty}^{\infty} x_m[n,\nu] p_\nu(t - nT + \Psi),
	\end{align}
	where $T$ is the symbol duration, and $\Psi$ a random variable, which is independent of all other sources of randomness and uniformly distributed on the interval $[{0,T}[$, that is introduced to make the transmit signals stationary.  For later use, the vector notation $\symbf{x}[n,\nu] \triangleq \bigl(x_1[n,\nu], \ldots, x_M[n,\nu]\bigr)^\tr$ is introduced.
	
	The array serves $K$ users whose channel impulse responses and large-scale fading are denoted by $\{\symbf{h}_k(\tau)\}$ and $\{\beta_k\}$, where $k = 1, \ldots, K$ is the user index.  The receive filters employed by the users are assumed to be matched to the transmit pulses $\{p_\nu(\tau)\}$ of the pulse amplitude modulation.  The effect of the channel and how the symbols transmitted with pulse $\nu$ affect the signals received through receive filter $\nu'$ is given by the impulse response:
	\begin{align}\label{eq:channel1929283}
		\symbf{h}_k[\ell, \nu, \nu'] \triangleq \Bigl(p_\nu(\tau) \star \symbf{h}_{\symfrak{x}_k}(\tau) \star p_{\nu'}^*(-\tau)\Bigr)(\ell T), \quad k=1,\ldots,K,
	\end{align}
	where $\star$ denotes convolution and $\ell$ is the tap index.  
	
	For common \OFDM \cite{proakis2002communication, stuber2001principles}, the pulses in \eqref{eq:10981238844}, or \emph{subcarriers} in the jargon of \OFDM, are given by:
	\begin{align}\label{eq:pam9283ds25676}
	p_\nu(\tau) = \left( p(t) e^{j 2 \pi t \nu f_0} \star z(t) \right)(\tau), \quad \nu = 0, \ldots, N-1,
	\end{align}
	where $f_0$ is the subcarrier spacing and $p(\tau)$ is a common pulse shape.  Since the pulse $p(\tau)$ usually is chosen as a time-limited pulse shape, the transmit signal is not strictly bandlimited.  To mitigate the out-of-band radiation caused by the pulse, different types of sidelobe suppression methods can be applied.  Here, the filter $z(\tau)$ is used to limit the frequency content of transmit signal to a given frequency band.  It will be chosen as an ideal lowpass filter in the examples shown in later sections to make the transmit signal strictly bandlimited prior to amplification.  Note that the pulse $p(\tau)$ is a baseband signal and that the low-pass filter $z(\tau)$ is the same for all pulses, e.g.\ its cutoff frequency does not depend on the subcarrier index.  In the theoretical analysis, however, we assume that pulses are unaffected by the filter, i.e.\ that $z(\tau)  = \delta(\tau)$ is an all-pass filter.  
	
	Besides filtering, which is discussed in \cite{faulkner2000effect}, there are other ways to suppress the sidelobes.  For example, sidelobes can be suppressed by precoding the symbols and making the subcarriers correlated \cite{cosovic2006subcarrier, tom2013mask} and by using pulses other than rectangular \cite{tan2004reduced}.  The different sidelobe suppression methods differ a bit in the amount of intercarrier interference they cause or how much spectral resources they occupy, the effect on the spectrum and the amplifier distortion, which is the main focus of this paper, is similar to the effect of filtering the signal by an ideal low-pass filter however.  For the sake of clarity and generality of the discussion and not to rely on any specific sidelobe suppression technique, we therefore use a low-pass filter when sidelobe suppression is discussed.
	
	To avoid interference, the pulse $p(\tau)e^{j2\pi \tau \nu f_0}$ has to be orthogonal to all other pulses $\{p(\tau - nT)e^{j2\pi \tau \nu' f_0}, (n,\nu') \neq (0, \nu)\}$ and their time shifts.  The rectangle pulse is one choice that fulfills the orthogonality requirement and that also achieves the smallest possible subcarrier spacing $f_0 = 1/T$:
	\begin{align}\label{eq:rect_pulse}
		p(\tau) = \frac{1}{\sqrt{T}} \operatorname{rect}\left(\frac{\tau}{T}\right),
	\end{align}
	where $\operatorname{rect}(\tau) = 1$ when $0 \leq \tau < 1$ and zero otherwise.  Other pulse shapes can also be used, but they would require a larger subcarrier spacing $f_0$ for the same symbol period $T$, which reduces the amount of subcarriers that fit in a given frequency band.
	
	A cyclic prefix that is longer than the delay spread of the channel is assumed.  It ensures that there is no intersymbol interference when pure \OFDM, $z(\tau) = \delta(\tau)$, is used.  When a sufficiently long cyclic prefix is inserted in the transmission and removed in the detection, the effective channel coefficient of the signal transmitted on subcarrier $\nu$ is given by
	\begin{align}
		\symbf{h}_k[\nu] \triangleq \sum_{\ell=0}^{\infty} \symbf{h}_k[\ell,\nu,\nu],
	\end{align}
	where $\{\symbf{h}_k[\ell,\nu,\nu']\}$ are given in \eqref{eq:channel1929283}.  Note that $\sum_{\ell} \symbf{h}_k[\ell,\nu,\nu'] = 0$ for $\nu\neq\nu'$ with a cyclic prefix.  No notation for the impulse responses for which $\nu\neq\nu'$ is therefore introduced.
	
	The data symbol that is to be transmitted on subcarrier $\nu$ to user $k$ is denoted $s_{k}[n,\nu]$ and its power is normalized such that $\Exp[|s_{k}[n,\nu]|^2] = 1$.  Since the effective channel $\symbf{h}_k[\nu]$ of a given subcarrier is frequency flat, the frequency response of the $\nu$\mbox{-}th subcarrier is constant $\symbfsf{H}_\nu[\theta] \triangleq (\symbf{h}_1[\nu], \ldots, \symbf{h}_K[\nu])^\tr$ over the normalized frequency $\theta$.  The data symbols $\symbf{s}[n, \nu] = \bigl(s_1[n,\nu], \ldots, s_K[n,\nu]\bigr)^\tr$ are precoded individually for each subcarrier by the precoder $\symbf{W}_\nu[0] = w(\symbfsf{H}_\nu[\theta])$, which is frequency flat and a function \mbox{$w:\symbb{C}^{K\times M} \to \symbb{C}^{M\times K}$} of the channel.  Some common precoders will be defined at the end of this section.  The digital signals for subcarrier $\nu$ are, therefore, given by:
	\begin{align}\label{eq:precoding9283838}
	\symbf{x}[n,\nu] = \symbf{W}_\nu[0] \symbf{D}^{1/2}_{\symbf{\xi}} \symbf{s}[n,\nu].
	\end{align}
	The diagonal matrix $\symbf{D}_{\symbf{\xi}} \triangleq \operatorname{diag}(\xi_1, \ldots, \xi_K)$ contains the relative power allocations of each user, which are chosen such that
	\begin{align}
		\sum_{k=1}^{K} \xi_k \leq 1.
	\end{align}

	\subsection{Single-Carrier Transmission}
	The use of just one pulse in \eqref{eq:pam9283ds25676}, i.e.\ $N=1$, is called \emph{single-carrier transmission}.  Since using one pulse over the same effective bandwidth as multi-carrier transmission gives a much shorter symbol period,  the relative time duration of the pulse can be made longer, which means that bandlimited pulses are feasible, e.g.\ a root-raised cosine can be used.
	
	The impulse response of the discrete-time channel is given by $\symbf{h}_k[\ell,0,0]$, in the same notation as in \eqref{eq:channel1929283}, and the frequency response at the normalized frequency $\theta$ is given by:
	\begin{align}\label{eq:SC_channel}
		\symbfsf{h}_k(\theta) \triangleq \sum_{\ell=-\infty}^{\infty} \symbf{h}_k[\ell,0,0] e^{-j2\pi\theta\ell}.
	\end{align}
	Note that the impulse response of the continuous-time channel has a finite support $\sigma_\tau$, sometimes referred to as \emph{delay spread}, and that the pulse $p_\nu(\tau)$ quickly falls off to zero.  The sum in \eqref{eq:SC_channel} therefore has a finite number $L \triangleq \sigma_\tau / T$ of significant terms.
	
	The data symbols $\symbf{s}[n, 0] \triangleq (s_1[n,0], \ldots, s_K[n,0])^\tr$ are precoded and the discrete-time transmit signal is given by:
	\begin{align}\label{eq:precoding_SC_192992}
		\symbf{x}[n,0] = \left( \symbf{W}_0[\ell] \star \symbf{D}^{1/2}_{\symbf{\xi}} \symbf{s}[\ell,0] \right)[n],
	\end{align}
	where the impulse response of the frequency-selective precoder is:
	\begin{align}
		\symbf{W}_0[\ell] \triangleq \int_{0}^{1} \symbfsf{W}_0[\theta] e^{j2\pi\ell\theta} \symrm{d}\theta.
	\end{align}
	The precoder $\symbfsf{W}_0[\theta] = w(\symbfsf{H}_0[\theta])$ is chosen as a function of the channel $\symbfsf{H}_0[\theta] \triangleq (\symbfsf{h}_1(\theta), \ldots, \symbfsf{h}_K(\theta))^\tr$.    
	
	By inserting a cyclic prefix also in the single-carrier transmission, i.e.\ by letting
	\begin{align}
	\symbf{x}[n,0] = \symbf{x}[n+N,0], \quad\text{for } n < 0,
	\end{align}
	and viewing the transmission only during the symbol periods $n = 0, \ldots, N - 1$, the received signal in \eqref{eq:rx_signal1929292} can be seen as a circular convolution.  This can simplify the equalization since it can be done in the frequency domain symbol-per-symbol as if the individual channels were frequency flat.  This is the idea used in single-carrier transmission with frequency-domain equalization \cite{falconer2002frequency} and another type of \OFDM \cite{li1997effects, tse2005fundamentals} (referred to as \OFDM type~2 here), both of whose transmit signals can be described in this framework.  
	
\begin{figure}
		\centering
		\def\sincsqr[#1]{(sin((#1)*180)/((#1)*pi))^2}
		\def\sinc[#1]{sin((#1)*180)/((#1)*pi)}
		\newsavebox{\graphone}
		\newsavebox{\graphtwo}
		\savebox\graphone{\begin{tikzpicture}
			\begin{axis}[
			xlabel={normalized frequency $fT$},
			ylabel={\rule{0pt}{0pt}\clap{power spectral density [dB]}\rule{0pt}{0pt}},
			xmin=-10,
			xmax=100,
			ymin=-60,
			y filter/.code={\pgfmathparse{10*log10(\pgfmathresult)}},
			]
			\addplot[no marks, samples at={-10.0001, -9.9376,..., 100}] {\sincsqr[x-60]};
			\addplot[densely dotted, line width=.4pt, line join=round, no marks, samples at={-10.0001, -9.9001,..., 100}] {\sincsqr[x] + \sincsqr[x-1] + \sincsqr[x-2] + \sincsqr[x-3] + \sincsqr[x-4] + \sincsqr[x-5] + \sincsqr[x-6] + \sincsqr[x-7] + \sincsqr[x-8] + \sincsqr[x-9] + \sincsqr[x-10] + \sincsqr[x-11] + \sincsqr[x-12] + \sincsqr[x-13] + \sincsqr[x-14] + \sincsqr[x-15] + \sincsqr[x-16] + \sincsqr[x-17] + \sincsqr[x-18] + \sincsqr[x-19] + \sincsqr[x-20] + \sincsqr[x-21] + \sincsqr[x-22] + \sincsqr[x-23] + \sincsqr[x-24] + \sincsqr[x-25] + \sincsqr[x-26] + \sincsqr[x-27] + \sincsqr[x-28] + \sincsqr[x-29] + \sincsqr[x-30] + \sincsqr[x-31] + \sincsqr[x-32] + \sincsqr[x-33] + \sincsqr[x-34] + \sincsqr[x-35] + \sincsqr[x-36] + \sincsqr[x-37] + \sincsqr[x-38] + \sincsqr[x-39] + \sincsqr[x-40] + \sincsqr[x-41] + \sincsqr[x-42] + \sincsqr[x-43] + \sincsqr[x-44] + \sincsqr[x-45] + \sincsqr[x-46] + \sincsqr[x-47] + \sincsqr[x-48] + \sincsqr[x-49] + \sincsqr[x-50] + \sincsqr[x-51] + \sincsqr[x-52] + \sincsqr[x-53] + \sincsqr[x-54] + \sincsqr[x-55] + \sincsqr[x-56] + \sincsqr[x-57] + \sincsqr[x-58] + \sincsqr[x-59] + \sincsqr[x-60] + \sincsqr[x-61] + \sincsqr[x-62] + \sincsqr[x-63]};
			\end{axis}
			\end{tikzpicture}}
		\savebox\graphtwo{\begin{tikzpicture}
			\begin{axis}[
			xlabel={normalized frequency $fT$},
			ylabel={\rule{0pt}{0pt}\clap{power spectral density [dB]}\rule{0pt}{0pt}},
			xmin=-.15625,
			xmax=1.5625,
			ymin=-60,
			y filter/.code={\pgfmathparse{10*log10(\pgfmathresult)}},
			]
			\addplot[no marks, samples at={0.00001,0.000400625,...,1}] {(\sinc[(64*x-60)]+\sinc[(64*x+4)])^2};
			\addplot[no marks, domain=-1000:0, samples=2] {0};
			\addplot[no marks, domain=1:1000, samples=2] {0};
			\addplot[densely dotted, no marks, line width=.4pt, domain=64:1000] coordinates {(0,.000001) (0,1) (1,1) (1,.000001)};
			\end{axis}
			\end{tikzpicture}}
		
		\ifdraft
		\rule{0pt}{0pt}\clap{\usebox{\graphone}\usebox{\graphtwo}}\rule{0pt}{0pt}
		\else
		\usebox{\graphone}
		\usebox{\graphtwo}
		\fi
		\caption{The power spectral density of \OFDM with rectangular pulses and \OFDM type~2 with a cardinal sine pulse $p(\tau) = \operatorname{sinc}(\tau / T)$.  The dotted curve shows the signal with 64 active tones, and the solid the signal with a single non-zero tone.}
		\label{fig:one_tone_OFDM_spect}
	\end{figure}
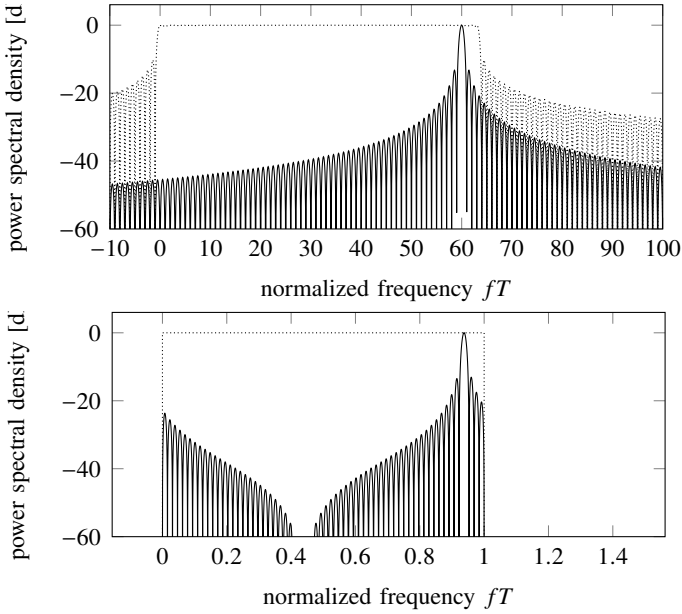

	To demonstrate the difference between \OFDM and \OFDM type~2, an example of their power spectral densities is shown in Figure~\ref{fig:one_tone_OFDM_spect}.  In contrast to the subcarriers in \OFDM, which have spectra with infinite bandwidth, the spectrum in \OFDM type~2 is bandlimited.  The spectrum of a signal from \OFDM type~2, where all frequency-domain symbols, except one, are set to zero, consists of a cardinal sine that is aliased and windowed.  A rigorous description of the two types of \OFDM is given in \cite{stuber2001principles}.  When a system with linear hardware is studied without consideration of the bandwidth of the actual continuous-time signal, the distinction between \OFDM and \OFDM type~2 is of little importance, since both transmission methods result in $N$ parallel, interference-free channels.  Here, we make the distinction because the sidelobe levels of the former has to be taken into account when studying the out-of-band radiation.
	
	\subsection{Common Precoders}
	Common precoders are the maximum-ratio, the zero-forcing and the $\lambda$\mbox{-}regularized zero-forcing precoder.  They are given by the following expressions in the same order:
	\begin{align}
		w(\symbfsf{H}) &= \alpha \symbfsf{H}^\conjtr,\label{eq:MR_919192}\\
		w(\symbfsf{H}) &= \alpha \symbfsf{H}^\conjtr \left(\symbfsf{H}\symbfsf{H}^\conjtr\right)^{-1},\label{eq:ZF_919192}\\
		w(\symbfsf{H}) &= \alpha \symbfsf{H}^\conjtr \left(\symbfsf{H} \symbfsf{H}^\conjtr + \lambda \symbf{I}_K\right)^{-1},\label{eq:RZF_919192}
	\end{align}
	where $\alpha$ is a constant used for power normalization that is chosen such that
	\begin{align}\label{eq:power_constraint129933}
		\sum_{m=1}^{M} \Exp\left[\left| \bigl[w(\symbfsf{H})\bigr]_{m,k} \right|^2\right] = \frac{1}{N}, \quad \forall k.
	\end{align}
	The regularization parameter $\lambda$ is used to obtain a balanced performance between array gain and interference suppression.  An introduction to different linear precoding techniques is given in \cite{marzetta2016fundamentals}.
	
	\section{Nonlinear Amplification}\label{sec:9829919992}
	In this section, the cross-correlation of the amplified transmit signals will be derived.  The amplified signals will also be partitioned into a desired term and distortion that is uncorrelated to the desired term.   The transfer function of the nonlinear amplifier is modeled by a memory polynomial \cite{ghannouchi2009behavioral} of order $\Pi$ with kernels $\{b_\varpi(\tau)\}$, where the amplifier output is assumed to be given by
	\begin{align}\label{eq:polynomial_model891291}
		\symcal{A}\left(x(t)\right) = \sum_{\varpi\in[1,\Pi]:\text{odd}} \int_{-\infty}^{\infty} b_{\varpi}(t - \tau) x(\tau) \left| x(\tau) \right|^{\varpi - 1} \symrm{d}\tau.
	\end{align}
	It is noted that only odd powers are included in the sum.  The model is the baseband representation of a special case of the general Volterra model \cite{schetzen1980volterra, mollen2017nonlinearInThesis}, where the off-diagonal kernels are set to zero.  
	
	Because of multiple carriers in \eqref{eq:10981238844}, multiuser precoding in \eqref{eq:precoding9283838} and \eqref{eq:precoding_SC_192992}, and the central limit theorem, the distribution of the digital transmit signals $\symbf{x}[n,\nu]$ is close to circularly symmetric Gaussian.  Note that this is true independently of whether \OFDM or single-carrier transmission is used and independently of the order of the symbol constellations used for the data symbols $\symbf{s}[n,\nu]$ when either the number of users or number of filter taps in the precoding is large \cite{mollen2016waveforms}.  We therefore assume that the digital transmit signals, and thus the analog transmit signals, are circularly symmetric Gaussian.
		
	To facilitate the analysis of the second-order statistics of the amplifier output, the following subset of the complex Itô generalization of the Hermite polynomials \cite{ito1952complex, ismail2015complex}:
	\begin{align}
	H_\varpi(x) \triangleq \sum_{i=0}^{\frac{\varpi-1}{2}} (-1)^i i! \binom{\frac{\varpi+1}{2}}{i} \binom{\frac{\varpi-1}{2}}{i} x |x|^{\frac{\varpi-1}{2}-i},\quad \varpi = 1, 3, 5, \ldots,
	\end{align}
	is used to rephrase the polynomial model as a Hermite expansion \cite{mollen2017nonlinearInThesis}:
	\begin{align}\label{eq:hermite_model}
		\symcal{A}\left(x(t)\right) = \sum_{\varpi\in[1,\Pi]:\text{odd}} \int_{-\infty}^{\infty} a_{\varpi}(t - \tau) \sigma_x^\varpi H_\varpi\left(\frac{x(\tau)}{\sigma_{x}}\right) \symrm{d}\tau,
	\end{align}	
	where $\sigma_x$ is the square root of the power of $x(t)$.  It is noted that the input signal has been rescaled by $1/\sigma_x$ so that the argument to the polynomial has unit power.  Furthermore the kernels $\{a_\varpi(\tau)\}$ are normalized by $\sigma_x^\varpi$.  This is to make the expressions that will be derived in the following sections of the paper easier to write.
	
	The new kernels $\{a_\varpi(\tau)\}$ are given as linear combinations of the original kernels $\{b_\varpi(\tau)\}$.  For example when $\Pi = 9$, the kernels for antenna $m$ in the system that we study are given by:
	\begin{align}
		a_{1m}(\tau) &= b_{1}(\tau) {+} 2 \sigmaxm^2 \! b_{3}(\tau) {+} 6 \sigmaxm^4 \! b_{5}(\tau) {+} 24 \sigmaxm^6 \! b_{7}(\tau) {+} 120 \sigmaxm^8 \! b_{9}(\tau)\\
		a_{3m}(\tau) &= b_{3}(\tau) + 6 \sigmaxm^2 b_{5}(\tau) + 36 \sigmaxm^4 b_{7}(\tau) + 240 \sigmaxm^6 b_{9}(\tau)\label{eq:81183368292}\\
		a_{5m}(\tau) &= b_{5}(\tau) + 12 \sigmaxm^2 b_{7}(\tau) + 120 \sigmaxm^4 b_{9}(\tau)\\
		a_{7m}(\tau) &= b_{7}(\tau) + 20 \sigmaxm^2 b_{9}(\tau)\\
		a_{9m}(\tau) &= b_{9}(\tau),
	\end{align}
	where $\sigmaxm^2$ is the power of the transmit signal $x_m(t)$.  This is easily obtained from Table~\ref{tab:hermite_polynomials} and \ref{tab:hermite_expansion_of_polynomials}, where a few of the Hermite functions are given.
	
	\begin{table}
		\centering
		\ifdraft
		\def\arraystretch{.65}
		\hspace{-1em}\mbox{\begin{minipage}{.5\linewidth}
			\caption{Complex Itô generalization of the Hermite polynomials\label{tab:hermite_polynomials}}
			\begin{tabular}{r@{\,}c@{\,}l}
				$H_1(x)$ & $=$ & $x$ \\
				$H_3(x)$ &$ = $&$x|x|^2 - 2x$\\
				$H_5(x)$ &$=$& $x|x|^4 - 6x|x|^2 + 6x$\\
				$H_7(x)$ & $=$ & $x|x|^6 -12 x|x|^4 + 36 x|x|^2 -24 x$\\
				$H_9(x)$ & $=$ & $x|x|^8 - 20 x|x|^6 + 120 x|x|^4 - 240 x|x|^2 + 120 x$\\
			\end{tabular}
		\end{minipage}%
		\begin{minipage}{.5\linewidth}
			\caption{Complex polynomials as generalized Hermite polynomials\label{tab:hermite_expansion_of_polynomials}}
			\begin{tabular}{r@{\,}c@{\,}l}
				$x$ &$=$ &$H_1(x)$\\
				$x|x|^2$ &$=$&$ H_3(x) + 2 H_1(x)$\\
				$x|x|^4$&$ = $&$H_5(x) + 6H_3(x) + 6 H_1(x)$\\
				$x|x|^6$ & $=$ & $H_7(x) + 12 H_5(x) + 36 H_3(x) + 24 H_1(x)$\\
				$x|x|^8$ & $=$ & $H_9(x) + 20 H_7(x) + 120 H_5(x) + 240 H_3(x) +120 H_1(x)$\\
			\end{tabular}
		\end{minipage}}
		\else
		\caption{Complex Itô generalization of the Hermite polynomials\label{tab:hermite_polynomials}}
		\begin{tabular}{r@{\,}c@{\,}l}
			$H_1(x)$ & $=$ & $x$ \\
			$H_3(x)$ &$ = $&$x|x|^2 - 2x$\\
			$H_5(x)$ &$=$& $x|x|^4 - 6x|x|^2 + 6x$\\
			$H_7(x)$ & $=$ & $x|x|^6 -12 x|x|^4 + 36 x|x|^2 -24 x$\\
			$H_9(x)$ & $=$ & $x|x|^8 - 20 x|x|^6 + 120 x|x|^4 - 240 x|x|^2 + 120 x$\\
			&$\vdots$ &\\
		\end{tabular}
		
		\vspace{\floatsep}
		
		\caption{Complex polynomials as generalized Hermite polynomials\label{tab:hermite_expansion_of_polynomials}}
		\begin{tabular}{r@{\,}c@{\,}l}
			$x$ &$=$ &$H_1(x)$\\
			$x|x|^2$ &$=$&$ H_3(x) + 2 H_1(x)$\\
			$x|x|^4$&$ = $&$H_5(x) + 6H_3(x) + 6 H_1(x)$\\
			$x|x|^6$ & $=$ & $H_7(x) + 12 H_5(x) + 36 H_3(x) + 24 H_1(x)$\\
			$x|x|^8$ & $=$ & $H_9(x) + 20 H_7(x) + 120 H_5(x) + 240 H_3(x) +120 H_1(x)$\\
			& $\vdots$&\\
		\end{tabular}
		\fi
	\end{table}		

	The Hermite functions are orthogonal in the sense that, for two jointly Gaussian random variables $X, Y  \sim \CN(0,1)$, the following holds \cite{mollen2017nonlinearInThesis}:
	\ifdraft
	\begin{align}\label{eq:hermite_property}
		\Exp\bracket*{H_\varpi(X) H^*_{\varpi'}(Y)} = \left(\frac{\varpi+1}{2}\right)!\, \left(\frac{\varpi-1}{2}\right)!\, \Exp\bracket*{XY^*} \left|\Exp\bracket*{XY^*}\right|^{\varpi-1} \delta[\varpi-\varpi'].
	\end{align}
	\else
	\begin{align}
		&\Exp\bracket*{H_\varpi(X) H^*_{\varpi'}(Y)}\notag\\
		&\quad= \left(\frac{\varpi+1}{2}\right)!\, \left(\frac{\varpi-1}{2}\right)!\, \Exp\bracket*{XY^*} \left|\Exp\bracket*{XY^*}\right|^{\varpi-1} \delta[\varpi-\varpi'].\label{eq:hermite_property}
	\end{align}
	\fi
	Thus, all the terms in the Hermite expansion in \eqref{eq:hermite_model} are mutually orthogonal.  The amplified signal can therefore be partitioned as:
	\begin{align}\label{eq:partitioning192893881}
	y_m(t) = u_m(t) + d_m(t),
	\end{align}
	where the linear term $u_m(t)$ and the distortion $d_m(t)$ are given by:
	\begin{align}
	u_m(t) &\triangleq \int_{-\infty}^{\infty} a_{1m}(t-\tau) x_m(\tau) \symrm{d}\tau,\\
	d_m(t) &\triangleq \sum_{\varpi\in[3,\Pi]:\text{odd}} \int_{-\infty}^{\infty} a_{\varpi m}(t-\tau) \sigma^\varpi_{x_m} H_\varpi\left(\frac{x_m(\tau)}{\sigma_{x_m}}\right)\symrm{d}\tau.
	\end{align}
	By virtue of the orthogonality property in \eqref{eq:hermite_property} and because a convolution is a deterministic linear transformation and all moments of $x_m$ are finite, these two terms are uncorrelated:
	\begin{align}
		\Exp\left[u_m(t)d_{m'}^*(t-\tau)\right] = 0, \quad\forall m, m', \tau.
	\end{align}
	The partitioning in \eqref{eq:partitioning192893881} can also be obtained using Bussgang's theorem.  The Hermite expansion, however, simplifies the derivation of the cross-correlation of the output signals, which is easily obtained from the orthogonality property in \eqref{eq:hermite_property}.  
	
	If the input signals are Gaussian stationary random processes with cross-correlations
	\begin{align}
		R_{x_mx_{m'}}(\tau) &\triangleq \Exp\left[x_m(t) x^*_{m'}(t-\tau)\right],
	\end{align}
	then the amplified signals are weak-sense stationary processes, whose cross-correlations are:
	\begin{align}
		R_{y_my_{m'}}(\tau) &\triangleq \Exp\bracket*{y_m(t) y^*_{m'}(t-\tau)} \\
		&= \sum_{\varpi\in[1,\Pi]:\text{odd}} \left( a_{\varpi m}(t) \star a^*_{\varpi m'}(-t) \star R^{(\varpi)}_{x_m x_{m'}}(t) \right)(\tau),
	\end{align}
	where the individual cross-correlations are
	\begin{align}
		R^{(\varpi)}_{x_m x_{m'}}(\tau) = \left(\frac{\varpi+1}{2}\right)!\, \left(\frac{\varpi-1}{2}\right)!\, R_{x_mx_{m'}}(\tau) \left|R_{x_mx_{m'}}(\tau)\right|^{\varpi-1}.
	\end{align}
	Equivalently, these expressions can be studied in the frequency domain in terms of the cross-spectrum $S_{x_mx_{m'}}(f)$, the Fourier transform of the cross-correlation $R_{x_mx_{m'}}(\tau)$.  The cross-spectra of the amplified signals are given by:
	\begin{align}\label{eq:amp_sig_psd}
		S_{y_my_{m'}}(f) = \sum_{\varpi\in[1,\Pi]:\text{odd}} A_{\varpi m}(f) A_{\varpi m'}^*(f) S^{(\varpi)}_{x_m x_{m'}}(f),
	\end{align}
	where $\{A_{\varpi m}(f)\}$ are the Fourier transforms of the kernels and the individual cross-spectra:
	\ifdraft
	\begin{align}
		S^{(\varpi)}_{x_m x_{m'}}(f) = \left(\frac{\varpi{+}1}{2}\right)!\, \left(\frac{\varpi{-}1}{2}\right)!\, \Bigl( \smash[b]{\underbrace{S_{x_mx_{m'}}(\varphi) \star \cdots \star S_{x_mx_{m'}}(\varphi)}_{\frac{\varpi+1}{2}\,\text{factors}} \star \underbrace{S^*_{x_mx_{m'}}(-\varphi) \star \cdots \star S^*_{x_mx_{m'}}(-\varphi)}_{\frac{\varpi - 1}{2}\,\text{factors}} \Bigr)(f)}.
	\end{align}
	\else
	\begin{align}
		S^{(\varpi)}_{x_m x_{m'}}(f) &= \left(\frac{\varpi+1}{2}\right)!\, \left(\frac{\varpi-1}{2}\right)!\, \Bigl( \underbrace{S_{x_mx_{m'}}(\varphi) \star \cdots \star S_{x_mx_{m'}}(\varphi)}_{\frac{\varpi+1}{2}\,\text{factors}}\notag\\
		&\quad \star \underbrace{S^*_{x_mx_{m'}}(-\varphi) \star \cdots \star S^*_{x_mx_{m'}}(-\varphi)}_{\frac{\varpi - 1}{2}\,\text{factors}} \Bigr)(f).
	\end{align}
	\fi
	It also follows that the spectral densities of the linearly amplified signal and of the uncorrelated distortion terms in \eqref{eq:partitioning192893881} are given by:
	\begin{align}
		S_{u_mu_{m'}}(f) &= A_{1 m}(f) A_{1 m'}^*(f) S_{x_mx_{m'}}(f)\label{eq:918183}\\
		S_{d_md_{m'}}(f) &= \sum_{\varpi\in[3,\Pi]:\text{odd}} A_{\varpi m}(f) A_{\varpi m'}^*(f) S^{(\varpi)}_{x_mx_{m'}}(f).
	\end{align}
	
	\section{Reciprocity Calibration}\label{sec:reciprocity_filter1923883902}
	In massive \MIMO, the full channel is only estimated in the uplink.  For the downlink, the uplink channel estimate is used for the precoding and any differences between the uplink and downlink channels are adjusted for by a calibration filter after the precoding.  Because the difference between the uplink and downlink channels mostly stems from difference in the hardware of the transmitter chains, the calibration filter can be computed based on calibration pilots that are sent from each antenna and received by the other antennas of the array, which is the common approach to learning the calibration filter \cite{vieira2017reciprocity}.  Assuming that the amplifier is the dominant source of the reciprocity error, we, here, propose to compute the reciprocity filter by using the Hermite expansion of the amplifier nonlinearity.  The influence of the amplifiers on the downlink channel is described by the linear impulse response $a_{1m}(\tau)$.  With knowledge of the amplifier characteristics, the reciprocity filter thus can be computed without transmitted calibration pilots.

	
	\section{Radiated Power Spectral Density Pattern}
	The vector $\symbf{s}[n,\nu]$ consists of the $K$ symbols that are transmitted to the users at time $n$ on pulse $\nu$.  It is assumed that the data symbols $\symbf{s}[n,\nu]$ are circularly symmetric and i.i.d.\ over both $n$ and $\nu$:
	\begin{align}
		\Exp\bracket*{\symbf{s}[n,\nu] \symbf{s}^\conjtr[n',\nu']} = \delta[n-n'] \delta[\nu-\nu'] \symbf{I}_K, \quad \forall n, n', \nu, \nu'.
	\end{align}
	The precoded digital transmit signal of the $\nu$\mbox{-}th pulse, then has the power spectral density given by	
	\begin{align}\label{eq:digital_tx_PSD199283}
		\symbf{S}^{(\nu)}_{\symbf{xx}}[\theta] = 
		\symbfsf{W}_\nu[\theta] \symbf{D}_{\symbf{\xi}} \symbfsf{W}_\nu^\conjtr[\theta],
	\end{align}
	where 
	the frequency response of the precoder is
	\begin{align}
		\symbfsf{W}_\nu[\theta] \triangleq \sum_{\ell=-\infty}^\infty \symbf{W}_\nu[\ell] e^{-j2\pi\ell\theta}.
	\end{align}
	Note that, in the case of \OFDM, the frequency responses are flat and constant over $\theta$.
	
	The pulse-amplitude modulated analog transmit signal has the operational power spectral density
	\begin{align}\label{eq:PAM_tx_signal_PSD9919001}
		\symbf{S}_{\symbf{xx}}(f) = \frac{1}{NT} \sum_{\nu=0}^{N-1} \left| \symsfit{p}_\nu(f) \right|^2 \symbf{S}^{(\nu)}_{\symbf{xx}}[fT],
	\end{align}
	where $\symsfit{p}_\nu(f) \triangleq \int_{-\infty}^{\infty} p_\nu(\tau)e^{-j2\pi\tau t}\symrm{d}\tau$ is the Fourier transform of pulse $p_\nu(\tau)$.  It is assumed that all pulses have the same energy.  The normalization by $N$ is to ensure that the power is the same, independently of the number of pulses.  
	
	The power spectral density of the amplified transmit signal that was given in \eqref{eq:amp_sig_psd} is written in matrix notation as follows:
	\begin{align}
		\symbf{S}_{\symbf{yy}}(f) = \sum_{\varpi\in[1,\Pi]:\text{odd}} \symbf{A}^\conjtr_\varpi(f) \symbf{S}^{(\varpi)}_{\symbf{xx}}(f) \symbf{A}_\varpi(f),
	\end{align}
	where $\symbf{A}_\varpi(f) \triangleq \operatorname{diag}(A_{\varpi 1}(f), \ldots, A_{\varpi M}(f))$, the $\varpi$\mbox{-}th order modulation term is given by
	\ifdraft
	\begin{align}\label{eq:020011992915}
	\symbf{S}^{(\varpi)}_{\symbf{xx}}(f) = \left(\frac{\varpi+1}{2}\right)!\, \left(\frac{\varpi-1}{2}\right)!\, \smash[b]{\Bigl( \underbrace{\symbf{S}_{\symbf{xx}}(\varphi) \circledstar \cdots \circledstar \symbf{S}_{\symbf{xx}}(\varphi)}_{\frac{\varpi+1}{2}\,\text{factors}} \circledstar \underbrace{\symbf{S}^*_{\symbf{xx}}(-\varphi) \circledstar \cdots \circledstar \symbf{S}^*_{\symbf{xx}}(-\varphi)}_{\frac{\varpi - 1}{2}\,\text{factors}} \Bigr)(f)},
	\end{align}
	\else
	\begin{align}
		\symbf{S}^{(\varpi)}_{\symbf{xx}}(f) &= \left(\frac{\varpi+1}{2}\right)!\, \left(\frac{\varpi-1}{2}\right)!\, \Bigl( \underbrace{\symbf{S}_{\symbf{xx}}(\varphi) \circledstar \cdots \circledstar \symbf{S}_{\symbf{xx}}(\varphi)}_{\frac{\varpi+1}{2}\,\text{factors}} \notag\\
		&\quad\circledstar \underbrace{\symbf{S}^*_{\symbf{xx}}(-\varphi) \circledstar \cdots \circledstar \symbf{S}^*_{\symbf{xx}}(-\varphi)}_{\frac{\varpi - 1}{2}\,\text{factors}} \Bigr)(f),\label{eq:020011992915}
	\end{align}
	\fi
	and $\circledstar$ denotes elementwise convolution.  Since the diagonal elements describe the power radiated from the individual antennas, the total power density transmitted at any frequency $f$ is given by:
	\begin{align}
		S_\text{tx}(f) = \operatorname{tr}(\symbf{S}_{\symbf{yy}}(f)).
	\end{align}
	
	To distinguish the desired signal from the distortion, it is convenient to use the partitioning of the transmit signal from \eqref{eq:partitioning192893881}.  Since the desired signal and distortion terms are uncorrelated, the power spectral density of the amplified transmit signal is naturally partitioned as follows: 
	\begin{align}\label{eq:psd_partitioning_192929}
		\symbf{S}_{\symbf{yy}}(f) = \symbf{S}_{\symbf{uu}}(f) + \symbf{S}_{\symbf{dd}}(f),
	\end{align}
	where the spectra of the linearly amplified term $\symbf{u}(t) \triangleq (u_1(t), \ldots, u_M(t))^\tr$ and the uncorrelated distortion $\symbf{d}(t) \triangleq (d_1(t), \ldots, d_M(t))^\tr$ are given by:
	\begin{align}
		\symbf{S}_{\symbf{uu}}(f) &= \symbf{A}^\conjtr_1(f) \symbf{S}_{\symbf{xx}}(f) \symbf{A}_1(f)\label{eq:928381818999}\\
		\symbf{S}_{\symbf{dd}}(f) &= \sum_{\varpi\in[3,\Pi]:\text{odd}} \symbf{A}^\conjtr_\varpi(f) \symbf{S}^{(\varpi)}_{\symbf{xx}}(f) \symbf{A}_\varpi(f).\label{eq:51628826687881}
	\end{align}
	
	In the frequency domain, the channel to location $\symfrak{x}$ is described by its transfer function:
	\begin{align}
	\symbfsf{h}_{\symfrak{x}}(f) \triangleq \int_{-\infty}^{\infty} \symbf{h}_{\symfrak{x}}(\tau) e^{-j2\pi\tau f},
	\end{align}
	and the operational power spectral density of the received signal in \eqref{eq:rx_signal1929292} is given by
	\begin{align}
		S_\symfrak{x}(f) = \beta_\symfrak{x} \symbfsf{h}^\conjtr_\symfrak{x}(f) \symbf{S}_{\symbf{yy}}(f) \symbfsf{h}_\symfrak{x}(f).
	\end{align}
	Using the partitioning in \eqref{eq:psd_partitioning_192929}, the operational power spectral densities of the linearly amplified signal and the uncorrelated distortion are then given by:
	\begin{align}
		S^{\text{lin}}_{\symfrak{x}}(f) &= \beta_\symfrak{x} \symbfsf{h}^\conjtr_\symfrak{x}(f) \symbf{S}_{\symbf{uu}}(f) \symbfsf{h}_\symfrak{x}(f),\\
		S^{\text{dist}}_{\symfrak{x}}(f) &= \beta_\symfrak{x} \symbfsf{h}^\conjtr_\symfrak{x}(f) \symbf{S}_{\symbf{dd}}(f) \symbfsf{h}_\symfrak{x}(f).
	\end{align}
	We note that the linear part has the same bandwidth as the signal input to the amplifier.
	
	\section{Distortion Directivity and Measures of Out-of-Band Radiation}
	The radiated distortion from the nonlinear amplifier is beamformed.  The directions and beamforming gain of the distortion are given by the power spectral density matrix $\symbf{S}_{\symbf{dd}}(f)$ and its eigenvectors and eigenvalues.  A measure of the directivity of the distortion at frequency $f$ can be defined as the power of the signal in the strongest direction (assuming that the channel vector is normalized such that its energy is $\beta_\symfrak{x}\|\symbf{h}_\symfrak{x}\|^2 = M$) over the radiated power:	
	\begin{align}
		G_\text{max}(f) \triangleq \frac{M \rho(\symbf{S}_{\symbf{dd}}(f))}{\symbf{S}_{\text{tx}}(f)},\label{eq:0020011}
	\end{align}
	where $\rho(\cdot)$ denotes the largest eigenvalue of a positive semi-definite matrix.  The factor $M$ in the numerator is the average channel power normalized by the large-scale path loss.  Note that $G_\text{max}(f) \geq \SI{0}{dBi}$ with equality only if the distortion is perfectly omnidirectional, i.e.\ all eigenvalues of $\symbf{S}_{\symbf{dd}}(f)$ are equal.
	
	The dimension of the correlation matrix $\symbf{S}_{\symbf{dd}}(f)$ is equal to the number of antennas, $M$.  When this number is large and there is only one (or a few) large eigenvalues, the maximum beamforming gain might be a pessimistic measure of the impact of the distortion.  With high probability the channel of a victim will not be in the subspace spanned by the large eigenvalues, at least not at all frequencies in the band.  A victim that is located at the position $\symfrak{x}$, is operating in the right adjacent band and is using the receive filter $\symsfit{p}_\text{v}(f)$, will pick up the following amount of distortion:
	\begin{align}
		D_\symfrak{x} \triangleq \beta_\symfrak{x} \int_{B/2}^{3B/2} |\symsfit{p}_\text{v}(f)|^2 \symbfsf{h}^\conjtr_\symfrak{x}(f) \symbf{S}_{\symbf{dd}}(f) \symbfsf{h}_\symfrak{x}(f) \symrm{d}f,
	\end{align}
	were $B$ is the width of the band.  By treating the location $\symfrak{x}$ of the victim as random, the complimentary cumulative distribution of the normalized adjacent-distortion power is given by:
	\begin{align}\label{eq:919u891991}
		F(p) \triangleq \Pr\left(\frac{D_\symfrak{x}}{\beta_\symfrak{x} \|\symbfsf{h}_\symfrak{x}\|^2} \geq p \right).
	\end{align}
	Given a realistic distribution of $\symfrak{x}$, the distribution of the distortion that is actually picked up can give a more complete picture of the directivity of the distortion than the maximum gain.
	
	Traditionally, the distortion that is emitted outside the allocated band is measured by the adjacent-channel-leakage ratio (\ACLR), which is the ratio between the leaked power that is radiated in the adjacent band and the useful radiated power inside the allocated band:
	\begin{align}\label{eq:939391812}
		\ACLR = \frac{\max\left\{\int_{-3B/2}^{B/2} S_\text{tx}(f) \symrm{d}f, \int_{B/2}^{3B/2} S_\text{tx}(f) \symrm{d}f\right\}}{\int_{-B/2}^{B/2}S_\text{tx}(f) \symrm{d}f}.
	\end{align}
	In a legacy system, where the radiation pattern of the signal is practically independent of the frequency, this measure makes sense, because the received power ratio at any point is the same as the transmitted.  With an array, however, the useful signal obtains an array gain that might be different from the array gain of the received disturbing power in the adjacent band.  The ratio between the two received powers is therefore different from the transmitted power ratio.  This is illustrated in the following example, where the array gives the in-band signal a gain of \SI{20}{dBi} and the distortion is assumed to be isotropic, i.e.\ to have an array gain of \SI{0}{dBi}.
	\begin{example}\label{ex:83828828122}
		Consider the two systems in Table~\ref{tab:link_budget}.  Both systems are required to serve their users with a received \SNR greater than \SI{0}{dB}.  To do that, the single-antenna transmitter has to transmit \SI{40}{dBm} of power.  The large array, however, has an array gain and, even when the transmit power has to be split among ten users, the array only has to emit \SI{30}{dBm} to achieve the target.  Further, assume that the single-antenna transmitter has a good \ACLR of \SI{-45}{dB} and the large array a somewhat poorer \ACLR of \SI{-35}{dB}.  Despite this, the power emitted in the adjacent band by the two transmitters is the same.  Since the distortion is close to isotropic when there are multiple served users, the power received by a victim receiver in the adjacent band is the same too in the two systems.
	\end{example}
	
	\begin{table}
		\centering
		\ifdraft
		\def\arraystretch{.65}
		\fi
		\caption{Link Budgets for Example~\ref{ex:83828828122}}
		\label{tab:link_budget}
		\begin{tabular}{lrr}
			\toprule
			& 1 antenna & 100 antennas\\
			\midrule
			transmit power & \SI{40}{dBm} & \SI{30}{dBm}\\
			array gain & \SI{0}{dBi} & \SI{20}{dBi}\\
			max.\ path loss & \SI{-140}{\decibel} & \SI{-140}{dB}\\
			noise power & \SI{-100}{dBm} & \SI{-100}{dBm}\\
			nr.\ users & 1 user & 10 users\\
			worst receive \SNR & \SI{0}{dB} & \SI{0}{dB}\\
			\ACLR & \SI{-45}{dB} & \SI{-35}{dB}\\
			radiated adjacent-band power & \SI{-5}{dBm} & \SI{-5}{dBm}\\
			\bottomrule
		\end{tabular}
	\end{table}
	
	Example~\ref{ex:83828828122} shows that the \ACLR in \eqref{eq:939391812} is not a fair measure of out-of-band radiation, because it does not account for the differences in array gain.  An alternative way to measure the out-of-band power is to define the minimum useful power, the lowest of the received powers at the served users, as:
	\ifdraft
	\vspace{-1ex}
	\fi
	\begin{align}
		P_\text{useful} \triangleq \min\left\{ P = \int_{-B/2}^{B/2} S^{\text{lin}}_{\symfrak{x}_k}(f) \,\symrm{d}f : k=1, \ldots, K \right\}
	\end{align}
	and to look at the leaked power in the adjacent channel with respect to reference point $\symfrak{x}_\text{ref}$:
	\begin{align}
		P_\text{leak} \triangleq \max\left\{\int_{-3B/2}^{-B/2} S_{\symfrak{x}_\text{ref}}(f) \,\symrm{d}f, \int_{3B/2}^{B/2} S_{\symfrak{x}_\text{ref}}(f) \,\symrm{d}f\right\}.
	\end{align}
	In complete analogy to \eqref{eq:939391812}, an \emph{array \ACLR} can be defined as:	
	\begin{align}
		\text{array \ACLR} = \smash{\frac{P_\text{leak}}{P_\text{useful}}}.
	\end{align}
	
\begin{figure}
		\centering
		\ifdraft
		\mbox{\hspace{-3em}\includegraphics{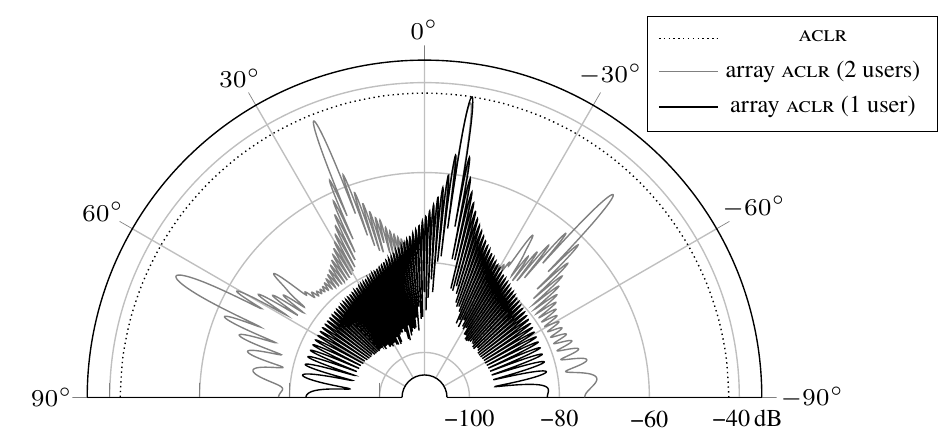}
		\begin{tikzpicture}
			\begin{semilogyaxis}[
			xlabel={array ACLRs [dB]},
			ylabel={probability},
			ymax=1.5,
			ymin=.004,
			]
			\addplot[
			densely dotted,
			color=black,
			no marks,
			] coordinates {
				(-42.33094104777792, .004)
				(-42.33094104777792, 1.5)};
			\addplot[
			no marks,
			color = black,
			densely dashed,
			] coordinates {
				( -63.8243924482 , 1.0 )
				( -63.4223566793 , 0.994444444444 )
				( -63.1153614537 , 0.988888888889 )
				( -62.6518261608 , 0.983333333333 )
				( -62.3491837456 , 0.977777777778 )
				( -62.0378294776 , 0.972222222222 )
				( -61.7174853928 , 0.966666666667 )
				( -61.3150973601 , 0.961111111111 )
				( -61.1202487697 , 0.955555555556 )
				( -60.8941106912 , 0.95 )
				( -60.6486162292 , 0.944444444444 )
				( -60.3406863986 , 0.938888888889 )
				( -60.0187955864 , 0.933333333333 )
				( -59.7844624252 , 0.927777777778 )
				( -59.4710536708 , 0.922222222222 )
				( -59.2300466372 , 0.916666666667 )
				( -58.9448727366 , 0.911111111111 )
				( -58.7741161815 , 0.905555555556 )
				( -58.6023286026 , 0.9 )
				( -58.5774005356 , 0.894444444444 )
				( -58.5385650241 , 0.888888888889 )
				( -58.4982284378 , 0.883333333333 )
				( -58.4565157501 , 0.877777777778 )
				( -58.3826769425 , 0.872222222222 )
				( -58.3126988626 , 0.866666666667 )
				( -58.2608859442 , 0.861111111111 )
				( -58.1853332937 , 0.855555555556 )
				( -58.1352753836 , 0.85 )
				( -58.0865659964 , 0.844444444444 )
				( -58.0530450695 , 0.838888888889 )
				( -58.0078571756 , 0.833333333333 )
				( -57.9445463658 , 0.827777777778 )
				( -57.8803553469 , 0.822222222222 )
				( -57.8275990416 , 0.816666666667 )
				( -57.7759397649 , 0.811111111111 )
				( -57.7209842639 , 0.805555555556 )
				( -57.6864575784 , 0.8 )
				( -57.6449256961 , 0.794444444444 )
				( -57.5031706213 , 0.788888888889 )
				( -57.3940248536 , 0.783333333333 )
				( -57.3343373011 , 0.777777777778 )
				( -57.245729476 , 0.772222222222 )
				( -57.1837686944 , 0.766666666667 )
				( -57.0703167585 , 0.761111111111 )
				( -57.004207805 , 0.755555555556 )
				( -56.9209345628 , 0.75 )
				( -56.8093517344 , 0.744444444444 )
				( -56.7365051747 , 0.738888888889 )
				( -56.6749974512 , 0.733333333333 )
				( -56.5890450743 , 0.727777777778 )
				( -56.540977749 , 0.722222222222 )
				( -56.4331446716 , 0.716666666667 )
				( -56.3884389663 , 0.711111111111 )
				( -56.3391863932 , 0.705555555556 )
				( -56.2968680271 , 0.7 )
				( -56.2644051273 , 0.694444444444 )
				( -56.1998398609 , 0.688888888889 )
				( -56.1357837289 , 0.683333333333 )
				( -56.0977083552 , 0.677777777778 )
				( -56.0525734665 , 0.672222222222 )
				( -56.0100815966 , 0.666666666667 )
				( -55.9651609885 , 0.661111111111 )
				( -55.9321760975 , 0.655555555556 )
				( -55.9025604059 , 0.65 )
				( -55.8623742414 , 0.644444444444 )
				( -55.8188165532 , 0.638888888889 )
				( -55.7874348473 , 0.633333333333 )
				( -55.7456121901 , 0.627777777778 )
				( -55.7218536011 , 0.622222222222 )
				( -55.6473884067 , 0.616666666667 )
				( -55.5878650151 , 0.611111111111 )
				( -55.5573822376 , 0.605555555556 )
				( -55.5154594614 , 0.6 )
				( -55.4463350444 , 0.594444444444 )
				( -55.4017802945 , 0.588888888889 )
				( -55.3514228449 , 0.583333333333 )
				( -55.3209364831 , 0.577777777778 )
				( -55.2975113154 , 0.572222222222 )
				( -55.2826574172 , 0.566666666667 )
				( -55.219166581 , 0.561111111111 )
				( -55.1803739263 , 0.555555555556 )
				( -55.1495371873 , 0.55 )
				( -55.1117113161 , 0.544444444444 )
				( -55.0723982564 , 0.538888888889 )
				( -55.0460150485 , 0.533333333333 )
				( -55.0113029939 , 0.527777777778 )
				( -54.9760036507 , 0.522222222222 )
				( -54.9541393085 , 0.516666666667 )
				( -54.917451461 , 0.511111111111 )
				( -54.8540476383 , 0.505555555556 )
				( -54.8243569484 , 0.5 )
				( -54.7671486254 , 0.494444444444 )
				( -54.7090192488 , 0.488888888889 )
				( -54.6508407532 , 0.483333333333 )
				( -54.564993539 , 0.477777777778 )
				( -54.5045930358 , 0.472222222222 )
				( -54.4704681412 , 0.466666666667 )
				( -54.4193773402 , 0.461111111111 )
				( -54.3683021832 , 0.455555555556 )
				( -54.3189412793 , 0.45 )
				( -54.2308439763 , 0.444444444444 )
				( -54.199277718 , 0.438888888889 )
				( -54.1393270464 , 0.433333333333 )
				( -54.0937700118 , 0.427777777778 )
				( -54.0074022862 , 0.422222222222 )
				( -53.9616078104 , 0.416666666667 )
				( -53.8756842418 , 0.411111111111 )
				( -53.7675071763 , 0.405555555556 )
				( -53.7191599674 , 0.4 )
				( -53.670739074 , 0.394444444444 )
				( -53.5602264007 , 0.388888888889 )
				( -53.5078481523 , 0.383333333333 )
				( -53.4463058413 , 0.377777777778 )
				( -53.405369069 , 0.372222222222 )
				( -53.3694332062 , 0.366666666667 )
				( -53.2895090364 , 0.361111111111 )
				( -53.2215633939 , 0.355555555556 )
				( -53.1739005804 , 0.35 )
				( -53.0580253502 , 0.344444444444 )
				( -52.998479453 , 0.338888888889 )
				( -52.9433781086 , 0.333333333333 )
				( -52.9042120139 , 0.327777777778 )
				( -52.8345265813 , 0.322222222222 )
				( -52.8020945933 , 0.316666666667 )
				( -52.7392600223 , 0.311111111111 )
				( -52.7015467724 , 0.305555555556 )
				( -52.6646597789 , 0.3 )
				( -52.6405315565 , 0.294444444444 )
				( -52.6016790989 , 0.288888888889 )
				( -52.5649911604 , 0.283333333333 )
				( -52.52246473 , 0.277777777778 )
				( -52.4595333167 , 0.272222222222 )
				( -52.3830731758 , 0.266666666667 )
				( -52.3499637354 , 0.261111111111 )
				( -52.2968985778 , 0.255555555556 )
				( -52.2683698803 , 0.25 )
				( -52.1660076594 , 0.244444444444 )
				( -52.0954915605 , 0.238888888889 )
				( -52.0495552414 , 0.233333333333 )
				( -52.0040682684 , 0.227777777778 )
				( -51.9660592299 , 0.222222222222 )
				( -51.9287813418 , 0.216666666667 )
				( -51.8215557877 , 0.211111111111 )
				( -51.723774029 , 0.205555555556 )
				( -51.6729957997 , 0.2 )
				( -51.6115633705 , 0.194444444444 )
				( -51.5614517577 , 0.188888888889 )
				( -51.4718809096 , 0.183333333333 )
				( -51.3899368119 , 0.177777777778 )
				( -51.2663204415 , 0.172222222222 )
				( -51.209141469 , 0.166666666667 )
				( -51.1152582787 , 0.161111111111 )
				( -51.0750299887 , 0.155555555556 )
				( -51.0093093939 , 0.15 )
				( -50.9581908662 , 0.144444444444 )
				( -50.9355869972 , 0.138888888889 )
				( -50.9123064614 , 0.133333333333 )
				( -50.8694081035 , 0.127777777778 )
				( -50.8038550626 , 0.122222222222 )
				( -50.7638572307 , 0.116666666667 )
				( -50.6764054134 , 0.111111111111 )
				( -50.581635689 , 0.105555555556 )
				( -50.5026341589 , 0.1 )
				( -50.4532746891 , 0.0944444444444 )
				( -50.3411983761 , 0.0888888888889 )
				( -50.2702754887 , 0.0833333333333 )
				( -50.208145122 , 0.0777777777778 )
				( -50.1290594248 , 0.0722222222222 )
				( -49.9573106665 , 0.0666666666667 )
				( -49.7228405695 , 0.0611111111111 )
				( -49.5163497853 , 0.0555555555556 )
				( -49.2423986901 , 0.05 )
				( -49.0414186883 , 0.0444444444444 )
				( -48.84617969 , 0.0388888888889 )
				( -48.6471710377 , 0.0333333333333 )
				( -48.5564995122 , 0.0277777777778 )
				( -48.4342996336 , 0.0222222222222 )
				( -48.2572409698 , 0.0166666666667 )
				( -48.1542508933 , 0.0111111111111 )
				( -48.0032038192 , 0.00555555555556 )
				};
			\addplot[
			no marks,
			] coordinates {
				(-97.013349025,1)
				(-96.1402592696,0.9994444444)
				(-95.364423271,0.9938888889)
				(-94.7795264436,0.9883333333)
				(-94.6132287453,0.9827777778)
				(-94.437586348,0.9772222222)
				(-94.2064452801,0.9716666667)
				(-94.0064860509,0.9661111111)
				(-93.8630077869,0.9605555556)
				(-93.742117952,0.955)
				(-93.553541456,0.9494444444)
				(-93.4239021561,0.9438888889)
				(-93.2754630956,0.9383333333)
				(-93.1513269833,0.9327777778)
				(-92.8903541267,0.9272222222)
				(-92.6947913729,0.9216666667)
				(-92.5343091856,0.9161111111)
				(-92.331343498,0.9105555556)
				(-92.1321144673,0.905)
				(-92.02657158,0.8994444444)
				(-91.8395515664,0.8938888889)
				(-91.5189577913,0.8883333333)
				(-91.3131197258,0.8827777778)
				(-90.9908151948,0.8772222222)
				(-90.8689847175,0.8716666667)
				(-90.732948799,0.8661111111)
				(-90.614492613,0.8605555556)
				(-90.3892572023,0.855)
				(-90.1305812361,0.8494444444)
				(-89.9099556538,0.8438888889)
				(-89.7835991256,0.8383333333)
				(-89.6034195948,0.8327777778)
				(-89.4377710359,0.8272222222)
				(-89.3289366919,0.8216666667)
				(-89.048417725,0.8161111111)
				(-88.8892658059,0.8105555556)
				(-88.6607265109,0.805)
				(-88.5439923998,0.7994444444)
				(-88.4032073035,0.7938888889)
				(-88.2210884723,0.7883333333)
				(-88.0901891032,0.7827777778)
				(-87.9113239731,0.7772222222)
				(-87.7456264527,0.7716666667)
				(-87.5805376777,0.7661111111)
				(-87.4991590293,0.7605555556)
				(-87.385508837,0.755)
				(-87.2515223582,0.7494444444)
				(-87.156113639,0.7438888889)
				(-86.8875905166,0.7383333333)
				(-86.8257662308,0.7327777778)
				(-86.6920518563,0.7272222222)
				(-86.6300175187,0.7216666667)
				(-86.5465236587,0.7161111111)
				(-86.3917997281,0.7105555556)
				(-86.2282301101,0.705)
				(-86.0845459878,0.6994444444)
				(-85.9720848139,0.6938888889)
				(-85.8398516084,0.6883333333)
				(-85.7147064757,0.6827777778)
				(-85.6057559647,0.6772222222)
				(-85.4661386917,0.6716666667)
				(-85.3671121952,0.6661111111)
				(-85.3168353836,0.6605555556)
				(-85.254293677,0.655)
				(-85.1733422144,0.6494444444)
				(-85.1006257423,0.6438888889)
				(-85.0412906291,0.6383333333)
				(-84.9455079065,0.6327777778)
				(-84.8330158079,0.6272222222)
				(-84.7145851893,0.6216666667)
				(-84.6149843587,0.6161111111)
				(-84.53391185,0.6105555556)
				(-84.4652234747,0.605)
				(-84.403751946,0.5994444444)
				(-84.3265338165,0.5938888889)
				(-84.2563121081,0.5883333333)
				(-84.1951872593,0.5827777778)
				(-84.129789586,0.5772222222)
				(-84.0347656042,0.5716666667)
				(-83.9458717873,0.5661111111)
				(-83.8803035701,0.5605555556)
				(-83.8316753441,0.555)
				(-83.7764814498,0.5494444444)
				(-83.7236503224,0.5438888889)
				(-83.6756766292,0.5383333333)
				(-83.6206611712,0.5327777778)
				(-83.5564974251,0.5272222222)
				(-83.4994913426,0.5216666667)
				(-83.4323376917,0.5161111111)
				(-83.3676396988,0.5105555556)
				(-83.2948046399,0.505)
				(-83.2434390318,0.4994444444)
				(-83.1964262327,0.4938888889)
				(-83.1625879993,0.4883333333)
				(-83.0995819134,0.4827777778)
				(-83.0086384099,0.4772222222)
				(-82.977485083,0.4716666667)
				(-82.935192547,0.4661111111)
				(-82.8999762319,0.4605555556)
				(-82.8664596211,0.455)
				(-82.8198633751,0.4494444444)
				(-82.7843421442,0.4438888889)
				(-82.7551741392,0.4383333333)
				(-82.7011922968,0.4327777778)
				(-82.6885499246,0.4272222222)
				(-82.6636300502,0.4216666667)
				(-82.6368257285,0.4161111111)
				(-82.6212810104,0.4105555556)
				(-82.5976180083,0.405)
				(-82.5753328062,0.3994444444)
				(-82.5525888485,0.3938888889)
				(-82.5229781996,0.3883333333)
				(-82.4964158513,0.3827777778)
				(-82.4714489414,0.3772222222)
				(-82.4445612529,0.3716666667)
				(-82.4167326714,0.3661111111)
				(-82.3998934378,0.3605555556)
				(-82.3631339645,0.355)
				(-82.303263106,0.3494444444)
				(-82.2572036472,0.3438888889)
				(-82.2077785472,0.3383333333)
				(-82.1121718639,0.3327777778)
				(-82.0380422419,0.3272222222)
				(-81.9466794334,0.3216666667)
				(-81.9086367978,0.3161111111)
				(-81.8141724458,0.3105555556)
				(-81.7324348011,0.305)
				(-81.6263933631,0.2994444444)
				(-81.491707979,0.2938888889)
				(-81.3971842691,0.2883333333)
				(-81.3306718649,0.2827777778)
				(-81.2099603209,0.2772222222)
				(-81.1243668883,0.2716666667)
				(-80.9847744046,0.2661111111)
				(-80.8203108031,0.2605555556)
				(-80.7514015808,0.255)
				(-80.5894883923,0.2494444444)
				(-80.3936726458,0.2438888889)
				(-80.309955847,0.2383333333)
				(-80.1774605309,0.2327777778)
				(-80.0252965114,0.2272222222)
				(-79.7933263315,0.2216666667)
				(-79.5764718089,0.2161111111)
				(-79.35984846,0.2105555556)
				(-79.2194847974,0.205)
				(-79.0858569996,0.1994444444)
				(-78.8603100285,0.1938888889)
				(-78.6156286649,0.1883333333)
				(-78.3653628932,0.1827777778)
				(-78.1294594632,0.1772222222)
				(-78.0001064188,0.1716666667)
				(-77.6602940926,0.1661111111)
				(-77.2851522891,0.1605555556)
				(-77.0556779928,0.155)
				(-76.8148631049,0.1494444444)
				(-76.4728842515,0.1438888889)
				(-76.2504802278,0.1383333333)
				(-76.0154757334,0.1327777778)
				(-75.3701941213,0.1272222222)
				(-75.0168158589,0.1216666667)
				(-74.7225095578,0.1161111111)
				(-74.2754119833,0.1105555556)
				(-73.9950772089,0.105)
				(-73.3695090241,0.0994444444)
				(-72.6951132227,0.0938888889)
				(-72.2497945494,0.0883333333)
				(-71.5614544039,0.0827777778)
				(-70.9117162479,0.0772222222)
				(-70.3502692101,0.0716666667)
				(-69.7677559892,0.0661111111)
				(-69.0429465161,0.0605555556)
				(-67.8307567014,0.055)
				(-67.0795558509,0.0494444444)
				(-65.5133037631,0.0438888889)
				(-64.5047721756,0.0383333333)
				(-62.9126725322,0.0327777778)
				(-60.4758652274,0.0272222222)
				(-57.5254837567,0.0216666667)
				(-55.6603188449,0.0161111111)
				(-44.8997903954,0.0105555556)
				(-42.347533449432134,0.005)
				};
			\addplot[
			no marks,
			color = gray,
			] coordinates {
				( -78.966358726 , 1.0 )
				( -78.8476503774 , 0.994444444444 )
				( -78.7609851425 , 0.988888888889 )
				( -78.6758950743 , 0.983333333333 )
				( -78.6269658213 , 0.977777777778 )
				( -78.5265854067 , 0.972222222222 )
				( -78.4110320237 , 0.966666666667 )
				( -78.3664803889 , 0.961111111111 )
				( -78.3118802687 , 0.955555555556 )
				( -78.2638904412 , 0.95 )
				( -78.2024970552 , 0.944444444444 )
				( -78.0952482271 , 0.938888888889 )
				( -78.0592290025 , 0.933333333333 )
				( -77.9936438832 , 0.927777777778 )
				( -77.9413578295 , 0.922222222222 )
				( -77.8836995419 , 0.916666666667 )
				( -77.8066971745 , 0.911111111111 )
				( -77.7593152716 , 0.905555555556 )
				( -77.6912417577 , 0.9 )
				( -77.6649712604 , 0.894444444444 )
				( -77.6215652403 , 0.888888888889 )
				( -77.5693932157 , 0.883333333333 )
				( -77.51153497 , 0.877777777778 )
				( -77.4468077292 , 0.872222222222 )
				( -77.3836818995 , 0.866666666667 )
				( -77.3371962089 , 0.861111111111 )
				( -77.285397882 , 0.855555555556 )
				( -77.2285095863 , 0.85 )
				( -77.185404696 , 0.844444444444 )
				( -77.1163060065 , 0.838888888889 )
				( -77.0700426387 , 0.833333333333 )
				( -77.0283244576 , 0.827777777778 )
				( -76.9588619841 , 0.822222222222 )
				( -76.9360804156 , 0.816666666667 )
				( -76.8906319667 , 0.811111111111 )
				( -76.8450377952 , 0.805555555556 )
				( -76.7908933323 , 0.8 )
				( -76.7651502459 , 0.794444444444 )
				( -76.7142202089 , 0.788888888889 )
				( -76.6706233552 , 0.783333333333 )
				( -76.6467292864 , 0.777777777778 )
				( -76.5928121252 , 0.772222222222 )
				( -76.5652012842 , 0.766666666667 )
				( -76.5146213723 , 0.761111111111 )
				( -76.4707747528 , 0.755555555556 )
				( -76.4340788811 , 0.75 )
				( -76.3788100528 , 0.744444444444 )
				( -76.3405288557 , 0.738888888889 )
				( -76.3099939683 , 0.733333333333 )
				( -76.279018331 , 0.727777777778 )
				( -76.236898537 , 0.722222222222 )
				( -76.1840617111 , 0.716666666667 )
				( -76.1363711239 , 0.711111111111 )
				( -76.0752886911 , 0.705555555556 )
				( -76.0192125578 , 0.7 )
				( -75.9860587149 , 0.694444444444 )
				( -75.9351803619 , 0.688888888889 )
				( -75.896909639 , 0.683333333333 )
				( -75.8221761126 , 0.677777777778 )
				( -75.7930800839 , 0.672222222222 )
				( -75.7590388723 , 0.666666666667 )
				( -75.7139214309 , 0.661111111111 )
				( -75.6551330547 , 0.655555555556 )
				( -75.5946874744 , 0.65 )
				( -75.5453556862 , 0.644444444444 )
				( -75.4886429184 , 0.638888888889 )
				( -75.422614481 , 0.633333333333 )
				( -75.3716376045 , 0.627777777778 )
				( -75.3054057757 , 0.622222222222 )
				( -75.2618805593 , 0.616666666667 )
				( -75.203168721 , 0.611111111111 )
				( -75.1459278385 , 0.605555555556 )
				( -75.1065946616 , 0.6 )
				( -75.0361749144 , 0.594444444444 )
				( -74.9838601272 , 0.588888888889 )
				( -74.9341876706 , 0.583333333333 )
				( -74.8665578579 , 0.577777777778 )
				( -74.8015540841 , 0.572222222222 )
				( -74.7308796297 , 0.566666666667 )
				( -74.6474258467 , 0.561111111111 )
				( -74.603209986 , 0.555555555556 )
				( -74.5164557139 , 0.55 )
				( -74.4582736104 , 0.544444444444 )
				( -74.3990076422 , 0.538888888889 )
				( -74.3534601406 , 0.533333333333 )
				( -74.306113111 , 0.527777777778 )
				( -74.2387234239 , 0.522222222222 )
				( -74.1891283547 , 0.516666666667 )
				( -74.1207731339 , 0.511111111111 )
				( -74.0466705744 , 0.505555555556 )
				( -73.9971643471 , 0.5 )
				( -73.9534568646 , 0.494444444444 )
				( -73.8915149551 , 0.488888888889 )
				( -73.8295573766 , 0.483333333333 )
				( -73.7621139558 , 0.477777777778 )
				( -73.6183953469 , 0.472222222222 )
				( -73.5433157844 , 0.466666666667 )
				( -73.4845822034 , 0.461111111111 )
				( -73.4001004907 , 0.455555555556 )
				( -73.2853542665 , 0.45 )
				( -73.2009574982 , 0.444444444444 )
				( -73.1131894091 , 0.438888888889 )
				( -73.0437692188 , 0.433333333333 )
				( -72.9844867307 , 0.427777777778 )
				( -72.948480983 , 0.422222222222 )
				( -72.8466823504 , 0.416666666667 )
				( -72.7461095126 , 0.411111111111 )
				( -72.6444358626 , 0.405555555556 )
				( -72.5516137115 , 0.4 )
				( -72.4607709354 , 0.394444444444 )
				( -72.3874445125 , 0.388888888889 )
				( -72.2987790387 , 0.383333333333 )
				( -72.1752533751 , 0.377777777778 )
				( -72.0678152968 , 0.372222222222 )
				( -71.9269548774 , 0.366666666667 )
				( -71.8432763012 , 0.361111111111 )
				( -71.8044685432 , 0.355555555556 )
				( -71.7432520946 , 0.35 )
				( -71.6528149812 , 0.344444444444 )
				( -71.6050196492 , 0.338888888889 )
				( -71.5269366978 , 0.333333333333 )
				( -71.4759028183 , 0.327777777778 )
				( -71.4201359686 , 0.322222222222 )
				( -71.3518782915 , 0.316666666667 )
				( -71.2946838558 , 0.311111111111 )
				( -71.2300943062 , 0.305555555556 )
				( -71.0120842777 , 0.3 )
				( -70.7790387875 , 0.294444444444 )
				( -70.5626703867 , 0.288888888889 )
				( -70.3464279932 , 0.283333333333 )
				( -70.226774941 , 0.277777777778 )
				( -70.1216282733 , 0.272222222222 )
				( -70.0052471483 , 0.266666666667 )
				( -69.8662744484 , 0.261111111111 )
				( -69.6630757289 , 0.255555555556 )
				( -69.5728643136 , 0.25 )
				( -69.4301610322 , 0.244444444444 )
				( -69.3316134606 , 0.238888888889 )
				( -69.2672666765 , 0.233333333333 )
				( -69.1306461753 , 0.227777777778 )
				( -68.7585255384 , 0.222222222222 )
				( -68.6419910448 , 0.216666666667 )
				( -68.5037302843 , 0.211111111111 )
				( -68.259696761 , 0.205555555556 )
				( -67.9274756813 , 0.2 )
				( -67.7239676923 , 0.194444444444 )
				( -67.5158099349 , 0.188888888889 )
				( -67.2690231982 , 0.183333333333 )
				( -67.071030425 , 0.177777777778 )
				( -66.9540663436 , 0.172222222222 )
				( -66.8275492204 , 0.166666666667 )
				( -66.6550457751 , 0.161111111111 )
				( -66.5591119598 , 0.155555555556 )
				( -66.4116463985 , 0.15 )
				( -66.0928299042 , 0.144444444444 )
				( -65.6180983813 , 0.138888888889 )
				( -65.1043664525 , 0.133333333333 )
				( -64.7224822537 , 0.127777777778 )
				( -64.2555221679 , 0.122222222222 )
				( -63.7876135686 , 0.116666666667 )
				( -63.379738574 , 0.111111111111 )
				( -62.9060803178 , 0.105555555556 )
				( -62.6149117885 , 0.1 )
				( -62.3010035616 , 0.0944444444444 )
				( -62.0703585467 , 0.0888888888889 )
				( -61.8880027767 , 0.0833333333333 )
				( -61.1423817431 , 0.0777777777778 )
				( -59.4816669613 , 0.0722222222222 )
				( -58.235447403 , 0.0666666666667 )
				( -57.7350423768 , 0.0611111111111 )
				( -57.2685692929 , 0.0555555555556 )
				( -54.6724163005 , 0.05 )
				( -52.8690941391 , 0.0444444444444 )
				( -51.6768822448 , 0.0388888888889 )
				( -50.2376073923 , 0.0333333333333 )
				( -49.5138478272 , 0.0277777777778 )
				( -48.8314880157 , 0.0222222222222 )
				( -48.5140004668 , 0.0166666666667 )
				( -46.4393700047 , 0.0111111111111 )
				( -44.336081879 , 0.00555555555556 )
				};
			\node[anchor=north east, rotate=-42, inner sep=0pt] at (axis cs: -78,.09){1~user};
			\node[anchor=north east, rotate=-36, inner sep=0pt] at (axis cs: -57,.1){2~users};
			\node[anchor=north east, rotate=-65, inner sep=0pt] at (axis cs: -45,.08){10~users};
			\end{semilogyaxis}
		\end{tikzpicture}
	}
		\else
		\includegraphics[width=\linewidth]{array_ACLR_LOS}
		\begin{tikzpicture}
		\begin{semilogyaxis}[
		xlabel={array ACLRs [dB]},
		ylabel={probability},
		ymax=1.5,
		ymin=.004,
		]
		\addplot[
		densely dotted,
		color=black,
		no marks,
		] coordinates {
			(-42.33094104777792, .004)
			(-42.33094104777792, 1.5)};
		\addplot[
		no marks,
		color = black,
		densely dashed,
		] coordinates {
			( -63.8243924482 , 1.0 )
			( -63.4223566793 , 0.994444444444 )
			( -63.1153614537 , 0.988888888889 )
			( -62.6518261608 , 0.983333333333 )
			( -62.3491837456 , 0.977777777778 )
			( -62.0378294776 , 0.972222222222 )
			( -61.7174853928 , 0.966666666667 )
			( -61.3150973601 , 0.961111111111 )
			( -61.1202487697 , 0.955555555556 )
			( -60.8941106912 , 0.95 )
			( -60.6486162292 , 0.944444444444 )
			( -60.3406863986 , 0.938888888889 )
			( -60.0187955864 , 0.933333333333 )
			( -59.7844624252 , 0.927777777778 )
			( -59.4710536708 , 0.922222222222 )
			( -59.2300466372 , 0.916666666667 )
			( -58.9448727366 , 0.911111111111 )
			( -58.7741161815 , 0.905555555556 )
			( -58.6023286026 , 0.9 )
			( -58.5774005356 , 0.894444444444 )
			( -58.5385650241 , 0.888888888889 )
			( -58.4982284378 , 0.883333333333 )
			( -58.4565157501 , 0.877777777778 )
			( -58.3826769425 , 0.872222222222 )
			( -58.3126988626 , 0.866666666667 )
			( -58.2608859442 , 0.861111111111 )
			( -58.1853332937 , 0.855555555556 )
			( -58.1352753836 , 0.85 )
			( -58.0865659964 , 0.844444444444 )
			( -58.0530450695 , 0.838888888889 )
			( -58.0078571756 , 0.833333333333 )
			( -57.9445463658 , 0.827777777778 )
			( -57.8803553469 , 0.822222222222 )
			( -57.8275990416 , 0.816666666667 )
			( -57.7759397649 , 0.811111111111 )
			( -57.7209842639 , 0.805555555556 )
			( -57.6864575784 , 0.8 )
			( -57.6449256961 , 0.794444444444 )
			( -57.5031706213 , 0.788888888889 )
			( -57.3940248536 , 0.783333333333 )
			( -57.3343373011 , 0.777777777778 )
			( -57.245729476 , 0.772222222222 )
			( -57.1837686944 , 0.766666666667 )
			( -57.0703167585 , 0.761111111111 )
			( -57.004207805 , 0.755555555556 )
			( -56.9209345628 , 0.75 )
			( -56.8093517344 , 0.744444444444 )
			( -56.7365051747 , 0.738888888889 )
			( -56.6749974512 , 0.733333333333 )
			( -56.5890450743 , 0.727777777778 )
			( -56.540977749 , 0.722222222222 )
			( -56.4331446716 , 0.716666666667 )
			( -56.3884389663 , 0.711111111111 )
			( -56.3391863932 , 0.705555555556 )
			( -56.2968680271 , 0.7 )
			( -56.2644051273 , 0.694444444444 )
			( -56.1998398609 , 0.688888888889 )
			( -56.1357837289 , 0.683333333333 )
			( -56.0977083552 , 0.677777777778 )
			( -56.0525734665 , 0.672222222222 )
			( -56.0100815966 , 0.666666666667 )
			( -55.9651609885 , 0.661111111111 )
			( -55.9321760975 , 0.655555555556 )
			( -55.9025604059 , 0.65 )
			( -55.8623742414 , 0.644444444444 )
			( -55.8188165532 , 0.638888888889 )
			( -55.7874348473 , 0.633333333333 )
			( -55.7456121901 , 0.627777777778 )
			( -55.7218536011 , 0.622222222222 )
			( -55.6473884067 , 0.616666666667 )
			( -55.5878650151 , 0.611111111111 )
			( -55.5573822376 , 0.605555555556 )
			( -55.5154594614 , 0.6 )
			( -55.4463350444 , 0.594444444444 )
			( -55.4017802945 , 0.588888888889 )
			( -55.3514228449 , 0.583333333333 )
			( -55.3209364831 , 0.577777777778 )
			( -55.2975113154 , 0.572222222222 )
			( -55.2826574172 , 0.566666666667 )
			( -55.219166581 , 0.561111111111 )
			( -55.1803739263 , 0.555555555556 )
			( -55.1495371873 , 0.55 )
			( -55.1117113161 , 0.544444444444 )
			( -55.0723982564 , 0.538888888889 )
			( -55.0460150485 , 0.533333333333 )
			( -55.0113029939 , 0.527777777778 )
			( -54.9760036507 , 0.522222222222 )
			( -54.9541393085 , 0.516666666667 )
			( -54.917451461 , 0.511111111111 )
			( -54.8540476383 , 0.505555555556 )
			( -54.8243569484 , 0.5 )
			( -54.7671486254 , 0.494444444444 )
			( -54.7090192488 , 0.488888888889 )
			( -54.6508407532 , 0.483333333333 )
			( -54.564993539 , 0.477777777778 )
			( -54.5045930358 , 0.472222222222 )
			( -54.4704681412 , 0.466666666667 )
			( -54.4193773402 , 0.461111111111 )
			( -54.3683021832 , 0.455555555556 )
			( -54.3189412793 , 0.45 )
			( -54.2308439763 , 0.444444444444 )
			( -54.199277718 , 0.438888888889 )
			( -54.1393270464 , 0.433333333333 )
			( -54.0937700118 , 0.427777777778 )
			( -54.0074022862 , 0.422222222222 )
			( -53.9616078104 , 0.416666666667 )
			( -53.8756842418 , 0.411111111111 )
			( -53.7675071763 , 0.405555555556 )
			( -53.7191599674 , 0.4 )
			( -53.670739074 , 0.394444444444 )
			( -53.5602264007 , 0.388888888889 )
			( -53.5078481523 , 0.383333333333 )
			( -53.4463058413 , 0.377777777778 )
			( -53.405369069 , 0.372222222222 )
			( -53.3694332062 , 0.366666666667 )
			( -53.2895090364 , 0.361111111111 )
			( -53.2215633939 , 0.355555555556 )
			( -53.1739005804 , 0.35 )
			( -53.0580253502 , 0.344444444444 )
			( -52.998479453 , 0.338888888889 )
			( -52.9433781086 , 0.333333333333 )
			( -52.9042120139 , 0.327777777778 )
			( -52.8345265813 , 0.322222222222 )
			( -52.8020945933 , 0.316666666667 )
			( -52.7392600223 , 0.311111111111 )
			( -52.7015467724 , 0.305555555556 )
			( -52.6646597789 , 0.3 )
			( -52.6405315565 , 0.294444444444 )
			( -52.6016790989 , 0.288888888889 )
			( -52.5649911604 , 0.283333333333 )
			( -52.52246473 , 0.277777777778 )
			( -52.4595333167 , 0.272222222222 )
			( -52.3830731758 , 0.266666666667 )
			( -52.3499637354 , 0.261111111111 )
			( -52.2968985778 , 0.255555555556 )
			( -52.2683698803 , 0.25 )
			( -52.1660076594 , 0.244444444444 )
			( -52.0954915605 , 0.238888888889 )
			( -52.0495552414 , 0.233333333333 )
			( -52.0040682684 , 0.227777777778 )
			( -51.9660592299 , 0.222222222222 )
			( -51.9287813418 , 0.216666666667 )
			( -51.8215557877 , 0.211111111111 )
			( -51.723774029 , 0.205555555556 )
			( -51.6729957997 , 0.2 )
			( -51.6115633705 , 0.194444444444 )
			( -51.5614517577 , 0.188888888889 )
			( -51.4718809096 , 0.183333333333 )
			( -51.3899368119 , 0.177777777778 )
			( -51.2663204415 , 0.172222222222 )
			( -51.209141469 , 0.166666666667 )
			( -51.1152582787 , 0.161111111111 )
			( -51.0750299887 , 0.155555555556 )
			( -51.0093093939 , 0.15 )
			( -50.9581908662 , 0.144444444444 )
			( -50.9355869972 , 0.138888888889 )
			( -50.9123064614 , 0.133333333333 )
			( -50.8694081035 , 0.127777777778 )
			( -50.8038550626 , 0.122222222222 )
			( -50.7638572307 , 0.116666666667 )
			( -50.6764054134 , 0.111111111111 )
			( -50.581635689 , 0.105555555556 )
			( -50.5026341589 , 0.1 )
			( -50.4532746891 , 0.0944444444444 )
			( -50.3411983761 , 0.0888888888889 )
			( -50.2702754887 , 0.0833333333333 )
			( -50.208145122 , 0.0777777777778 )
			( -50.1290594248 , 0.0722222222222 )
			( -49.9573106665 , 0.0666666666667 )
			( -49.7228405695 , 0.0611111111111 )
			( -49.5163497853 , 0.0555555555556 )
			( -49.2423986901 , 0.05 )
			( -49.0414186883 , 0.0444444444444 )
			( -48.84617969 , 0.0388888888889 )
			( -48.6471710377 , 0.0333333333333 )
			( -48.5564995122 , 0.0277777777778 )
			( -48.4342996336 , 0.0222222222222 )
			( -48.2572409698 , 0.0166666666667 )
			( -48.1542508933 , 0.0111111111111 )
			( -48.0032038192 , 0.00555555555556 )
		};
			\addplot[
			no marks,
			] coordinates {
				(-97.013349025,1)
				(-96.1402592696,0.9994444444)
				(-95.364423271,0.9938888889)
				(-94.7795264436,0.9883333333)
				(-94.6132287453,0.9827777778)
				(-94.437586348,0.9772222222)
				(-94.2064452801,0.9716666667)
				(-94.0064860509,0.9661111111)
				(-93.8630077869,0.9605555556)
				(-93.742117952,0.955)
				(-93.553541456,0.9494444444)
				(-93.4239021561,0.9438888889)
				(-93.2754630956,0.9383333333)
				(-93.1513269833,0.9327777778)
				(-92.8903541267,0.9272222222)
				(-92.6947913729,0.9216666667)
				(-92.5343091856,0.9161111111)
				(-92.331343498,0.9105555556)
				(-92.1321144673,0.905)
				(-92.02657158,0.8994444444)
				(-91.8395515664,0.8938888889)
				(-91.5189577913,0.8883333333)
				(-91.3131197258,0.8827777778)
				(-90.9908151948,0.8772222222)
				(-90.8689847175,0.8716666667)
				(-90.732948799,0.8661111111)
				(-90.614492613,0.8605555556)
				(-90.3892572023,0.855)
				(-90.1305812361,0.8494444444)
				(-89.9099556538,0.8438888889)
				(-89.7835991256,0.8383333333)
				(-89.6034195948,0.8327777778)
				(-89.4377710359,0.8272222222)
				(-89.3289366919,0.8216666667)
				(-89.048417725,0.8161111111)
				(-88.8892658059,0.8105555556)
				(-88.6607265109,0.805)
				(-88.5439923998,0.7994444444)
				(-88.4032073035,0.7938888889)
				(-88.2210884723,0.7883333333)
				(-88.0901891032,0.7827777778)
				(-87.9113239731,0.7772222222)
				(-87.7456264527,0.7716666667)
				(-87.5805376777,0.7661111111)
				(-87.4991590293,0.7605555556)
				(-87.385508837,0.755)
				(-87.2515223582,0.7494444444)
				(-87.156113639,0.7438888889)
				(-86.8875905166,0.7383333333)
				(-86.8257662308,0.7327777778)
				(-86.6920518563,0.7272222222)
				(-86.6300175187,0.7216666667)
				(-86.5465236587,0.7161111111)
				(-86.3917997281,0.7105555556)
				(-86.2282301101,0.705)
				(-86.0845459878,0.6994444444)
				(-85.9720848139,0.6938888889)
				(-85.8398516084,0.6883333333)
				(-85.7147064757,0.6827777778)
				(-85.6057559647,0.6772222222)
				(-85.4661386917,0.6716666667)
				(-85.3671121952,0.6661111111)
				(-85.3168353836,0.6605555556)
				(-85.254293677,0.655)
				(-85.1733422144,0.6494444444)
				(-85.1006257423,0.6438888889)
				(-85.0412906291,0.6383333333)
				(-84.9455079065,0.6327777778)
				(-84.8330158079,0.6272222222)
				(-84.7145851893,0.6216666667)
				(-84.6149843587,0.6161111111)
				(-84.53391185,0.6105555556)
				(-84.4652234747,0.605)
				(-84.403751946,0.5994444444)
				(-84.3265338165,0.5938888889)
				(-84.2563121081,0.5883333333)
				(-84.1951872593,0.5827777778)
				(-84.129789586,0.5772222222)
				(-84.0347656042,0.5716666667)
				(-83.9458717873,0.5661111111)
				(-83.8803035701,0.5605555556)
				(-83.8316753441,0.555)
				(-83.7764814498,0.5494444444)
				(-83.7236503224,0.5438888889)
				(-83.6756766292,0.5383333333)
				(-83.6206611712,0.5327777778)
				(-83.5564974251,0.5272222222)
				(-83.4994913426,0.5216666667)
				(-83.4323376917,0.5161111111)
				(-83.3676396988,0.5105555556)
				(-83.2948046399,0.505)
				(-83.2434390318,0.4994444444)
				(-83.1964262327,0.4938888889)
				(-83.1625879993,0.4883333333)
				(-83.0995819134,0.4827777778)
				(-83.0086384099,0.4772222222)
				(-82.977485083,0.4716666667)
				(-82.935192547,0.4661111111)
				(-82.8999762319,0.4605555556)
				(-82.8664596211,0.455)
				(-82.8198633751,0.4494444444)
				(-82.7843421442,0.4438888889)
				(-82.7551741392,0.4383333333)
				(-82.7011922968,0.4327777778)
				(-82.6885499246,0.4272222222)
				(-82.6636300502,0.4216666667)
				(-82.6368257285,0.4161111111)
				(-82.6212810104,0.4105555556)
				(-82.5976180083,0.405)
				(-82.5753328062,0.3994444444)
				(-82.5525888485,0.3938888889)
				(-82.5229781996,0.3883333333)
				(-82.4964158513,0.3827777778)
				(-82.4714489414,0.3772222222)
				(-82.4445612529,0.3716666667)
				(-82.4167326714,0.3661111111)
				(-82.3998934378,0.3605555556)
				(-82.3631339645,0.355)
				(-82.303263106,0.3494444444)
				(-82.2572036472,0.3438888889)
				(-82.2077785472,0.3383333333)
				(-82.1121718639,0.3327777778)
				(-82.0380422419,0.3272222222)
				(-81.9466794334,0.3216666667)
				(-81.9086367978,0.3161111111)
				(-81.8141724458,0.3105555556)
				(-81.7324348011,0.305)
				(-81.6263933631,0.2994444444)
				(-81.491707979,0.2938888889)
				(-81.3971842691,0.2883333333)
				(-81.3306718649,0.2827777778)
				(-81.2099603209,0.2772222222)
				(-81.1243668883,0.2716666667)
				(-80.9847744046,0.2661111111)
				(-80.8203108031,0.2605555556)
				(-80.7514015808,0.255)
				(-80.5894883923,0.2494444444)
				(-80.3936726458,0.2438888889)
				(-80.309955847,0.2383333333)
				(-80.1774605309,0.2327777778)
				(-80.0252965114,0.2272222222)
				(-79.7933263315,0.2216666667)
				(-79.5764718089,0.2161111111)
				(-79.35984846,0.2105555556)
				(-79.2194847974,0.205)
				(-79.0858569996,0.1994444444)
				(-78.8603100285,0.1938888889)
				(-78.6156286649,0.1883333333)
				(-78.3653628932,0.1827777778)
				(-78.1294594632,0.1772222222)
				(-78.0001064188,0.1716666667)
				(-77.6602940926,0.1661111111)
				(-77.2851522891,0.1605555556)
				(-77.0556779928,0.155)
				(-76.8148631049,0.1494444444)
				(-76.4728842515,0.1438888889)
				(-76.2504802278,0.1383333333)
				(-76.0154757334,0.1327777778)
				(-75.3701941213,0.1272222222)
				(-75.0168158589,0.1216666667)
				(-74.7225095578,0.1161111111)
				(-74.2754119833,0.1105555556)
				(-73.9950772089,0.105)
				(-73.3695090241,0.0994444444)
				(-72.6951132227,0.0938888889)
				(-72.2497945494,0.0883333333)
				(-71.5614544039,0.0827777778)
				(-70.9117162479,0.0772222222)
				(-70.3502692101,0.0716666667)
				(-69.7677559892,0.0661111111)
				(-69.0429465161,0.0605555556)
				(-67.8307567014,0.055)
				(-67.0795558509,0.0494444444)
				(-65.5133037631,0.0438888889)
				(-64.5047721756,0.0383333333)
				(-62.9126725322,0.0327777778)
				(-60.4758652274,0.0272222222)
				(-57.5254837567,0.0216666667)
				(-55.6603188449,0.0161111111)
				(-44.8997903954,0.0105555556)
				(-42.347533449432134,0.005)
			};
			\addplot[
			no marks,
			color = gray,
			] coordinates {
				( -78.966358726 , 1.0 )
				( -78.8476503774 , 0.994444444444 )
				( -78.7609851425 , 0.988888888889 )
				( -78.6758950743 , 0.983333333333 )
				( -78.6269658213 , 0.977777777778 )
				( -78.5265854067 , 0.972222222222 )
				( -78.4110320237 , 0.966666666667 )
				( -78.3664803889 , 0.961111111111 )
				( -78.3118802687 , 0.955555555556 )
				( -78.2638904412 , 0.95 )
				( -78.2024970552 , 0.944444444444 )
				( -78.0952482271 , 0.938888888889 )
				( -78.0592290025 , 0.933333333333 )
				( -77.9936438832 , 0.927777777778 )
				( -77.9413578295 , 0.922222222222 )
				( -77.8836995419 , 0.916666666667 )
				( -77.8066971745 , 0.911111111111 )
				( -77.7593152716 , 0.905555555556 )
				( -77.6912417577 , 0.9 )
				( -77.6649712604 , 0.894444444444 )
				( -77.6215652403 , 0.888888888889 )
				( -77.5693932157 , 0.883333333333 )
				( -77.51153497 , 0.877777777778 )
				( -77.4468077292 , 0.872222222222 )
				( -77.3836818995 , 0.866666666667 )
				( -77.3371962089 , 0.861111111111 )
				( -77.285397882 , 0.855555555556 )
				( -77.2285095863 , 0.85 )
				( -77.185404696 , 0.844444444444 )
				( -77.1163060065 , 0.838888888889 )
				( -77.0700426387 , 0.833333333333 )
				( -77.0283244576 , 0.827777777778 )
				( -76.9588619841 , 0.822222222222 )
				( -76.9360804156 , 0.816666666667 )
				( -76.8906319667 , 0.811111111111 )
				( -76.8450377952 , 0.805555555556 )
				( -76.7908933323 , 0.8 )
				( -76.7651502459 , 0.794444444444 )
				( -76.7142202089 , 0.788888888889 )
				( -76.6706233552 , 0.783333333333 )
				( -76.6467292864 , 0.777777777778 )
				( -76.5928121252 , 0.772222222222 )
				( -76.5652012842 , 0.766666666667 )
				( -76.5146213723 , 0.761111111111 )
				( -76.4707747528 , 0.755555555556 )
				( -76.4340788811 , 0.75 )
				( -76.3788100528 , 0.744444444444 )
				( -76.3405288557 , 0.738888888889 )
				( -76.3099939683 , 0.733333333333 )
				( -76.279018331 , 0.727777777778 )
				( -76.236898537 , 0.722222222222 )
				( -76.1840617111 , 0.716666666667 )
				( -76.1363711239 , 0.711111111111 )
				( -76.0752886911 , 0.705555555556 )
				( -76.0192125578 , 0.7 )
				( -75.9860587149 , 0.694444444444 )
				( -75.9351803619 , 0.688888888889 )
				( -75.896909639 , 0.683333333333 )
				( -75.8221761126 , 0.677777777778 )
				( -75.7930800839 , 0.672222222222 )
				( -75.7590388723 , 0.666666666667 )
				( -75.7139214309 , 0.661111111111 )
				( -75.6551330547 , 0.655555555556 )
				( -75.5946874744 , 0.65 )
				( -75.5453556862 , 0.644444444444 )
				( -75.4886429184 , 0.638888888889 )
				( -75.422614481 , 0.633333333333 )
				( -75.3716376045 , 0.627777777778 )
				( -75.3054057757 , 0.622222222222 )
				( -75.2618805593 , 0.616666666667 )
				( -75.203168721 , 0.611111111111 )
				( -75.1459278385 , 0.605555555556 )
				( -75.1065946616 , 0.6 )
				( -75.0361749144 , 0.594444444444 )
				( -74.9838601272 , 0.588888888889 )
				( -74.9341876706 , 0.583333333333 )
				( -74.8665578579 , 0.577777777778 )
				( -74.8015540841 , 0.572222222222 )
				( -74.7308796297 , 0.566666666667 )
				( -74.6474258467 , 0.561111111111 )
				( -74.603209986 , 0.555555555556 )
				( -74.5164557139 , 0.55 )
				( -74.4582736104 , 0.544444444444 )
				( -74.3990076422 , 0.538888888889 )
				( -74.3534601406 , 0.533333333333 )
				( -74.306113111 , 0.527777777778 )
				( -74.2387234239 , 0.522222222222 )
				( -74.1891283547 , 0.516666666667 )
				( -74.1207731339 , 0.511111111111 )
				( -74.0466705744 , 0.505555555556 )
				( -73.9971643471 , 0.5 )
				( -73.9534568646 , 0.494444444444 )
				( -73.8915149551 , 0.488888888889 )
				( -73.8295573766 , 0.483333333333 )
				( -73.7621139558 , 0.477777777778 )
				( -73.6183953469 , 0.472222222222 )
				( -73.5433157844 , 0.466666666667 )
				( -73.4845822034 , 0.461111111111 )
				( -73.4001004907 , 0.455555555556 )
				( -73.2853542665 , 0.45 )
				( -73.2009574982 , 0.444444444444 )
				( -73.1131894091 , 0.438888888889 )
				( -73.0437692188 , 0.433333333333 )
				( -72.9844867307 , 0.427777777778 )
				( -72.948480983 , 0.422222222222 )
				( -72.8466823504 , 0.416666666667 )
				( -72.7461095126 , 0.411111111111 )
				( -72.6444358626 , 0.405555555556 )
				( -72.5516137115 , 0.4 )
				( -72.4607709354 , 0.394444444444 )
				( -72.3874445125 , 0.388888888889 )
				( -72.2987790387 , 0.383333333333 )
				( -72.1752533751 , 0.377777777778 )
				( -72.0678152968 , 0.372222222222 )
				( -71.9269548774 , 0.366666666667 )
				( -71.8432763012 , 0.361111111111 )
				( -71.8044685432 , 0.355555555556 )
				( -71.7432520946 , 0.35 )
				( -71.6528149812 , 0.344444444444 )
				( -71.6050196492 , 0.338888888889 )
				( -71.5269366978 , 0.333333333333 )
				( -71.4759028183 , 0.327777777778 )
				( -71.4201359686 , 0.322222222222 )
				( -71.3518782915 , 0.316666666667 )
				( -71.2946838558 , 0.311111111111 )
				( -71.2300943062 , 0.305555555556 )
				( -71.0120842777 , 0.3 )
				( -70.7790387875 , 0.294444444444 )
				( -70.5626703867 , 0.288888888889 )
				( -70.3464279932 , 0.283333333333 )
				( -70.226774941 , 0.277777777778 )
				( -70.1216282733 , 0.272222222222 )
				( -70.0052471483 , 0.266666666667 )
				( -69.8662744484 , 0.261111111111 )
				( -69.6630757289 , 0.255555555556 )
				( -69.5728643136 , 0.25 )
				( -69.4301610322 , 0.244444444444 )
				( -69.3316134606 , 0.238888888889 )
				( -69.2672666765 , 0.233333333333 )
				( -69.1306461753 , 0.227777777778 )
				( -68.7585255384 , 0.222222222222 )
				( -68.6419910448 , 0.216666666667 )
				( -68.5037302843 , 0.211111111111 )
				( -68.259696761 , 0.205555555556 )
				( -67.9274756813 , 0.2 )
				( -67.7239676923 , 0.194444444444 )
				( -67.5158099349 , 0.188888888889 )
				( -67.2690231982 , 0.183333333333 )
				( -67.071030425 , 0.177777777778 )
				( -66.9540663436 , 0.172222222222 )
				( -66.8275492204 , 0.166666666667 )
				( -66.6550457751 , 0.161111111111 )
				( -66.5591119598 , 0.155555555556 )
				( -66.4116463985 , 0.15 )
				( -66.0928299042 , 0.144444444444 )
				( -65.6180983813 , 0.138888888889 )
				( -65.1043664525 , 0.133333333333 )
				( -64.7224822537 , 0.127777777778 )
				( -64.2555221679 , 0.122222222222 )
				( -63.7876135686 , 0.116666666667 )
				( -63.379738574 , 0.111111111111 )
				( -62.9060803178 , 0.105555555556 )
				( -62.6149117885 , 0.1 )
				( -62.3010035616 , 0.0944444444444 )
				( -62.0703585467 , 0.0888888888889 )
				( -61.8880027767 , 0.0833333333333 )
				( -61.1423817431 , 0.0777777777778 )
				( -59.4816669613 , 0.0722222222222 )
				( -58.235447403 , 0.0666666666667 )
				( -57.7350423768 , 0.0611111111111 )
				( -57.2685692929 , 0.0555555555556 )
				( -54.6724163005 , 0.05 )
				( -52.8690941391 , 0.0444444444444 )
				( -51.6768822448 , 0.0388888888889 )
				( -50.2376073923 , 0.0333333333333 )
				( -49.5138478272 , 0.0277777777778 )
				( -48.8314880157 , 0.0222222222222 )
				( -48.5140004668 , 0.0166666666667 )
				( -46.4393700047 , 0.0111111111111 )
				( -44.336081879 , 0.00555555555556 )
			};
			\draw[->] (axis cs:-75, .03) node[anchor=east] {1, 2, 10 users} -- (axis cs:-45,.1);
		\end{semilogyaxis}
		\end{tikzpicture}
		\fi
		\caption{\ifdraft{}Left\else{}Above\fi: The array \ACLR as measured at different angles to a uniform linear array with 100 antennas that serves different number of users when the emitted signal has \num{-42}\,dB \ACLR.  \ifdraft{}Right\else{}Below{}\fi: The distribution of the array \ACLR if the reference point is a considered random and its angle to the array is uniformly distributed on the interval \SI{-90}{\degree} to \SI{90}{\degree}.  The amplifiers are backed off 8\,dB from the one-dB compression point and all users are served with the same power.}
		\label{fig:arrayACLR}
	\end{figure}

	The array \ACLR depends on the location of the reference
	point.  In many cases, however, the out-of-band radiation is
	isotropic, as in
	Figure~\ref{fig:inband_radiation_pattern_OFDM}.  Then, the
	reference point matters little.  In other cases, it might be
	desirable to treat the reference point as a stochastic
	variable and estimate the distribution of the array \ACLR, to
	obtain a percentile, as was discussed in connection
	to \eqref{eq:919u891991}.  This is illustrated for a uniform
	linear array and line-of-sight propagation in
	Figure~\ref{fig:arrayACLR}.  It can be seen that the
	array \ACLR is much smaller than the \ACLR most of the time.
	Only in the worst case is the array \ACLR equal to the \ACLR,
	which happens when a single user is served in a narrow beam
	towards the served user.
	
	The advantages of the array \ACLR are: (i) It is easy to measure and
	a standardized test can be set up in a reverberation
	chamber \cite{holloway2006use}. (ii) It is a
	generalization of the classical \ACLR to arrays.  How to
	fairly measure out-of-band radiation from large arrays is also
	discussed in \cite{UGUSGC14, mollen2016OOB, federal2016fcc},
	where other measures are proposed and evaluated.
	
	\section{Case Studies} To draw conclusions about the
	directivity of the distortion and to illustrate the derived
	power spectral densities, some case studies are provided in
	this section, see Table~\ref{tab:CaseStudies}.  The first
	three cases study single-carrier transmission to show that the
	distortion practically is omnidirectional when there are
	multiple users or multiple channel taps.  The extension
	to \OFDM is straightforward, albeit cumbersome, and the
	results are the same.
	
	\begin{table}
		\centering
		\caption{Case Studies per Section}
		\label{tab:CaseStudies}
		\begin{tabular}{rll}
			\toprule
			& single-carrier & \OFDM\\
			\midrule
			frequency-flat fading & \ref{sec:case_study_ff_fading}, \ref{sec:9398821} & \ref{sec:893858190111223}, \ref{sec:9885728}\\
			frequency-selective fading & \ref{sec:028857612} & {*}\\
			\bottomrule
		\end{tabular}
		
		\textsuperscript{*} Section~\ref{sec:893858190111223} discusses how the results from D carries over to \OFDM when all users are served on all subcarriers.
	\end{table}
	
	The last cases are about \OFDM transmission and how subcarrier-specific beamforming affects the beamforming of the distortion.  The carrier frequency and beamforming direction of the intermodulation products are given and the relation to the carrier frequency and beamforming directions of the subcarriers is given.  It turns out this relation is intricate and hard to interpret intuitively.  Therefore, a special case is studied, where only two subcarriers are active.  This results in a “spatial” two-tone test, for which the frequencies and beamforming directions of the intermodulation products are derived.
	
	In the case study with two active subcarriers, the signal is not Gaussian unless the transmitted symbols are Gaussian.  Strictly speaking, a non-Gaussian distribution would require a different set of orthogonal basis polynomials.  We conjecture, however, that qualitatively the final results and conclusions will be the same, though the coefficients $\{a_{\varpi m}\}$ may be different.
	
	The main results in this section are: The number of distortion directions (or beamforming modes) grows as the cube of the number of significant users $K^3$ and the square of the number of significant channel taps $L^2$.  When the number of directions is greater than the number of antennas $M$, the distortion becomes omnidirectional.  The amount of distortion received by the served user scales as $M/K^2$, if the amplifiers are operated at the same input power and the power allocation to each user is proportional to $1/K$, until it saturates at approximately $\beta_\symfrak{x}\operatorname{tr}(\symbf{S}_{\symbf{dd}}(f))$.  All results are obtained from the mathematical formulas stated and derived in the previous sections.
	
	The effect of the reciprocity filter is to adjust for the differences in amplification between antennas and focus the beam of the desired signal $\symbf{u}(t)$.  In the study of the distortion, the reciprocity filter is neglected for clarity and $A_{1m}(f) = 1$ for all antennas $m$.  
	
	\subsection{Random Channel Generation}\label{sec:channel_generation}
	To illustrate the behavior of the distortion in the following sections, the channel model explained in this section will be used.  The theoretical results, however, are general and do not rely on the following assumed channel model.
	
	It will be assumed that the receivers are much farther away from the array than the aperture of the transmitter.  Then the propagating waves are approximately planar and the frequency response of the channel from the linear array to user $k$ is given by:
	\begin{align}\label{eq:088110087120781290}
		\symsfit{h}_{km}(f) = \frac{1}{\sqrt{V}} \sum_{v=1}^{V} e^{-j2\pi f (\tau_{kv} + \Delta_m \sin\theta_{kv} / c)},
	\end{align}
	where $\tau_{kv}$ is the delay of the signal from the reference antenna to user $k$ associated with propagation path $v$, the angle of departure $\theta_{kv}$ of path $v$ to user $k$ and the distance $\Delta_m$ between the reference antenna and antenna $m$.  The delays are assumed to be uniformly distributed between $0$ and the delay spread $\sigma_\tau$.
	
	The channel response in \eqref{eq:088110087120781290} will be used to model isotropic fading by assuming that the number of paths is large ($V=60$) and that the angle of departure $\theta_{kv}$ of each path is uniformly distributed over $[-\pi/2,\pi/2]$ and independent between different paths.  Different values of the delay spread will be used to model different degrees of frequency selectiveness.
	
	The same channel response \eqref{eq:088110087120781290} will also be used to model line-of-sight propagation.  Then there is one tap $V = 1$ and the delay spread is set to $\sigma_\tau = \SI{0.5}{\nano\second}$, which is the reciprocal of a carrier frequency of \SI{2}{GHz}, to model the randomness of the phase of the channel due to differences in propagation distance.  
	
	\subsection{Frequency-Flat Fading and Single-Carrier Transmission}\label{sec:case_study_ff_fading}
	A single-carrier scenario with one pulse, $N=1$, is considered.  It is assumed that the spectrum of the discrete channel to user $k$ is flat, i.e.\ $\symbfsf{h}_k[\theta]$ is constant over $\theta$.  Further, it is assumed that the same precoder is used at all frequencies, i.e.\ that $\symbfsf{W}_0[\theta]$ and $\symbf{S}^{(0)}_{\symbf{xx}}[\theta]$ are constant over $\theta$.  Because the precoding matrix is frequency flat, the third-degree term of the distortion, the first term in \eqref{eq:51628826687881} \ifdraft$\symbf{S}^{(3)}_{\symbf{xx}}(f) = 2 \Bigl( \symbf{S}_{\symbf{xx}}(\varphi) \circledstar \symbf{S}_{\symbf{xx}}(\varphi) \circledstar \symbf{S}^*_{\symbf{xx}}(-\varphi) \Bigr)\!(f),$\ \else
	\begin{align}
		\symbf{S}^{(3)}_{\symbf{xx}}(f) &= 2 \Bigl( \symbf{S}_{\symbf{xx}}(\varphi) \circledstar \symbf{S}_{\symbf{xx}}(\varphi) \circledstar \symbf{S}^*_{\symbf{xx}}(-\varphi) \Bigr)\!(f),
	\end{align}\fi
	which often dominates the distortion, is:
	\ifdraft
	\begin{align}
		\symbf{S}^{(3)}_{\symbf{xx}}(f) = \frac{2}{T^3} \Bigl(|\symsfit{p}_0(\varphi)|^2 \star |\symsfit{p}_0(\varphi)|^2 \star |\symsfit{p}_0(-\varphi)|^2 \Bigr)\!(f) \symbf{S}^{(0)}_{\symbf{xx}}[fT] \odot \symbf{S}^{(0)}_{\symbf{xx}}[fT] \odot {\symbf{S}^{(0)}_{\symbf{xx}}}^*[-fT],
	\end{align}
	\else
	\begin{align}
		\symbf{S}^{(3)}_{\symbf{xx}}(f) &= \frac{2}{T^3} \Bigl(|\symsfit{p}_0(\varphi)|^2 \star |\symsfit{p}_0(\varphi)|^2 \star |\symsfit{p}_0(-\varphi)|^2 \Bigr)\!(f)\notag\\
		&\quad\times \symbf{S}^{(0)}_{\symbf{xx}}[fT] \odot \symbf{S}^{(0)}_{\symbf{xx}}[fT] \odot {\symbf{S}^{(0)}_{\symbf{xx}}}^*[-fT],
	\end{align}
	\fi
	where $\odot$ stands for elementwise product (Hadamard product).  The beamforming of the third-degree term of the distortion is thus determined by $\symbf{S}^{(0)}_{\symbf{xx}}[\theta] \odot \symbf{S}^{(0)}_{\symbf{xx}}[\theta] \odot {\symbf{S}^{(0)}_{\symbf{xx}}}^*[\theta]$, a product of the matrix in \eqref{eq:digital_tx_PSD199283}, which is constant over $\theta$.
	
	To study this third-degree term, the $(m,m')$\mbox{-}th term of the matrix $\symbf{A}^\conjtr_{3}(f) \symbf{S}^{(3)}_{\symbf{xx}}(f) \symbf{A}_{3}(f)$ is investigated closer.  It is given by:
	\ifdraft
	\begin{align}\label{eq:12020100}
		&S^{(3)}_{x_mx_{m'}}(f)\notag\\
		&=\!\! \smash[b]{\frac{2}{T^3} \Bigl(\! \abs{\symsfit{p}_0(\varphi)}^2 {\star} \abs{\symsfit{p}_0(\varphi)}^2 {\star} \abs{\symsfit{p}_0(-\varphi)}^2 \Bigr)\!(f) \!\sum_{\mathclap{k=1}}^{K} \sum_{\mathclap{k'=1}}^{K} \sum_{\mathclap{k''=1}}^{K}} \! \xi_k \xi_{k'} \xi_{k''} A_{3m}(f) w_{mk} w_{mk'} w^*_{mk''} \left(A_{3m'}(f)\!w_{m'k} w_{m'k'} w^*_{m'k''}\!\right)^{\!\!*}\!\!,
	\end{align}
	\else
	\begin{align}
		S^{(3)}_{x_mx_{m'}}(f) &= \frac{2}{T^3} \Bigl( \abs{\symsfit{p}_0(\varphi)}^2 \star \abs{\symsfit{p}_0(\varphi)}^2 \star \abs{\symsfit{p}_0(-\varphi)}^2 \Bigr)(f)\notag\\
		&\hspace{-5em}\times\!\sum_{k=1}^{K} \sum_{k'=1}^{K} \sum_{k''=1}^{K}\! \xi_k \xi_{k'} \xi_{k''} A_{3m}(f) w_{mk} w_{mk'} w^*_{mk''} \! \left(\!A_{3m'}(f) w_{m'k} w_{m'k'} w^*_{m'k''}\!\right)^*\!\!,\label{eq:12020100}
	\end{align}
	\fi
	where $w_{mk}$ is the $(m,k)$\mbox{-}th element of the frequency-flat precoding matrix $\symbfsf{W}_0[\theta]$.  We compare the structure of this term and the corresponding term of the linearly amplified signal in \eqref{eq:918183}:
	\begin{align}\label{eq:9190128022}
	S_{u_mu_{m'}}(f) = \frac{1}{T} \abs{\symsfit{p}_0(f)}^2 \sum_{k=1}^{K}  \xi_k w_{mk} w^*_{m'k},
	\end{align}
	which we know is beamformed in the directions given by the precoding vectors:
	\ifdraft
	\begin{align}\label{eq:8191820093}
		\left\{(w_{1k}, \ldots, w_{Mk})^\tr : k=1,\ldots,K \right\}.
	\end{align}
	\else
	\begin{align}\label{eq:8191820093}
		\left\{\begin{pmatrix}
		w_{1k}\\
		\vdots\\
		w_{Mk}
		\end{pmatrix} : k=1,\ldots,K \right\}.
	\end{align}
	\fi
	The beamforming directions of the linear term are thus given by the terms that show up as conjugated pairs in the sum in \eqref{eq:9190128022}.  In the same way, the beamforming directions of the third-degree distortion term are given by:
	\ifdraft
	\begin{align}\label{eq:01902803001012}
	\left\{(A_{3,1}(f) w_{1k} w_{1k'} w^*_{1k''}, \ldots, A_{3M}(f) w_{Mk} w_{Mk'} w^*_{Mk''})^\tr : k,k',k''=1,\ldots,K \right\}.
	\end{align}
	\else
	\begin{align}\label{eq:01902803001012}
	\left\{\begin{pmatrix}
	w_{1k} w_{1k'} w^*_{1k''}\\
	\vdots\\
	w_{Mk} w_{Mk'} w^*_{Mk''}
	\end{pmatrix} : k,k',k''=1,\ldots,K \right\}.
	\end{align}
	\fi
	By counting the number of vectors in this set, it is seen that the distortion is beamformed in more directions than the linearly amplified signal.  Note that the directions in \eqref{eq:01902803001012} that are given by $(k,k',k'') = (k_0,k'_0,k_0'')$ and $(k,k',k'') = (k'_0,k_0,k_0'')$ are identical for all choices of $(k_0,k'_0,k''_0)$.  Straightforward combinatorial arguments give the following conclusion.
	\begin{theorem}\label{the:nr_dist_directions}
		In general, the number of vectors in \eqref{eq:01902803001012}, and thus the number of directions of the third-degree term, is at most $(K^3 + K^2)/2$.  
	\end{theorem}
	
\begin{figure}
		\centering
		\includegraphics{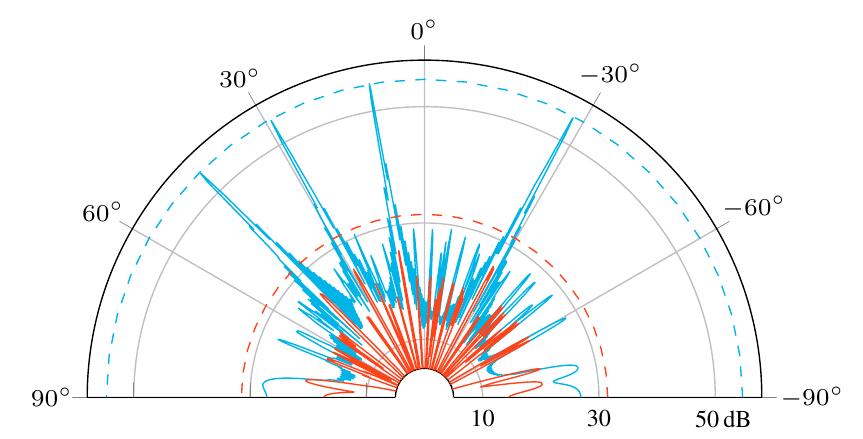}
		\caption{The radiation pattern at $f=0$ and $f=B$ in a single-carrier system with $K=4$ users and $M = 300$ antennas in line-of-sight.  Even though it is difficult to count the number of directions, in which the distortion at $f=B$ is beamformed, because the beams partly overlap, it can be seen that the predicted number $(K^3+K^2)/2 = 40$ is reasonable.}
		\label{fig:rad_pattern_four_users}
	\end{figure}

	Thus in a scenario with four users, $K=4$, the distortion should be radiated in approximately $(K^3 + K^2)/2 = 40$ directions.  Figure~\ref{fig:rad_pattern_four_users} shows such a scenario in a line-of-sight setting.  Even though many of the lobes partly overlap, a count shows that the number is reasonable.
	
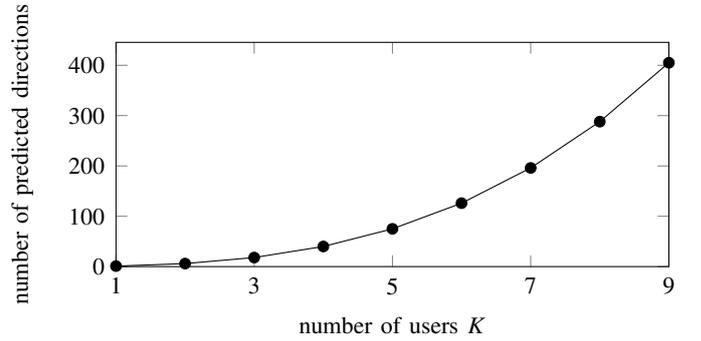
\begin{figure}
		\centering
		\begin{tikzpicture}
			\begin{axis}[
			xlabel={number of users $K$},
			ylabel={number of predicted directions},
			xmin=1,
			xmax=9,
			ymin=0,
			xtick={1,3,5,7,9},
			]
			\addplot[
			samples at={1,2,3,4,5,6,7,8,9},
			mark=*,
			] {(x^3+x^2)/2};
			\end{axis}
		\end{tikzpicture}
		\caption{The maximum number of directions of the third-degree term of the distortion, $(K^3+K^2)/2$, for different number of served users $K$.}
		\label{fig:nr_pred_dist_directions}
	\end{figure}

	Since the signal space is $M$ dimensional, the uncorrelated distortion can only be omnidirectional if the number of directions is greater than the number of dimensions, i.e.\ when $(K^3 + K^2)/2 > M$.  This number is shown in Figure~\ref{fig:nr_pred_dist_directions}.  For example, for an array with $M=100$ antennas, the distortion becomes omnidirectional at $K \geq 6$ users. 
	
	\begin{remark}\label{rem:38229222231196}
		The directions of the third-degree distortion are affected by the amplifier characteristics and operating point of the amplifiers given by the diagonal third-degree Hermite matrix $\symbf{A}_3(f)$, as is seen in \eqref{eq:51628826687881}.  It can be seen in \eqref{eq:81183368292} that the diagonal elements in $\symbf{A}_3(f)$ are non-zero for a system that is not perfectly linear and that the matrix thus has full rank and does not affect the \emph{number} of directions of the distortion.  In general, the diagonal elements in $\symbf{A}_3(f)$ are different and the amplifier characteristics affect the direction of the distortion.  In the special case, where the powers of the input signals all are equal, the diagonal elements in $\symbf{A}_3(f)$ are equal too and the amplifier characteristics do not affect the directions of the distortion.  This can happen if the channel coefficients between the array and the user all have the same modulus and maximum-ratio precoding is used, e.g., when there is only one strong propagation path between the array and each user. 
	\end{remark}
	
	As can be seen in \eqref{eq:12020100}, the beamforming directions of the third-degree distortion term are scaled by $\xi_k\xi_{k'}\xi_{k''}$.  If all users are allocated the same power, i.e.\ if $\xi_k$ is the same for all $k$, only then will all the $(K^3+K^2)/2$ directions be significant.  If the power allocation is not uniform, then only the directions, for which $\xi_k\xi_{k'}\xi_{k''}$ is large, are significant.  To approximate the number of directions in this case, we can assume that $\xi_k = 0$ for non-significant users, i.e.\ users $k$ whose power allocation $\xi_{k} \ll \max\{\xi_{k'}\}$.  The remaining $K'$ users then give rise to $(K'^3 + K'^2)/2$ distortion directions, and $(K'^3 + K'^2)/2 > M$ is a necessary requirement for the distortion to be omnidirectional.

	Furthermore, if there is a single dominant user, i.e.\ a user $k$ such that $\xi_k \gg \xi_{k'}$ for all $k' \neq k$, the distortion is mostly directed in one direction, given by $(A_{3,1}(f)w_{1k} |w_{1k}|^2, \ldots, A_{3M}(f) w_{Mk} |w_{Mk}|^2)^\tr\!$, which is similar to the direction of the dominant user $(w_{1k}, \ldots, w_{Mk})^\tr$.

\begin{remark}
	In the following, we will argue that the distortion power in the strongest direction scales approximately as $M/K^2$.  For simplicity, the influence of the amplifier characteristics given by the matrices $\{\symbf{A}_\varpi(f)\}$ on the directions of the beamforming is neglected.  As noted in Remark~\ref{rem:38229222231196}, this effect can be neglected when the transmit powers at the different antennas are close to equal, for example, when a line-of-sight channel is considered.  
	
	For the indices $(k,k',k'') = (k_0,k_0',k_0')$, $k_0' = 1, \ldots, K$, each coefficient of the beamforming vector \ifdraft $(A_{3,1}(f)w_{1k}w_{1k'}w^*_{1k''}, \ldots, A_{3M}(f)w_{Mk}w_{Mk'}w^*_{Mk''})^\tr$ in \eqref{eq:01902803001012} \else in \eqref{eq:01902803001012},
	\begin{align}\label{eq:19383839191000}
	(A_{3,1}(f)w_{1k}w_{1k'}w^*_{1k''}, \ldots, A_{3M}(f)w_{Mk}w_{Mk'}w^*_{Mk''})^\tr,
	\end{align}\fi%
	shares the same relative phases as the linearly amplified term that is beamformed in the direction $(w_{1k_0}, \ldots, w_{Mk_0})^\tr$, assuming that $\{A_{3m}(f)\}$ have the same phase for all antennas $m$.  Thus, the array gain in the direction of user $k_0$ is the same for the third-degree distortion and the linearly amplified signal, whose array gain scales linearly with the number of antennas $M$.  Furthermore, there are at least $K$ distortion terms that build up constructively at each user $k_0$.  If we assume uniform power allocation, i.e.\ $\xi_k=\xi_{k'}=\xi_{k''} = 1/K$, then the distortion power of one of the terms in the sum \eqref{eq:12020100} decreases as $\xi_k\xi_{k'}\xi_{k''} = 1/K^3$ as $K$ grows.  Because there are $K$ of these terms that build up constructively at each user, with an array gain that is proportional to $M$, the received distortion power is proportional to $M/K^2$ for different number of antennas $M$ and users $K$.  This proportionality only applies as the number of directions is significantly smaller than the signal space, i.e.\ when $(K^3+K^2)/2 \ll M$.  When the number of distortion directions increases and approaches the dimension of the space, the distortion becomes omnidirectional and the distortion power stops decreasing and approaches the constant level $\beta_\symfrak{x} S_\text{tx}(f)$.
\end{remark}

\begin{remark}\label{rem:2771002}
	A consequence of the fact that the received distortion power at the served user scales as $M/K^2$ when all amplifiers are operated at the same power level, is that the received distortion power does not vanish in the limit of infinite number of antennas and a fixed number of users, which is a scenario where $(K^3+K^2)/2 \ll M$ holds.  The received \SINR after \textsc{iq} demodulation is then limited by the ratio between power of the transmitted linear term and the transmitted distortion.  Since this ratio commonly is tens of decibels, this limitation might be of little practical consequence however.
\end{remark}

\begin{remark}
	The direction of the distortion in \eqref{eq:01902803001012} is a function of the precoding weights.  With knowledge of the nonlinearity characteristics $\{A_{\varpi m}(f)\}$, it is therefore possible to steer the distortion away from the served user, i.e.\ make the distortion vector \eqref{eq:01902803001012} orthogonal to the channel of the user.  With such distortion steering, the scaling of the received distortion power in Remark~\ref{rem:2771002} would be different, and the distortion would not necessary upper bound the received \SINR in the limit of infinite number of antennas.  Distortion steering would, however, reduce the array gain of the desired signal and require knowledge of the nonlinearity coefficients.  Distortion steering is further complicated by the fact that the coefficients $\{A_{\varpi m}(f)\}$ depend on the per-antenna transmit power and thus the precoding weights.  Nevertheless, such distortion steering would improve performance, especially in a system where most of the transmit power is beamformed towards one user and a significant amount of distortion is radiated in the direction of the users that are served with little power.  
\end{remark}

	If there is only one user and maximum-ratio precoding is used, the precoding weights are $w_{m1} = \symsfit{h}^*_{1m}[\theta]$, where $\symsfit{h}_{1m}[\theta]$ is the $m$\mbox{-}th element of the channel vector $\symbfsf{h}_{1}[\theta]$.  The only direction of the third-degree distortion term is then
	\begin{align}
		(A_{3,1}(f) \symsfit{h}^*_{11}[\theta] |\symsfit{h}_{11}[\theta]|^2, \ldots, A_{3M}(f) \symsfit{h}^*_{1M}[\theta] |\symsfit{h}_{1M}[\theta]|^2)^\tr.\label{eq:26891309}
	\end{align}
	When the coefficients $\{A_{3m}(f)\}$ have the same phases for all antennas $m$, the elements of this vector have the same relative phases as the linearly amplified term, which is beamformed in the direction given by $(\symsfit{h}^*_{11}[\theta], \ldots, \symsfit{h}^*_{1M}[\theta])^\tr$.  The radiation pattern of the distortion is therefore similar to the radiation pattern of the desired signal: the distortion builds up constructively at the served user and destructively in almost all other directions.
	
	\subsection{Narrowband Line-of-Sight and Maximum-Ratio Precoding}\label{sec:9398821}
	For simplicity of the exposition, in this section, where line-of-sight propagation will be investigated, we assume that the array is uniform with antenna spacing $\Delta$.  We also use the narrowband assumption, i.e.\ assume that the channel response to user $k$, who stands at an angle $\theta_k$ to the array, is frequency flat and given by:
	\ifdraft
	\begin{align}\label{eq:LoS_channel10292929}
	\symbfsf{h}_{k}[\theta] = (e^{j \phi_k}, e^{j 2 \phi_k}, \ldots, e^{j M \phi_k})^\tr, \quad \forall \theta,
	\end{align}
	\else
	\begin{align}\label{eq:LoS_channel10292929}
		\symbfsf{h}_{k}[\theta] = \begin{pmatrix}
		e^{j \phi_k}\\
		e^{j 2 \phi_k}\\
		\vdots\\
		e^{j M \phi_k}
		\end{pmatrix}, \quad \forall \theta,
	\end{align}
	\fi
	where $\phi_k \triangleq -2 \pi \sin(\theta_k) \Delta / \lambda$ and $\lambda = c/f_\text{c}$ is the wavelength of the carrier frequency $f_\text{c}$.  The illustrations are however generated without the narrowband assumption, using the channel described in Section~\ref{sec:channel_generation}. 
	
	If maximum-ratio transmission is used, the $(m,m')$\mbox{-}th element in the linear part of the radiation pattern is given by \eqref{eq:9190128022} as:
	\begin{align}
		S_{x_mx_{m'}}(f) = \frac{1}{T M} |\symsfit{p}_0(f)|^2 \sum_{k=1}^{K} \xi_k e^{j \phi_k (m' - m)}.
	\end{align}
	The $K$ beamforming directions are thus given by the phases $\{\phi_k : k=1,\ldots,K\}$ in the exponent.  This can be compared to the radiation pattern of the third-degree term of the uncorrelated distortion:
	\ifdraft
	\begin{align}
		S^{(3)}_{x_mx_{m'}}(f) \!=\! \smash[b]{\frac{A_{3m}(f)A^*_{3m'}(f)}{T^3 M^3} \Bigl(|\symsfit{p}_0(\varphi)|^2 \star |\symsfit{p}_0(\varphi)|^2 \star |\symsfit{p}_0(-\varphi)|^2 \Bigr)\!(f) \sum_{\mathclap{k=1}}^{K} \sum_{\mathclap{k'=1}}^{K} \sum_{\mathclap{k''=1}}^{K}} \xi_k \xi_{k'} \xi_{k''} e^{j (\phi_k + \phi_{k'} - \phi_{k''}) (m' - m)}.
	\end{align}
	\else
	\begin{align}
		S^{(3)}_{x_mx_{m'}}(f) &= \frac{A_{3m}(f)A_{3m'}(f)}{T^3 M^3} \Bigl(|\symsfit{p}_0(\varphi)|^2 \star |\symsfit{p}_0(\varphi)|^2 \star |\symsfit{p}_0(-\varphi)|^2 \Bigr)(f)\notag\\
		&\quad\times\sum_{k=1}^{K} \sum_{k'=1}^{K} \sum_{k''=1}^{K} \xi_k \xi_{k'} \xi_{k''} e^{j (\phi_k + \phi_{k'} - \phi_{k''}) (m' - m)}.
	\end{align}
	\fi
	Because the power of the transmit signals is the same at all antennas and all amplifiers are identical and operated with the same input power, the coefficients $\{A_{3m}(f)\}$ are the same for all antennas $m$ and do not affect the beamforming directions.  We see that the distortion is beamformed in more directions than the linearly amplified term, which is stated by the following theorem that also gives the beamforming directions of the distortion.  
	\begin{theorem}
		The third-degree distortion is beamformed in the $(K^3+K^2)/2 - K(K-1) = (K^3-K^2+2K)/2$ directions that are given by the phases $\{\phi_k + \phi_{k'} - \phi_{k''} : k,k',k''= 1, \ldots, K\}$.
	\end{theorem}
	\begin{IEEEproof}
		The phase $\phi_k + \phi_{k'} - \phi_{k''}$ is the same for $(k,k',k'') = (k_0,k'_0,k_0'')$ and $(k,k',k'') = (k'_0,k_0,k_0'')$ as in Theorem~\ref{the:nr_dist_directions}.  Additionally, the phase equals $\phi_{k_0}$ when $(k,k',k'') = (k_0,k_0',k_0')$ for all $k_0'$.
	\end{IEEEproof}
	It is noted that the original beamforming directions (given by $\{\phi_k\}$) of the linearly amplified term are among the directions of the distortion (obtained when $k'=k''$).  
	
	\begin{remark}\label{rem:rad_pattern_same_in_singleuserLOS}
		In the special case, where there is only a single user, $K=1$, it is evident from \eqref{eq:26891309} that the beamforming pattern of the distortion is identical to that of the linearly amplified term.  This is different from the general case studied in Section~\ref{sec:case_study_ff_fading}, where we only could conclude that the distortion would combine constructively at the served user if no attempt is made to steer it away.  
	\end{remark}
	
	A consequence of Remark~\ref{rem:rad_pattern_same_in_singleuserLOS} is that, in a comparison between a single-antenna transmitter and an antenna array, where the amplifiers have the same operating point as in the single-antenna transmitter, the amount of received distortion at the one served user is the \textit{same} in the two systems independently of the number of antennas in the array.  In other directions, however, barely any distortion is received from the array, which stands in contrast to the single-antenna array that radiates distortion in all directions.  This point was not correctly described in \cite{mollen2016OOB}, where it was claimed that the distortion always has an array gain smaller than the desired signal.
	
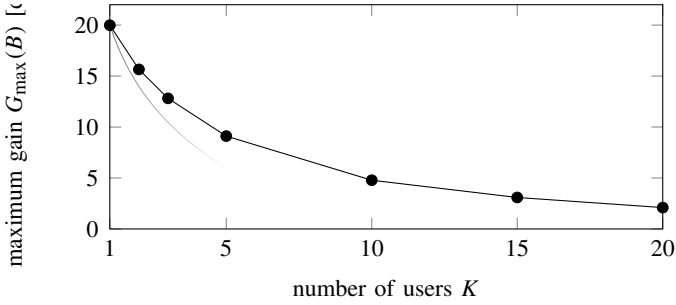
\begin{figure}
		\centering
		\begin{tikzpicture}
			\begin{axis}[
			ymin=0,
			xmin=1,
			xmax=20,
			xtick ={1,5,10,15,20},
			xlabel={number of users $K$},
			ylabel={\rule{0pt}{0pt}\clap{maximum gain $G_\text{max}(B)$ [dBi]}\rule{0pt}{0pt}},
			]
			\addplot[
			mark=*,
			]table[x=nruser,y=gain, row sep=\\]{
				nruser gain\\
				1	19.99067499\\
				2	15.65131957\\
				3	12.81271174\\
				5	9.1104536\\
				10	4.78170861\\
				15	3.08951599\\
				20	2.09818075\\
				};
  				
			\addplot[
			color=gray,
			path fading=east,
			domain=1:5,
			]{10*log10(100/x^2)};				
			\end{axis}
		\end{tikzpicture}
		\caption{The maximum gain of the distortion at $f = B$ (the center frequency of the adjacent band to the right) in a single-carrier system with $M = 100$ antennas that serves $K$ users over a line-of-sight channel.  The grey curve shows the approximation $M/K^2$ from Section~\ref{sec:case_study_ff_fading}, which only is applicable when $(K^3+K^2)/2 \ll M$, i.e.\ when $K<6$.  The amplifiers are operated \SI{7}{dB} below the one-dB compression point.}
		\label{fig:max_gain}
	\end{figure}

	Figure~\ref{fig:max_gain} shows how the maximum beamforming gain at the out-of-band frequency $f = B$ is changing as the number of users increases.  As expected, the signal becomes more and more omnidirectional as the number of users is increased, which is seen on the decreasing maximum beamforming gain.  The approximation $1/K^2$ obtained in Section~\ref{sec:case_study_ff_fading}, is seen to hold for small number of users.  For a signal space with $M=100$ dimensions, however, the approximation rapidly becomes loose as the number of users increases.
	
	\subsection{Frequency-Selective Fading}\label{sec:028857612}

	Next, a single-carrier scenario with a general
	frequency-selective channel is considered.  Many of the
	results from the frequency-flat scenario carry over to the
	frequency-selective case: the distortion is beamformed, the
	directions of the beamforming are functions of the beamforming
	directions of the input signal, and the number of directions
	grows with the number of input beamforming directions.  A
	difference, however, is that the out-of-band radiation is not
	necessarily beamformed to the served users, since their
	out-of-band channels are different from their in-band
	channels, and that the number of directions also scales with
	the number of significant taps in the precoding filter, which
	is approximately the same as the number of significant taps in
	the channel impulse response.
	
	By denoting column $k$ of the precoding matrix
	$\symbfsf{W}_0[\theta]$ by $\symbfsf{w}_{k}[\theta]$, the
	power spectral density of the third-degree term of the
	distortion can be written
	as: \ifdraft \begin{align} \symbf{S}^{(3)}_{\symbf{xx}}(f) &=
	2 \left(\symbf{S}_{\symbf{xx}}(\varphi) \circledstar \symbf{S}_{\symbf{xx}}(\varphi) \circledstar \symbf{S}^*_{\symbf{xx}}(-\varphi) \right)\!(f)\\
	&=
	2 \int_{-\infty}^{\infty} \int_{-\infty}^{\infty} \symbf{S}_{\symbf{xx}}(\varphi) \odot \symbf{S}_{\symbf{xx}}(\varphi') \odot \symbf{S}^*_{\symbf{xx}}(\varphi
	+ \varphi' - f) \, \symrm{d}\varphi \symrm{d}\varphi'\\
	&= \frac{2}{T^3} \iint_{\symcal{B}(f)}
	|\symsfit{p}_{0}(\varphi)|^2 |\symsfit{p}_{0}(\varphi')|^2
	|\symsfit{p}_{0}(\varphi {+} \varphi' {-}
	f)|^2 \sum_{k=1}^K\sum_{k'=1}^{K}\sum_{k''=1}^{K} \xi_k\xi_{k'}\xi_{k''}\notag\\
	&\quad\times \!\!\biggl(\symbfsf{w}_{k}[\varphi T]
	{\odot} \symbfsf{w}_{k'}[\varphi' T]
	{\odot} \symbfsf{w}_{k''}[(\varphi{+}\varphi'{-}f)T]\biggr)\biggl(\symbfsf{w}_{k}[\varphi
	T] {\odot} \symbfsf{w}_{k'}[\varphi' T]
	{\odot} \symbfsf{w}_{k''}[(\varphi{+}\varphi'{-}f)T]\biggr)^{\!\!\conjtr} \!\symrm{d}\varphi \symrm{d}\varphi' \end{align} \else \begin{align} \symbf{S}^{(3)}_{\symbf{xx}}(f)
	&=
	2 \left(\symbf{S}_{\symbf{xx}}(\varphi) \circledstar \symbf{S}_{\symbf{xx}}(\varphi) \circledstar \symbf{S}^*_{\symbf{xx}}(-\varphi) \right)\!(f)\\
	&=
	2 \int_{-\infty}^{\infty} \int_{-\infty}^{\infty} \symbf{S}_{\symbf{xx}}(\varphi) \odot \symbf{S}_{\symbf{xx}}(\varphi') \odot \symbf{S}^*_{\symbf{xx}}(\varphi
	+ \varphi' - f) \, \symrm{d}\varphi \symrm{d}\varphi'\\
	&= \frac{2}{T^3} \iint_{\symcal{B}(f)}
	|\symsfit{p}_{0}(\varphi)|^2 |\symsfit{p}_{0}(\varphi')|^2
	|\symsfit{p}_{0}(\varphi {+} \varphi' {-} f)|^2 \notag\\
	&\quad\times\sum_{k=1}^K\sum_{k'=1}^{K}\sum_{k''=1}^{K} \xi_k\xi_{k'}\xi_{k''}\notag\\
	&\quad\times \biggl(\symbfsf{w}_{k}[\varphi
	T] \odot \symbfsf{w}_{k'}[\varphi'
	T] \odot \symbfsf{w}_{k''}[(\varphi{+}\varphi'{-}f)T]\biggr)\notag\\
	&\quad\times\biggl(\symbfsf{w}_{k}[\varphi
	T] \odot \symbfsf{w}_{k'}[\varphi'
	T] \odot \symbfsf{w}_{k''}[(\varphi{+}\varphi'{-}f)T]\biggr)^\conjtr \symrm{d}\varphi \symrm{d}\varphi'
	\end{align}
	\fi
	The integration is done over the two-dimensional area defined by the set $\symcal{B}(f)$.  If we assume that the pulse $\symsfit{p}_0(\varphi)$ is bandlimited to $[-B/2,B/2]$, the set equals:
	\begin{align}
		\symcal{B}(f) &= \{(\varphi,\varphi'): \varphi \in [a,b], \varphi' \in [a',b']\},
	\end{align}
	where the end values depend on $f$.  For example for $f \in [B/2,3B/2]$, the end values are:
	\ifdraft
	\begin{align}
		a &= f-B &	b &= B/2\\
		a' &= \begin{cases}
		\varphi - B/2, &\text{if } \varphi > 0\\
		-B/2, &\text{if } \varphi \leq 0
		\end{cases} & b' &= \begin{cases}
		B/2, &\text{if } \varphi > 0\\
		\varphi + B/2, &\text{if } \varphi \leq 0
		\end{cases}
	\end{align}
	\else
	\begin{align}
		a &= f-B\\
		b &= B/2\\
		a' &= \begin{cases}
		\varphi - B/2, &\text{if } \varphi > 0\\
		-B/2, &\text{if } \varphi \leq 0
		\end{cases}\\
		b' &= \begin{cases}
		B/2, &\text{if } \varphi > 0\\
		\varphi + B/2, &\text{if } \varphi \leq 0
		\end{cases}
	\end{align}
	\fi
	and the area, over which is integrated, is 
	\ifdraft
	\begin{align}
		A(f) = \iint_{\symcal{B}(f)}\symrm{d}\varphi\symrm{d}\varphi' = \frac{15}{8} B^2 - 2Bf + \frac{1}{2} f^2, \quad \text{for } f \in [B/2,3B/2].
	\end{align}
	\else
	\begin{align}
		A(f) = \iint_{\symcal{B}(f)}\symrm{d}\varphi\symrm{d}\varphi' = \frac{15}{8} B^2 - 2Bf + \frac{1}{2} f^2, 
	\end{align}
	for $f \in [B/2,3B/2]$.
	\fi
	
	To approximate the number of directions at a given frequency
	$f$, it will be assumed that the directions of the integrand
	change smoothly over the area of integration and that
	coherence interval of these changes is $1/\sigma_\tau$.
        The integral can
	thus be considered as a sum of $A(f)\sigma_\tau$ 
	integrands.  Each integrand is a sum of matrices with rank
	one, which is similar to the sum \eqref{eq:12020100} that was
	studied for frequency-flat fading in
	Section~\ref{sec:case_study_ff_fading}.  As was concluded in
	that section, the number of unique terms in the sum is
	approximately $(K^3+K^2)/2$.  The total number of directions
	is therefore approximately equal to \ifdraft $A(f)\sigma_\tau
	(K^3+K^2)/2$.  \else\begin{align}A(f)\sigma_\tau
	(K^3+K^2)/2.\end{align}\fi
        If we write the bandwidth in terms
	of the excess bandwidth $\alpha$ as $B = \alpha / T$, the
	number of integrands is thus
	approximately
        \begin{align} A(f) \sigma_\tau =
	(\sigma_\tau/T)^2 \underbrace{\biggl(\frac{15}{8}\alpha^2 -
	2\frac{f\alpha}{B}
	+ \frac{f^2\alpha^2}{2B^2}\biggr)}_{\triangleq \upsilon(f)} =
	L^2 \upsilon(f),
        \end{align}
        which is proportional to the
	square of the number of significant taps in the channel $L
	= \sigma_\tau / T$.  Thus, each of the $(K^3+K^2)/2$ terms contributes to approximately $L^2$ directions.  The number
	of directions of the distortion at frequency $f$ is therefore
	upper bounded
	by
        \begin{align} \min\left\{M, \frac{K^3+K^2}{2}
	L^2 \upsilon(f) \right\}.
        \end{align}
        An increased number of
	channel taps, thus, makes the distortion more isotropic, which
	is summarized in the following theorem.
        \begin{theorem} A
	necessary condition for the distortion to behave
	omnidirectionally
	is
        \begin{align}\label{eq:omnidirectional_requirement19929} \frac{K^3
	+ K^2}{2} L^2 \upsilon(f) \geq M.  \end{align}
        \end{theorem}
	
	
	
	A practical phenomenon with a significant impact on the amount
	of distortion created by the amplifiers is the variation in
	transmit power at individual amplifiers across time.  In an
	environment with isotropic fading, the channel coefficients of
	individual channels will vary and a few antennas, for which
	the channel coefficients are good, will use very high
	transmit power compared to the average.  The effect of this is
	that a few power amplifiers will be operated close to, or even
	in, saturation, which cause a few antennas to emit much more
	distortion than the average and an increase in the total
	amount of radiated distortion.
	
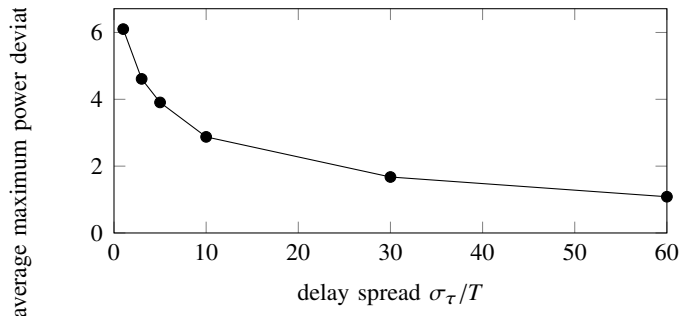
\begin{figure}
		\centering
		\begin{tikzpicture}
		\begin{axis}[
		xlabel={delay spread $\sigma_\tau/T$},
		ylabel={\rule{0pt}{0pt}\clap{average maximum power deviation [dB]}\rule{0pt}{0pt}},
		xmin=0,
		xmax=60,
		ymin=0,
		]
		\addplot[
		mark=*,
		]table[x=nruser,y=gain, row sep=\\]{
			nruser gain\\
			1	6.10076687\\
			3	4.61043211\\
			5	3.90630502\\
			10	2.87527319\\
			30	1.67481932\\
			60	1.08438073\\
		};			
		\end{axis}
		\end{tikzpicture}
		\caption{The difference between the average and maximum power of the transmit signals prior to amplification in an array with $M = 100$ antennas that serves $K = 1$ user over a channel with isotropic fading.  The definition of average maximum power deviation is given in \eqref{eq:ampd}.}
		\label{fig:power_variations}
	\end{figure}

	To illustrate this phenomenon, the transmit power of individual antennas was computed for many channel realizations.  The antenna with the highest transmit power during each channel realization has been compared to the average and the following \emph{average maximum power deviation} computed for different delay spreads:
	\begin{align}\label{eq:ampd}
		\Exp\left[\frac{\max_m\{\Exp[|x_m(t)|^2] \mid \symbf{H}]}{\Exp[|x_m(t)|^2 \mid \symbf{H}]}\right],
	\end{align}
	where $\Exp[\,\cdot \mid \symbf{H}]$ denotes expectation given a specific channel realization.  The outer expectation averages over channel realizations.  The average maximum power deviation is shown in Figure~\ref{fig:power_variations}, where it can be seen that, for channels with small delay spreads, the variations in power can be large---in this case up to \SI{6}{\decibel}.  

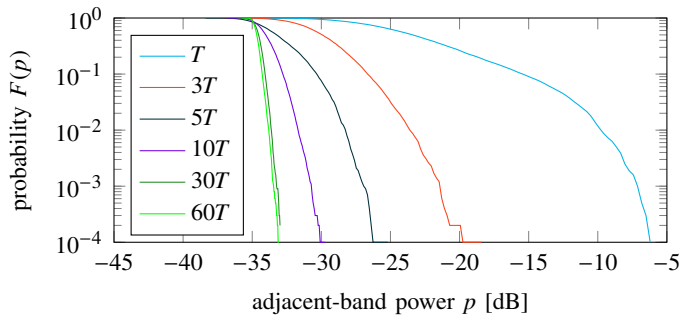
\begin{figure}
		\centering
		\begin{tikzpicture}
		\begin{semilogyaxis}[
		ymin=1e-4,
		ymax=1,
		xmin=-45,
		xmax=-5,
		xlabel={adjacent-band power $p$ [dB]},
		ylabel={probability $F(p)$},
		legend pos = south west,
		legend cell align=left,
		]
			
		\addplot[
		no marks,
		color=color1,
		]table[col sep=comma]{illustrationer/OOB_rad_CCDF_nr_taps_1.0.csv};
		\addlegendentry{$T$}
		
		\addplot[
		no marks,
		color=color2,
		]table[col sep=comma]{illustrationer/OOB_rad_CCDF_nr_taps_3.0.csv};
		\addlegendentry{$3T$}
		
		\addplot[
		no marks,
		color=color3,
		]table[col sep=comma]{illustrationer/OOB_rad_CCDF_nr_taps_5.0.csv};
		\addlegendentry{$5T$}
		
		\addplot[
		no marks,
		color=color4,
		]table[col sep=comma]{illustrationer/OOB_rad_CCDF_nr_taps_10.0.csv};
		\addlegendentry{$10T$}
		
		\addplot[
		no marks,
		color=color5,
		]table[col sep=comma]{illustrationer/OOB_rad_CCDF_nr_taps_30.0.csv};
		\addlegendentry{$30T$}
		
		\addplot[
		no marks,
		color=green,
		]table[col sep=comma]{illustrationer/OOB_rad_CCDF_nr_taps_60.0.csv};
		\addlegendentry{$60T$}		

		\end{semilogyaxis}
		\end{tikzpicture}
		\caption{Distribution of the normalized adjacent-distortion power from a uniform linear array with 100 antennas that are used to beamform a signal at an angle \SI{9}{\degree} off its normal.  The channel is assumed to be isotropic with delay spreads equal to different multiples of the symbol period $T$.  The amplifiers are backed off by \SI{8}{dB} from the one-dB compression point on average.}
		\label{fig:oob_ccdf}
	\end{figure}

	We have thus identified two phenomena connected to the delay spread: \ifdraft 1. The directivity of the distortion decreases with longer delay spreads. 2. The total amount of radiated distortion decreases with longer delay spreads.\ \else
	\begin{enumerate}
		\item The directivity of the distortion decreases with longer delay spreads.
		\item The total amount of radiated distortion decreases with longer delay spreads.
	\end{enumerate}
	\fi
	The combined effect of these phenomena can be seen in Figure~\ref{fig:oob_ccdf}, which shows the distribution \eqref{eq:919u891991} of the power received in the adjacent band.  It can be seen how the curves become more vertical as the delay spread increases; this is the effect of the lower directivity, which makes the received distortion power the same at all positions around the array.  It can also be seen how the curves move to the right as the delay spread decreases;\footnote{It should be noted that a line-of-sight channel does not result in variations in transmit power because all channel coefficients have the same modulus.} this is the effect of increased variations in transmit power caused by precoding and increased fading variations, which makes a few amplifiers operate much closer to saturation than on average and cause a high amount of distortion.

	From a distortion perspective, a long delay spread is thus beneficial since it reduces the power variations, which allows the amplifiers to be operated close to the chosen power level, and makes the distortion omnidirectional.  In an outdoor environment, a high delay spread is to be expected.  For example, if the maximum difference in length between two propagation paths is $d=\SI{1}{\kilo\metre}$, then the delay spread is approximately $\sigma_\tau \approx d/c \approx \SI{3}{\micro\second}$.  With a symbol period of $T=1/(\SI{20}{\mega\hertz})$, the delay spread is $\sigma_\tau \approx 67 T$.  In an indoor environment, however, the delay spread might be much shorter.

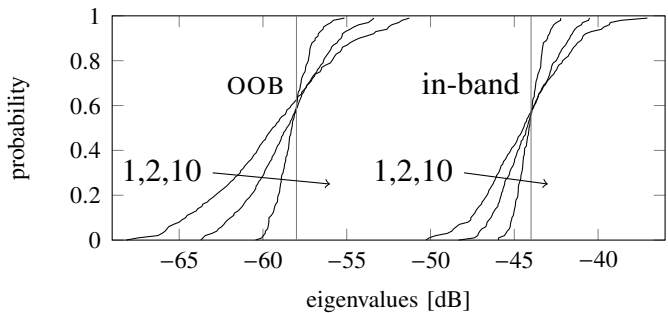
\begin{figure}
		\centering
		\begin{tikzpicture}
			\begin{axis}[
			xlabel={eigenvalues [dB]},
			ylabel={probability},
			ymin=0,
			ymax=1,
			xmin=-69,
			xmax=-36,
			]
			\addplot[
			color=black,
			no marks,
			] coordinates {
				( -68.1692469661 , 0.0 )
				( -67.4520367372 , 0.01 )
				( -66.2767936297 , 0.02 )
				( -66.0230927695 , 0.03 )
				( -65.8392311403 , 0.04 )
				( -65.7279187891 , 0.05 )
				( -65.6253166805 , 0.06 )
				( -65.342395388 , 0.07 )
				( -64.9974142409 , 0.08 )
				( -64.8412597298 , 0.09 )
				( -64.6168075951 , 0.1 )
				( -64.5284511145 , 0.11 )
				( -64.2536253008 , 0.12 )
				( -63.9852406771 , 0.13 )
				( -63.8927996052 , 0.14 )
				( -63.6496930508 , 0.15 )
				( -63.5958635208 , 0.16 )
				( -63.3164937085 , 0.17 )
				( -63.1924258629 , 0.18 )
				( -62.9956432844 , 0.19 )
				( -62.8449779259 , 0.2 )
				( -62.6833062998 , 0.21 )
				( -62.5185995422 , 0.22 )
				( -62.347600294 , 0.23 )
				( -62.1819951601 , 0.24 )
				( -62.0498275627 , 0.25 )
				( -61.930017586 , 0.26 )
				( -61.7505941571 , 0.27 )
				( -61.6703253942 , 0.28 )
				( -61.4990596029 , 0.29 )
				( -61.3593687088 , 0.3 )
				( -61.3251061042 , 0.31 )
				( -61.1738511912 , 0.32 )
				( -61.0931543101 , 0.33 )
				( -60.9733222293 , 0.34 )
				( -60.8547985391 , 0.35 )
				( -60.7018918951 , 0.36 )
				( -60.6204534309 , 0.37 )
				( -60.5404035005 , 0.38 )
				( -60.3999236552 , 0.39 )
				( -60.3426198438 , 0.4 )
				( -60.2614804683 , 0.41 )
				( -60.197740099 , 0.42 )
				( -60.0883663826 , 0.43 )
				( -59.9976273538 , 0.44 )
				( -59.8463434945 , 0.45 )
				( -59.804680919 , 0.46 )
				( -59.7215526494 , 0.47 )
				( -59.6220627263 , 0.48 )
				( -59.5152726125 , 0.49 )
				( -59.4298122501 , 0.5 )
				( -59.2584177619 , 0.51 )
				( -59.1925604584 , 0.52 )
				( -59.0520078124 , 0.53 )
				( -58.9977172856 , 0.54 )
				( -58.8872592129 , 0.55 )
				( -58.6759165338 , 0.56 )
				( -58.6164376441 , 0.57 )
				( -58.5716525151 , 0.58 )
				( -58.3926006735 , 0.59 )
				( -58.3194227042 , 0.6 )
				( -58.1898334346 , 0.61 )
				( -58.0730809503 , 0.62 )
				( -57.9492551662 , 0.63 )
				( -57.9237175558 , 0.64 )
				( -57.7060025901 , 0.65 )
				( -57.6817834484 , 0.66 )
				( -57.5562731745 , 0.67 )
				( -57.41918977 , 0.68 )
				( -57.378537394 , 0.69 )
				( -57.2257042127 , 0.7 )
				( -57.1123874794 , 0.71 )
				( -57.0564440674 , 0.72 )
				( -56.9326776626 , 0.73 )
				( -56.8884776435 , 0.74 )
				( -56.5993091989 , 0.75 )
				( -56.5225193533 , 0.76 )
				( -56.4854554088 , 0.77 )
				( -56.2653877713 , 0.78 )
				( -55.9778610481 , 0.79 )
				( -55.8836335564 , 0.8 )
				( -55.7987880094 , 0.81 )
				( -55.7244623321 , 0.82 )
				( -55.6186482471 , 0.83 )
				( -55.387473518 , 0.84 )
				( -55.3022406608 , 0.85 )
				( -55.2603504529 , 0.86 )
				( -54.9680759872 , 0.87 )
				( -54.8026489631 , 0.88 )
				( -54.5505318645 , 0.89 )
				( -54.2693041386 , 0.9 )
				( -53.8283958133 , 0.91 )
				( -53.4048836351 , 0.92 )
				( -53.2588613026 , 0.93 )
				( -52.6036928784 , 0.94 )
				( -52.5033996322 , 0.95 )
				( -52.3198953572 , 0.96 )
				( -51.8993581261 , 0.97 )
				( -51.6984366667 , 0.98 )
				( -51.2612258018 , 0.99 )
				};
			
			\addplot[no marks,] coordinates {
				( -50.2771761987 , 0.0 )
				( -50.0775147757 , 0.01 )
				( -49.6898367504 , 0.02 )
				( -48.9946253917 , 0.03 )
				( -48.7297540456 , 0.04 )
				( -48.6212253536 , 0.05 )
				( -48.4450259212 , 0.06 )
				( -48.2851459228 , 0.07 )
				( -47.7596324629 , 0.08 )
				( -47.6006273795 , 0.09 )
				( -47.5910656149 , 0.1 )
				( -47.4774114872 , 0.11 )
				( -47.3627124893 , 0.12 )
				( -47.2972345962 , 0.13 )
				( -47.2098374684 , 0.14 )
				( -47.0992199338 , 0.15 )
				( -46.9641195247 , 0.16 )
				( -46.9194457603 , 0.17 )
				( -46.8110210878 , 0.18 )
				( -46.7233410739 , 0.19 )
				( -46.6285228624 , 0.2 )
				( -46.5582438036 , 0.21 )
				( -46.530959729 , 0.22 )
				( -46.4398847684 , 0.23 )
				( -46.3618338617 , 0.24 )
				( -46.2685856128 , 0.25 )
				( -46.2174072542 , 0.26 )
				( -46.1434095951 , 0.27 )
				( -46.0850110303 , 0.28 )
				( -46.045582545 , 0.29 )
				( -45.9326349547 , 0.3 )
				( -45.8781548292 , 0.31 )
				( -45.827914637 , 0.32 )
				( -45.795034164 , 0.33 )
				( -45.679905028 , 0.34 )
				( -45.6489447952 , 0.35 )
				( -45.5323236391 , 0.36 )
				( -45.4607251746 , 0.37 )
				( -45.3992740238 , 0.38 )
				( -45.2815693681 , 0.39 )
				( -45.2183536149 , 0.4 )
				( -45.1543341295 , 0.41 )
				( -45.0267494694 , 0.42 )
				( -44.9300133459 , 0.43 )
				( -44.8559471639 , 0.44 )
				( -44.806589347 , 0.45 )
				( -44.7654978494 , 0.46 )
				( -44.642419467 , 0.47 )
				( -44.5714145089 , 0.48 )
				( -44.5190350775 , 0.49 )
				( -44.4503994968 , 0.5 )
				( -44.4011952029 , 0.51 )
				( -44.3671027779 , 0.52 )
				( -44.2711001371 , 0.53 )
				( -44.2504296368 , 0.54 )
				( -44.1566933528 , 0.55 )
				( -44.1059118015 , 0.56 )
				( -44.0066120143 , 0.57 )
				( -43.9734850396 , 0.58 )
				( -43.8972540964 , 0.59 )
				( -43.8240986923 , 0.6 )
				( -43.7612684399 , 0.61 )
				( -43.7134089046 , 0.62 )
				( -43.6339921202 , 0.63 )
				( -43.5646445521 , 0.64 )
				( -43.5411542505 , 0.65 )
				( -43.4216857069 , 0.66 )
				( -43.3917412836 , 0.67 )
				( -43.3335767046 , 0.68 )
				( -43.1794206588 , 0.69 )
				( -43.0907338021 , 0.7 )
				( -43.0225374634 , 0.71 )
				( -42.8905292356 , 0.72 )
				( -42.7972680737 , 0.73 )
				( -42.6446356969 , 0.74 )
				( -42.5847034252 , 0.75 )
				( -42.4713573648 , 0.76 )
				( -42.4207802401 , 0.77 )
				( -42.3459910673 , 0.78 )
				( -42.1624145936 , 0.79 )
				( -42.1111467971 , 0.8 )
				( -41.9671452581 , 0.81 )
				( -41.7802341111 , 0.82 )
				( -41.7208569729 , 0.83 )
				( -41.6143845849 , 0.84 )
				( -41.5463871192 , 0.85 )
				( -41.4546789437 , 0.86 )
				( -41.3191989588 , 0.87 )
				( -41.2198112615 , 0.88 )
				( -41.0221022296 , 0.89 )
				( -40.9393477818 , 0.9 )
				( -40.8996712215 , 0.91 )
				( -40.7486618222 , 0.92 )
				( -40.3923682533 , 0.93 )
				( -39.8033175304 , 0.94 )
				( -39.6569205816 , 0.95 )
				( -39.3386344187 , 0.96 )
				( -39.2639067528 , 0.97 )
				( -38.6364117232 , 0.98 )
				( -37.0546876935 , 0.99 )
			};
			\addplot[
			no marks,
			] coordinates {
				( -48.3165300715 , 0.0 )
				( -47.3798339087 , 0.01 )
				( -47.1866609461 , 0.02 )
				( -47.1350973532 , 0.03 )
				( -47.0548909109 , 0.04 )
				( -46.9324637916 , 0.05 )
				( -46.6976915068 , 0.06 )
				( -46.6488725875 , 0.07 )
				( -46.6238082116 , 0.08 )
				( -46.5061442261 , 0.09 )
				( -46.391646457 , 0.1 )
				( -46.3312357269 , 0.11 )
				( -46.2782069942 , 0.12 )
				( -46.1970827409 , 0.13 )
				( -46.108926603 , 0.14 )
				( -45.9825042201 , 0.15 )
				( -45.9669234116 , 0.16 )
				( -45.8941652415 , 0.17 )
				( -45.8626580517 , 0.18 )
				( -45.8249003413 , 0.19 )
				( -45.7803153034 , 0.2 )
				( -45.7609954111 , 0.21 )
				( -45.6895764868 , 0.22 )
				( -45.6713121818 , 0.23 )
				( -45.6155154624 , 0.24 )
				( -45.5757558087 , 0.25 )
				( -45.5654549003 , 0.26 )
				( -45.5002081164 , 0.27 )
				( -45.3917778271 , 0.28 )
				( -45.3067874961 , 0.29 )
				( -45.2795778774 , 0.3 )
				( -45.2400454436 , 0.31 )
				( -45.1791773585 , 0.32 )
				( -45.088868493 , 0.33 )
				( -45.0334419038 , 0.34 )
				( -44.992484996 , 0.35 )
				( -44.9806988899 , 0.36 )
				( -44.9184643046 , 0.37 )
				( -44.8692617705 , 0.38 )
				( -44.8374142908 , 0.39 )
				( -44.7815711151 , 0.4 )
				( -44.6884232384 , 0.41 )
				( -44.6693292523 , 0.42 )
				( -44.6051487519 , 0.43 )
				( -44.5958558857 , 0.44 )
				( -44.4823574531 , 0.45 )
				( -44.4554823904 , 0.46 )
				( -44.3331209864 , 0.47 )
				( -44.3151578687 , 0.48 )
				( -44.2960742047 , 0.49 )
				( -44.250722618 , 0.5 )
				( -44.2015583715 , 0.51 )
				( -44.1676452006 , 0.52 )
				( -44.1340811373 , 0.53 )
				( -44.076045164 , 0.54 )
				( -44.0544209628 , 0.55 )
				( -43.9967390636 , 0.56 )
				( -43.9517086858 , 0.57 )
				( -43.9451911959 , 0.58 )
				( -43.8523593158 , 0.59 )
				( -43.8200696917 , 0.6 )
				( -43.7240003471 , 0.61 )
				( -43.6733167129 , 0.62 )
				( -43.6281778192 , 0.63 )
				( -43.584439058 , 0.64 )
				( -43.5490527774 , 0.65 )
				( -43.4557946305 , 0.66 )
				( -43.3424214699 , 0.67 )
				( -43.2839669937 , 0.68 )
				( -43.2642548388 , 0.69 )
				( -43.1290089207 , 0.7 )
				( -43.0475517501 , 0.71 )
				( -42.9834302279 , 0.72 )
				( -42.9364741686 , 0.73 )
				( -42.8805243044 , 0.74 )
				( -42.8630894681 , 0.75 )
				( -42.8024535452 , 0.76 )
				( -42.7831745756 , 0.77 )
				( -42.7467058493 , 0.78 )
				( -42.6748074444 , 0.79 )
				( -42.4960769842 , 0.8 )
				( -42.4278949737 , 0.81 )
				( -42.3555220716 , 0.82 )
				( -42.3188507168 , 0.83 )
				( -42.2280474221 , 0.84 )
				( -42.1704384997 , 0.85 )
				( -42.062607925 , 0.86 )
				( -41.9373160755 , 0.87 )
				( -41.8608498151 , 0.88 )
				( -41.7928100137 , 0.89 )
				( -41.6907590941 , 0.9 )
				( -41.6517355736 , 0.91 )
				( -41.5401899274 , 0.92 )
				( -41.476385676 , 0.93 )
				( -41.382989861 , 0.94 )
				( -41.2474898498 , 0.95 )
				( -40.994919236 , 0.96 )
				( -40.928628598 , 0.97 )
				( -40.5713860439 , 0.98 )
				( -40.5059831483 , 0.99 )
				};
			\addplot[
			no marks,
			] coordinates {
				( -63.7141331182 , 0.0 )
				( -63.5798136574 , 0.01 )
				( -63.4465781456 , 0.02 )
				( -63.0422721315 , 0.03 )
				( -62.7557197443 , 0.04 )
				( -62.6585315512 , 0.05 )
				( -62.4617930717 , 0.06 )
				( -62.3327946122 , 0.07 )
				( -62.2375345177 , 0.08 )
				( -62.1045578932 , 0.09 )
				( -61.9316625211 , 0.1 )
				( -61.7492365656 , 0.11 )
				( -61.7108011085 , 0.12 )
				( -61.4883403851 , 0.13 )
				( -61.3577563047 , 0.14 )
				( -61.2343420276 , 0.15 )
				( -61.1444032077 , 0.16 )
				( -61.0687004046 , 0.17 )
				( -60.9619325883 , 0.18 )
				( -60.8069691253 , 0.19 )
				( -60.719169173 , 0.2 )
				( -60.6067255083 , 0.21 )
				( -60.5056756683 , 0.22 )
				( -60.4355881235 , 0.23 )
				( -60.4121697478 , 0.24 )
				( -60.3190639305 , 0.25 )
				( -60.2596538928 , 0.26 )
				( -60.1413503382 , 0.27 )
				( -60.0744148337 , 0.28 )
				( -59.9675588553 , 0.29 )
				( -59.9204654589 , 0.3 )
				( -59.8833769225 , 0.31 )
				( -59.8084506418 , 0.32 )
				( -59.7340458737 , 0.33 )
				( -59.6482213054 , 0.34 )
				( -59.583752597 , 0.35 )
				( -59.4750753686 , 0.36 )
				( -59.4583847886 , 0.37 )
				( -59.3695505376 , 0.38 )
				( -59.2892264339 , 0.39 )
				( -59.2444892535 , 0.4 )
				( -59.1907089866 , 0.41 )
				( -59.1368343742 , 0.42 )
				( -59.0314247441 , 0.43 )
				( -58.9768870252 , 0.44 )
				( -58.9170021148 , 0.45 )
				( -58.8929212328 , 0.46 )
				( -58.8012752171 , 0.47 )
				( -58.75867088 , 0.48 )
				( -58.6491030307 , 0.49 )
				( -58.5859341321 , 0.5 )
				( -58.4849262443 , 0.51 )
				( -58.3967523373 , 0.52 )
				( -58.3640946263 , 0.53 )
				( -58.2755523734 , 0.54 )
				( -58.2342328392 , 0.55 )
				( -58.1758562195 , 0.56 )
				( -58.1032581543 , 0.57 )
				( -58.0643725024 , 0.58 )
				( -57.964205486 , 0.59 )
				( -57.9241868896 , 0.6 )
				( -57.8725143077 , 0.61 )
				( -57.8273074405 , 0.62 )
				( -57.7659545731 , 0.63 )
				( -57.7074494481 , 0.64 )
				( -57.6743891026 , 0.65 )
				( -57.5659181542 , 0.66 )
				( -57.5279385285 , 0.67 )
				( -57.4405948734 , 0.68 )
				( -57.3800908385 , 0.69 )
				( -57.279908857 , 0.7 )
				( -57.1773017685 , 0.71 )
				( -57.1051765099 , 0.72 )
				( -57.0244341709 , 0.73 )
				( -56.912292175 , 0.74 )
				( -56.7953914943 , 0.75 )
				( -56.7491533227 , 0.76 )
				( -56.6276666391 , 0.77 )
				( -56.5579761031 , 0.78 )
				( -56.4161712484 , 0.79 )
				( -56.3870666587 , 0.8 )
				( -56.3328270903 , 0.81 )
				( -56.2340749014 , 0.82 )
				( -56.1404716476 , 0.83 )
				( -56.0433732597 , 0.84 )
				( -55.95243914 , 0.85 )
				( -55.9129423284 , 0.86 )
				( -55.7867228913 , 0.87 )
				( -55.5738887208 , 0.88 )
				( -55.4975273415 , 0.89 )
				( -55.3323425333 , 0.9 )
				( -55.2776080414 , 0.91 )
				( -55.2581805531 , 0.92 )
				( -55.0740365844 , 0.93 )
				( -54.8100466401 , 0.94 )
				( -54.5843849566 , 0.95 )
				( -54.2917250986 , 0.96 )
				( -53.7228659866 , 0.97 )
				( -53.5304294907 , 0.98 )
				( -53.373367275 , 0.99 )
				};
				\addplot[
				no marks,
				] coordinates {
					( -60.444018744 , 0.0 )
					( -60.0654953978 , 0.01 )
					( -59.9807771393 , 0.02 )
					( -59.8687721781 , 0.03 )
					( -59.8538495376 , 0.04 )
					( -59.7253970645 , 0.05 )
					( -59.693185562 , 0.06 )
					( -59.6680995191 , 0.07 )
					( -59.6511643922 , 0.08 )
					( -59.6448638994 , 0.09 )
					( -59.6415757954 , 0.1 )
					( -59.5002123092 , 0.11 )
					( -59.409670917 , 0.12 )
					( -59.3553175114 , 0.13 )
					( -59.3273014044 , 0.14 )
					( -59.2954298493 , 0.15 )
					( -59.2439013181 , 0.16 )
					( -59.1384471511 , 0.17 )
					( -59.1266795249 , 0.18 )
					( -59.1229281361 , 0.19 )
					( -59.0919970797 , 0.2 )
					( -59.0266635016 , 0.21 )
					( -58.9990261118 , 0.22 )
					( -58.9905910189 , 0.23 )
					( -58.9666242397 , 0.24 )
					( -58.9528052671 , 0.25 )
					( -58.9145782212 , 0.26 )
					( -58.9114715817 , 0.27 )
					( -58.8975279598 , 0.28 )
					( -58.8735883647 , 0.29 )
					( -58.7994437377 , 0.3 )
					( -58.7811363867 , 0.31 )
					( -58.7411594781 , 0.32 )
					( -58.7218245928 , 0.33 )
					( -58.7082411503 , 0.34 )
					( -58.6510909834 , 0.35 )
					( -58.6426264615 , 0.36 )
					( -58.6302870403 , 0.37 )
					( -58.5543017926 , 0.38 )
					( -58.5508384669 , 0.39 )
					( -58.5050526364 , 0.4 )
					( -58.4698254964 , 0.41 )
					( -58.4623279404 , 0.42 )
					( -58.4515694119 , 0.43 )
					( -58.4346952126 , 0.44 )
					( -58.4231963608 , 0.45 )
					( -58.4119191095 , 0.46 )
					( -58.3515510827 , 0.47 )
					( -58.3271923137 , 0.48 )
					( -58.3179353719 , 0.49 )
					( -58.2776833886 , 0.5 )
					( -58.2541736248 , 0.51 )
					( -58.2437008575 , 0.52 )
					( -58.225291296 , 0.53 )
					( -58.1703476184 , 0.54 )
					( -58.1393418183 , 0.55 )
					( -58.1230477457 , 0.56 )
					( -58.0582205627 , 0.57 )
					( -58.0323986226 , 0.58 )
					( -58.0130362565 , 0.59 )
					( -57.997131834 , 0.6 )
					( -57.9827779147 , 0.61 )
					( -57.9519302923 , 0.62 )
					( -57.9450205573 , 0.63 )
					( -57.9283606283 , 0.64 )
					( -57.9167804634 , 0.65 )
					( -57.8907574814 , 0.66 )
					( -57.8099152547 , 0.67 )
					( -57.7831382227 , 0.68 )
					( -57.7704606714 , 0.69 )
					( -57.729993889 , 0.7 )
					( -57.7177123659 , 0.71 )
					( -57.6945674435 , 0.72 )
					( -57.6413195695 , 0.73 )
					( -57.5921248369 , 0.74 )
					( -57.4940631057 , 0.75 )
					( -57.466281573 , 0.76 )
					( -57.4532534511 , 0.77 )
					( -57.4251656894 , 0.78 )
					( -57.3468681074 , 0.79 )
					( -57.2955923961 , 0.8 )
					( -57.2741205918 , 0.81 )
					( -57.2504846233 , 0.82 )
					( -57.2268249901 , 0.83 )
					( -57.2093040606 , 0.84 )
					( -57.2012345106 , 0.85 )
					( -57.1705004438 , 0.86 )
					( -57.1242413963 , 0.87 )
					( -57.0715023218 , 0.88 )
					( -56.89784001 , 0.89 )
					( -56.8778659652 , 0.9 )
					( -56.7108117812 , 0.91 )
					( -56.6977973454 , 0.92 )
					( -56.5779368553 , 0.93 )
					( -56.4191161835 , 0.94 )
					( -56.3837722606 , 0.95 )
					( -56.1873881554 , 0.96 )
					( -55.9379512467 , 0.97 )
					( -55.8019728704 , 0.98 )
					( -55.1356377372 , 0.99 )
					};
				\addplot[
				no marks,
				] coordinates {
					( -45.954987263 , 0.0 )
					( -45.9320232367 , 0.01 )
					( -45.5893890747 , 0.02 )
					( -45.5012206544 , 0.03 )
					( -45.418681533 , 0.04 )
					( -45.2966632457 , 0.05 )
					( -45.2429462244 , 0.06 )
					( -45.2312031013 , 0.07 )
					( -45.2084591681 , 0.08 )
					( -45.1607512185 , 0.09 )
					( -45.1418620211 , 0.1 )
					( -45.055136085 , 0.11 )
					( -45.0357313694 , 0.12 )
					( -45.0066956572 , 0.13 )
					( -44.9853022087 , 0.14 )
					( -44.9466855076 , 0.15 )
					( -44.9322910131 , 0.16 )
					( -44.92176932 , 0.17 )
					( -44.892039737 , 0.18 )
					( -44.8088921473 , 0.19 )
					( -44.7826431044 , 0.2 )
					( -44.7767438918 , 0.21 )
					( -44.7639434691 , 0.22 )
					( -44.761463552 , 0.23 )
					( -44.7553020905 , 0.24 )
					( -44.7189997534 , 0.25 )
					( -44.6768406373 , 0.26 )
					( -44.6268478774 , 0.27 )
					( -44.6071657223 , 0.28 )
					( -44.5982605322 , 0.29 )
					( -44.5588871637 , 0.3 )
					( -44.5301666828 , 0.31 )
					( -44.5055290561 , 0.32 )
					( -44.4961418218 , 0.33 )
					( -44.4889673219 , 0.34 )
					( -44.4777317076 , 0.35 )
					( -44.462008782 , 0.36 )
					( -44.457761751 , 0.37 )
					( -44.404454463 , 0.38 )
					( -44.3956336171 , 0.39 )
					( -44.3917940284 , 0.4 )
					( -44.3764005038 , 0.41 )
					( -44.2760958395 , 0.42 )
					( -44.2637841105 , 0.43 )
					( -44.2455786567 , 0.44 )
					( -44.2264352038 , 0.45 )
					( -44.2122649988 , 0.46 )
					( -44.1560193779 , 0.47 )
					( -44.1506895758 , 0.48 )
					( -44.1278175364 , 0.49 )
					( -44.122874273 , 0.5 )
					( -44.0824542934 , 0.51 )
					( -44.0574228436 , 0.52 )
					( -44.0523481858 , 0.53 )
					( -44.0476246254 , 0.54 )
					( -44.0337148982 , 0.55 )
					( -44.0259872326 , 0.56 )
					( -43.9948544617 , 0.57 )
					( -43.9706405524 , 0.58 )
					( -43.9434378547 , 0.59 )
					( -43.9294316603 , 0.6 )
					( -43.9233672268 , 0.61 )
					( -43.902700608 , 0.62 )
					( -43.8803969988 , 0.63 )
					( -43.808622926 , 0.64 )
					( -43.803829387 , 0.65 )
					( -43.7987858214 , 0.66 )
					( -43.762855128 , 0.67 )
					( -43.7481495699 , 0.68 )
					( -43.7289420629 , 0.69 )
					( -43.7159925107 , 0.7 )
					( -43.6925563579 , 0.71 )
					( -43.6451181517 , 0.72 )
					( -43.613772154 , 0.73 )
					( -43.5883527178 , 0.74 )
					( -43.5287855462 , 0.75 )
					( -43.4831660371 , 0.76 )
					( -43.4279566108 , 0.77 )
					( -43.4188618833 , 0.78 )
					( -43.4152987715 , 0.79 )
					( -43.4027853405 , 0.8 )
					( -43.3920027751 , 0.81 )
					( -43.3798909878 , 0.82 )
					( -43.356735403 , 0.83 )
					( -43.342432417 , 0.84 )
					( -43.3274292058 , 0.85 )
					( -43.3081084487 , 0.86 )
					( -43.2813592771 , 0.87 )
					( -43.2612271018 , 0.88 )
					( -43.0880324534 , 0.89 )
					( -43.0147488845 , 0.9 )
					( -42.9988426801 , 0.91 )
					( -42.901906273 , 0.92 )
					( -42.8588934541 , 0.93 )
					( -42.8150295413 , 0.94 )
					( -42.721474723 , 0.95 )
					( -42.6500580744 , 0.96 )
					( -42.4834646041 , 0.97 )
					( -42.2376798628 , 0.98 )
					( -42.2326955475 , 0.99 )
					};
				\addplot[
				no marks,
				color=gray,
				] coordinates {
					(-44,0)
					(-44,1)
				};
				\addplot[
				no marks,
				color=gray,
				] coordinates {
					(-58,0)
					(-58,1)
				};
				\node[anchor=south east] at (axis cs: -58,.6) {\textsc{oob}};
				\node[anchor=south east] at (axis cs: -44,.6) {in-band};
				\draw[->] (axis cs:-63, .3) node[anchor=east] {1,2,10} -- (axis cs:-56,.25);
				\draw[->] (axis cs:-48, .3) node[anchor=east] {1,2,10} -- (axis cs:-43,.25);
			\end{axis}
		\end{tikzpicture}
		\caption{The eigenvalue distribution of $\symbf{S}_{\symbf{dd}}(f)$ at $f=0$ and $f=1.22$ with an array with 100 antennas and a delay spread $\sigma_\tau = 60T$.  The per-antenna power of the distortion is marked with a vertical line.  In all cases, this power varied less than 0.5\,dB.  The amplifiers were backed off by 10\,dB on average.}
		\label{fig:82001}
	\end{figure}

	Another way to illustrate the directivity of the distortion is
	to show the eigenvalue distribution of the correlation matrix of the
	transmitted distortion $\symbf{S}_{\symbf{dd}}(f)$; see
        Figure~\ref{fig:82001}.  It can be seen that the
	worst direction has an array gain of 7\,dB with one user and
	2--3\,dB with ten users, c.f.\ \eqref{eq:0020011}.

	\subsection{OFDM in Line-of-Sight}\label{sec:893858190111223}
\begin{figure}
		\centering
		\ifdraft
		\includegraphics[scale=.8]{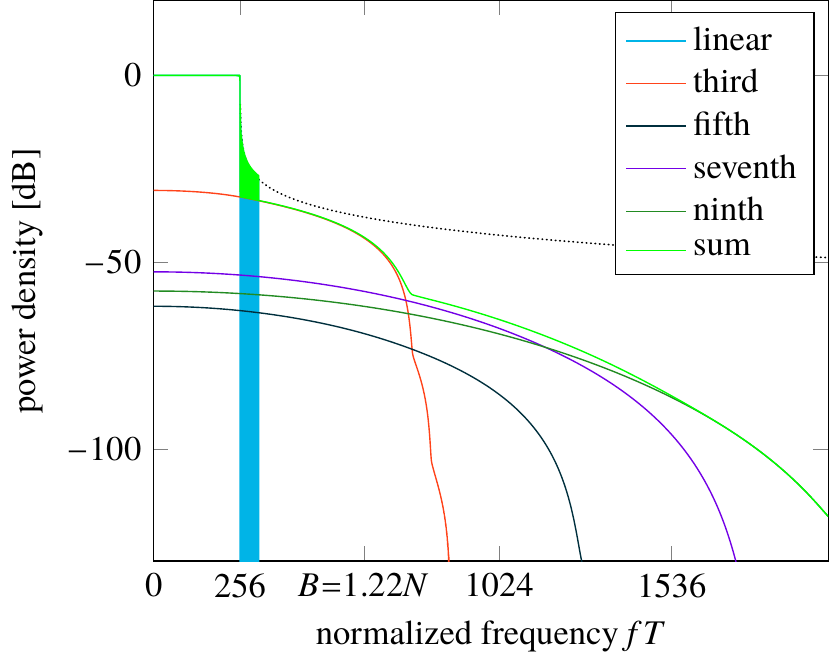}
		\else
		\includegraphics{illustrationer/tx_psds_OFDM.pdf}
		\fi
		\caption{The power spectral density of the precoded \OFDM signal transmitted from one of the 100 antennas in the array.  There are $N=512$ subcarriers and 10 served users.  Rectangular pulses as in \eqref{eq:rect_pulse} are used with $f_0 = 1/T$.  The \OFDM signal is filtered by an ideal bandpass filter of bandwidth $B=1.22Nf_0$.  The contour of the unfiltered, unamplified signal is drawn with a dotted line.  On average the amplifiers operate \SI{7}{\decibel} below the one-dB compression point.  The frequency $B$ is the measurement point used in Figure~\ref{fig:inband_radiation_pattern_OFDM}.  The power spectral density labeled “linear” is one of the diagonal elements in $\symbf{S}_{\symbf{uu}}(f)$ in \eqref{eq:928381818999}, and the “third”, “fifth”, \ldots, refer to the same diagonal element in the different terms in the sum $\symbf{S}_{\symbf{dd}}(f)$ in \eqref{eq:51628826687881}.}
		\label{fig:psd_ofdm}
	\end{figure}

\begin{figure}
		\centering
		\ifdraft
		\rule{0pt}{0pt}\clap{\mbox{\includegraphics{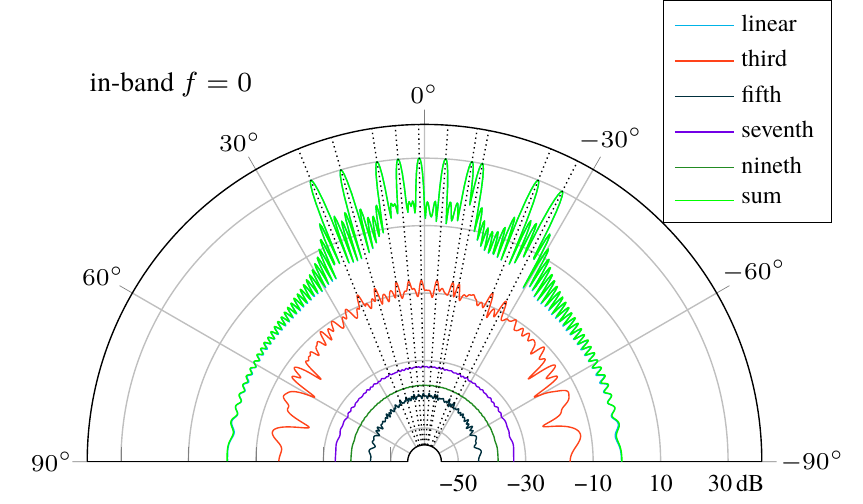}\includegraphics{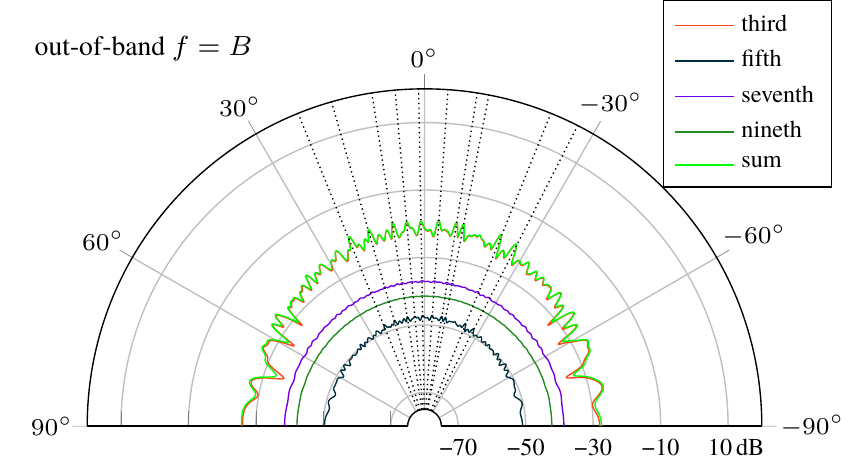}}}\rule{0pt}{0pt}
		\else
		\includegraphics{illustrationer/radpattern_IB_OFDM_10users.pdf}\par
		\bigskip
		\includegraphics{illustrationer/radpattern_OOB_OFDM_10users.pdf}
		\fi
		\caption{The radiation pattern from the same system that is studied in Figure~\ref{fig:psd_ofdm} at the frequencies $f=0$ (in-band) and $f=B$ (out-of-band).  The array has 100 antennas and transmits precoded \OFDM signals with $N=1024$ subcarriers to 10 users.  The allocated band has bandwidth $B = 1.22 N f_0$, where $f_0 = 1/T$.  }
		\label{fig:inband_radiation_pattern_OFDM}
	\end{figure}

	When each user is served over the whole spectrum, the \OFDM system behaves almost identically to the single-carrier system studied above.  Specifically, the criterion derived in \eqref{eq:omnidirectional_requirement19929} is then also applicable to \OFDM.  The transmitted power spectral density when $K=10$ users are served over the whole band are shown in Figure~\ref{fig:psd_ofdm} and the radiation patterns at the in-band frequency $f=0$ and the out-of-band frequency $f=B$ is shown in Figure~\ref{fig:inband_radiation_pattern_OFDM}.  An ideal low-pass filter has been used to make the input signal to the amplifier perfectly bandlimited to a band of width $1.22N/T$, i.e.\ to limit the excess bandwidth to 1.22.  It can be seen that, in the immediate adjacent band, the third-degree distortion term is dominant.  Only as one moves further away from the in-band signal in the spectrum, the higher-order terms become significant.  This is true both for the transmitted spectrum and the received one that can be seen in Figure~\ref{fig:inband_radiation_pattern_OFDM}.
	
	It can be seen that the third-degree distortion term is approximately \SI{30}{dB} below the linear signal for this particular back-off and amplifier.  This emission level happens to be similar to the out-of-band emission of the linear signal without sidelobe suppression (without the low-pass filter), which is shown as a dotted contour in Figure~\ref{fig:psd_ofdm}.  To maximize power efficiency, the back-off should be chosen such that the distortion is level with the sidelobes; and, to maximize spectral efficiency, the sidelobe level should be suppressed to meet the out-of-band radiation requirement (with some margin to accommodate for the distortion).
	
\begin{figure}
		\centering
		\ifdraft
		\rule{0pt}{0pt}\clap{\mbox{\includegraphics{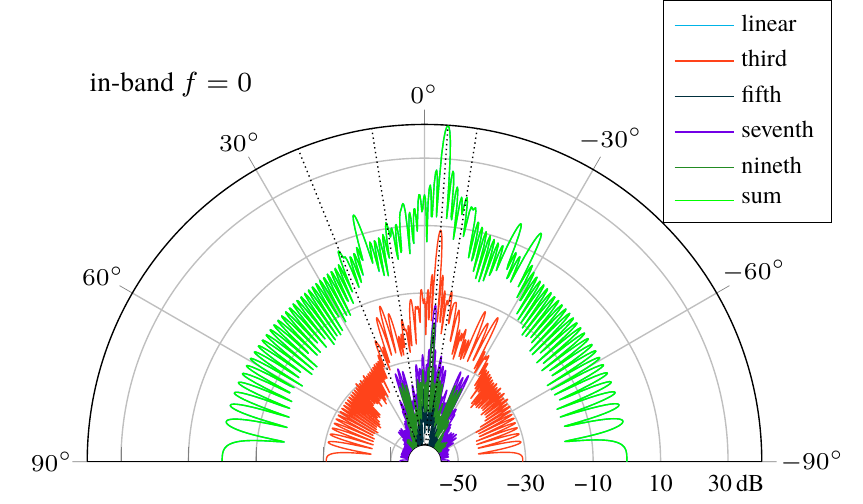}		\includegraphics{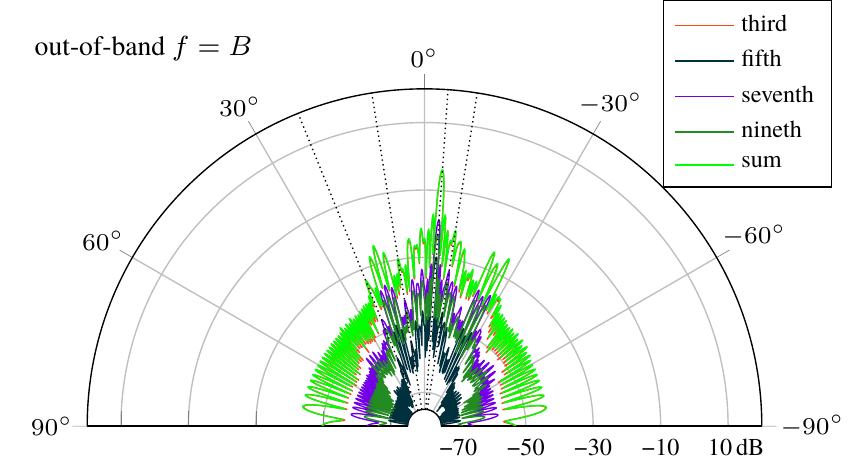}}}\rule{0pt}{0pt}
		\else
		\includegraphics{illustrationer/radpattern_IB_OFDM_4_distributed_users.pdf}\par
		\bigskip
		\includegraphics{illustrationer/radpattern_OOB_OFDM_4_distributed_users.pdf}
		\fi
		\caption{The radiation pattern at frequencies $f=0$ (in-band) and $f=B$ (out-of-band) of an array with 100 antennas transmitting a precoded \OFDM signal with $N=1024$ subcarriers that serves 4 users with different powers: \SI{-34}{dB}, \SI{-31}{dB}, \SI{-0.022}{dB}, \SI{-24}{dB} from left to right.  The allocated band has bandwidth $B = 1.22 N f_0$, where $f_0 = 1/T$.  The power amplifiers are, on average, operated \SI{7}{dB} from the one-dB compression point.}
		\label{fig:inband_radiation_pattern_OFDM_dist_users}
	\end{figure}

	In many multi-user scenarios, different beams can be radiated with very different powers.  This is illustrated in Figure~\ref{fig:inband_radiation_pattern_OFDM_dist_users}, where $K=4$ users are served but there is one dominant user whose beam is much stronger than the other beams.  In this case, it can be seen that the distortion behaves as if there were only one served user---it is highly directive and directed towards the dominant user.
	
	Instead of studying the case, where all users are served on all subcarriers, we study a scenario, where each user is served on only a subset of the available subcarriers.  Such a scenario might happen when there are users that continuously have to be served with a small data rate.  We denote the index set of users that are served on subcarrier $\nu$ by $\symcal{K}_{\nu}$.  Assume that all users are in line-of-sight, i.e.\ that the user channels are given by \eqref{eq:LoS_channel10292929}. Further, assume that maximum-ratio precoding is used, so that the precoding weights $w_{mk}[\nu] = e^{-jm\phi_k}$ for all subcarriers $\nu$.  The linearly amplified term, then, has the power spectral density:
	\begin{align}
		S_{x_mx_{m'}}(f) &= \frac{1}{NT} \sum_{\nu=0}^{N-1} |\symsfit{p}_\nu(f)|^2 \sum_{k\in\symcal{K}_\nu} \xi_k w_{mk}[\nu] w^*_{m'k}[\nu]\\
		&= \frac{1}{NT} \sum_{\nu=0}^{N-1} |\symsfit{p}_\nu(f)|^2 \sum_{k\in\symcal{K}_\nu} e^{j\phi_k(m'-m)}.
	\end{align}
	To alleviate the notation, the third-degree pulse is defined as:
	\begin{align}
		\symsfit{p}^{(3)}_\nu(f) \triangleq \Big( \abs*{\symsfit{p}(\varphi)}^2 \star \abs*{\symsfit{p}(\varphi)}^2 \star \abs*{\symsfit{p}(- \varphi)}^2 \Big)(f - \nu/T).
	\end{align}
	The third-degree term of the distortion is then:
	\ifdraft
	\begin{align}\label{eq:OFDM_thrid_order_dist_term919019}
		S^{(3)}_{x_mx_{m'}}(f) &= \frac{1}{N^3T^3} \sum_{\mathclap{\nu=-N+1}}^{\mathclap{2N-2}} \symsfit{p}^{(3)}_\nu(f) \sum_{\nu', \nu''} \sum_{k \in \symcal{K}_{\nu'}} \sum_{k' \in \symcal{K}_{\nu''}} \sum_{\mathclap{\hspace{3em}k'' \in \symcal{K}_{\nu' + \nu'' - \nu}}} \xi_{k} \xi_{k'} \xi_{k''} e^{j(\phi_k+\phi_{k'}-\phi_{k''})(m'-m)}.
	\end{align}
	\else
	\begin{align}
		S^{(3)}_{x_mx_{m'}}(f) &= \frac{1}{N^3T^3} \sum_{\mathclap{\nu=-N+1}}^{\mathclap{2N-2}} \symsfit{p}^{(3)}_\nu(f)\notag\\
		&\quad\times\!\! \sum_{\nu', \nu''} \sum_{k \in \symcal{K}_{\nu'}} \sum_{k' \in \symcal{K}_{\nu''}} \sum_{\mathclap{\hspace{3em}k'' \in \symcal{K}_{\nu' + \nu'' - \nu}}} \xi_{k} \xi_{k'} \xi_{k''} e^{j(\phi_k+\phi_{k'}-\phi_{k''})(m'-m)}.\label{eq:OFDM_thrid_order_dist_term919019}
	\end{align}
	\fi
	
	\begin{theorem}
		At a given tone $\nu$, the distortion is beamformed towards the directions given by $\phi_k + \phi_{k'} - \phi_{k''}$, where $(k,k',k'') \in \symcal{K}_{\nu'} \times \symcal{K}_{\nu'} \times \symcal{K}_{\nu'+\nu''-\nu}$, for some $\nu',\nu''= 0,\ldots,N-1$.  
	\end{theorem}
	Note that all beamforming directions of the linearly amplified signal at a given subcarrier are also present at the same subcarrier in the uncorrelated distortion.  For example, if $k_0 \in \symcal{K}_\nu$, then the pulse $\symsfit{p}^{(3)}_\nu(f)$ is beamformed, among other directions, in the direction given by $\phi_{k_0}$.
	
	\begin{remark}\label{rem:adjacent_subcarrier}
		Given a subcarrier $\nu$ and a user $k_0 \in \symcal{K}_\nu$, the pulse $\symsfit{p}^{(3)}_{\nu+n}(f)$ at an adjacent subcarrier, $n$ subcarriers away from $\nu$, will be beamformed in the direction given by $\phi_{k_0}$, if there exists a $\nu'=0,\ldots,N-1$ and a $k'_0$ such that $k'_0 \in \symcal{K}_{\nu'} \cap \symcal{K}_{\nu'-n}$.
	\end{remark}
	As a consequence of Remark~\ref{rem:adjacent_subcarrier}, if there is a user $k'_0$ who is served on all subcarriers $k'_0 \in \bigcap_{\nu=0}^{N-1} \symcal{K}_\nu$, then the uncorrelated distortion at all in-band subcarriers $\nu = 0,\ldots,N-1$ is beamformed in all directions \mbox{$\{\phi_k : k = 1,2, \ldots, K\}$}.  The strength of the beam in the direction given by $\phi_k$, however, depends on the number of summands in \eqref{eq:OFDM_thrid_order_dist_term919019} that correspond to that direction.  While this number is $\sum_{\nu'=0}^{N-1} |\symcal{K}_{\nu'}|$ at a frequency $\nu$ such that $k \in \symcal{K}_\nu$, it shrinks to
	\begin{align}
		\sum_{\nu'} \left|\{k : k \in \symcal{K}_{\nu'} \cap \symcal{K}_{\nu'-n}\} \right|
	\end{align}
	at frequencies $n$ subcarriers away from $\nu$.

	\subsection{Distortion-Aware Frequency Scheduling}
	As has been demonstrated in Section~\ref{sec:893858190111223}, it is possible to use the theory presented in this paper to predict the beamforming directions of the distortion that is created by the nonlinear amplifiers.  This could potentially be used to schedule users in frequency in such a way that the influence of the distortion is minimized.  For example, if a large piece of the spectrum is beamformed towards a single user, another user that has a similar channel should not be scheduled to use subcarriers close to that user.
	
	\subsection{Two Tones}\label{sec:9885728}
	Now assume that there are only two users, each allocated its own subcarrier: $\nu_1$ and $\nu_2$ respectively.  Then the third-degree term consists of eight terms (counted with multiplicities):
	\ifdraft
	\begin{align}
		S^{(3)}_{x_mx_{m'}}(f) &= \frac{1}{N^3T^3} \Big( 3 \symsfit{p}^{(3)}_{\nu_1}(f) e^{j\phi_1(m'-m)} + \symsfit{p}^{(3)}_{2\nu_1-\nu_2}(f) e^{j(2\phi_1-\phi_2)(m'-m)}\notag\\
		&\quad + \symsfit{p}^{(3)}_{2\nu_2-\nu_1}(f) e^{j(2\phi_2-\phi_1)(m'-m)} + 3 \symsfit{p}^{(3)}_{\nu_2}(f) e^{j\phi_2(m'-m)}\Big)
	\end{align}
	\else
	\begin{align}
		S^{(3)}_{x_mx_{m'}}(f) &= \frac{1}{N^3T^3} \Big( 3 \symsfit{p}^{(3)}_{\nu_1}(f) e^{j\phi_1(m'-m)} \notag\\
		&\quad+ \symsfit{p}^{(3)}_{2\nu_1-\nu_2}(f) e^{j(2\phi_1-\phi_2)(m'-m)}\notag\\
		&\quad + \symsfit{p}^{(3)}_{2\nu_2-\nu_1}(f) e^{j(2\phi_2-\phi_1)(m'-m)} \notag\\
		&\quad+ 3 \symsfit{p}^{(3)}_{\nu_2}(f) e^{j\phi_2(m'-m)}\Big)
	\end{align}
	\fi
	In a two-tone system, the frequencies and directions of the distortion are thus given by the following theorem.
	\begin{theorem}\label{the:two_tone}
		The third-degree distortion consists of four distortion terms pulse-shaped by $\symsfit{p}^{(3)}_{\nu}(f)$.  Two at the frequencies of the users, $\nu = \nu_1$ and $\nu_2$, and two intermodulation products at $\nu = 2\nu_1 - \nu_2$ and $2\nu_2-\nu_1$---one above $\max\{\nu_1,\nu_2\}$ and one below $\min\{\nu_1,\nu_2\}$.  They are beamformed in the directions of the two users $\phi_1$ and $\phi_2$ and in the directions given by $2\phi_1 - \phi_2$ and $2\phi_2 - \phi_1$ respectively.
	\end{theorem}
	
\begin{figure}
		\centering
		\ifdraft
		\includegraphics[scale=.8]{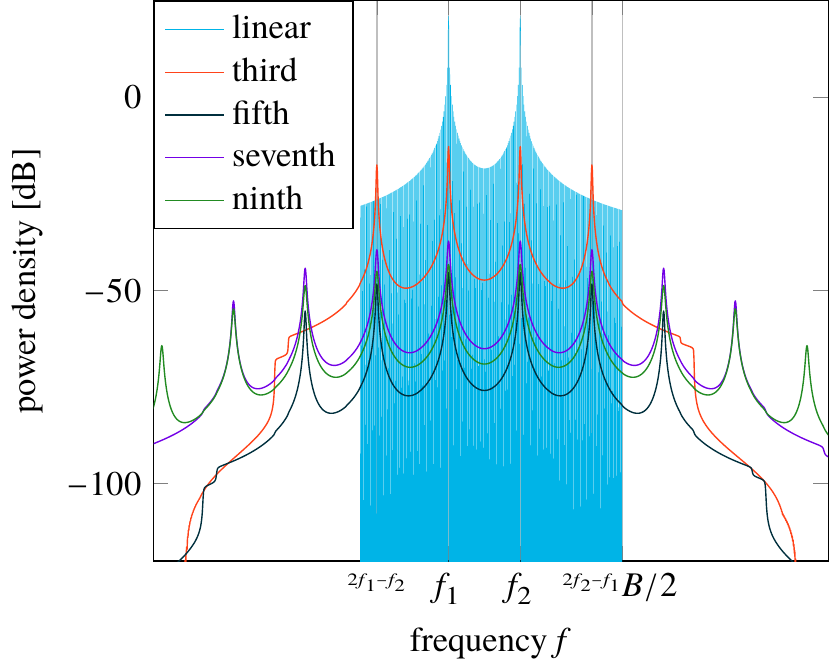}
		\else
		\includegraphics{tx_psds_two_tone.pdf}
		\fi
		\caption{The power spectral density of the transmitted signal at one antenna when two subcarriers are beamformed towards two different angles.  The signal is backed off by \SI{9}{\decibel} from the one-dB compression point.}
		\label{fig:two_tone_tx_psd}
	\end{figure}

\begin{figure}
		\newsavebox{\subfigone}\newsavebox{\subfigtwo}
		\savebox{\subfigone}{\includegraphics{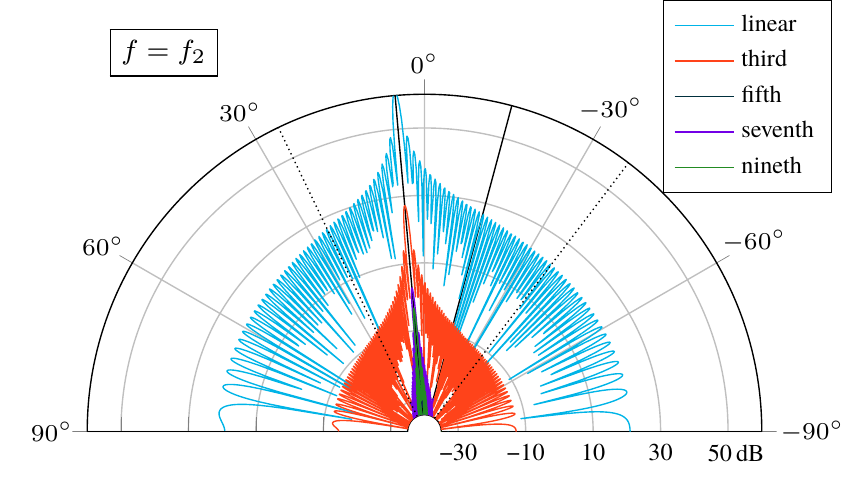}}
		\savebox{\subfigtwo}{\includegraphics{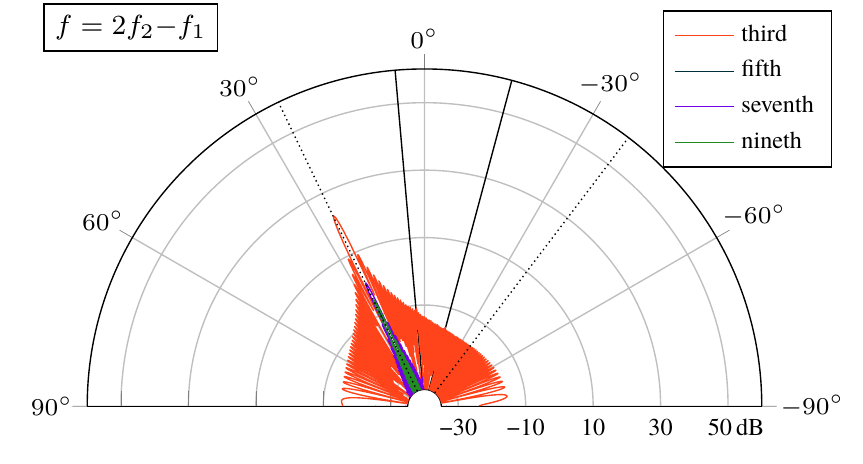}}
		\newlength{\figonelen}\newlength{\figtwolen}
		\settowidth{\figonelen}{\usebox{\subfigone}}
		\settowidth{\figtwolen}{\usebox{\subfigtwo}}
		\newsavebox{\boxone}\newsavebox{\boxtwo}
		\savebox{\boxone}{\parbox[b]{\figonelen}{\centering\usebox{\subfigone}\\\raisebox{\baselineskip}{\footnotesize{}(a) main frequency}}}
		\savebox{\boxtwo}{\parbox[b]{\figtwolen}{\centering\usebox{\subfigtwo}\\\raisebox{\baselineskip}{\footnotesize{}(b) intermodulation product}}}
		
		\centering
		\ifdraft
		\rule{0pt}{0pt}\clap{\mbox{\usebox{\boxone}\usebox{\boxtwo}}}\rule{0pt}{0pt}\\[-\baselineskip]
		\else
		\subfloat[main frequency]{\includegraphics{radpattern_two_tone_inband.pdf}}

		\subfloat[intermodulation product]{\includegraphics{radpattern_two_tone_intermodulation.pdf}}
		\fi
		\caption{The radiation pattern (a) at the carrier frequency $f=f_2$ of the pulse aimed at user~2 and (b) at the frequency $f=2f_2-f_1$ of the second intermodulation product.  Two pulses, with carrier frequencies $f_1=-50f_0 + f_c$ and $f_2=35f_0 + f_c$, are beamformed towards the angles $\theta_1=\SI{-15}{\degree}$ and $\theta_2=\SI{5}{\degree}$ (marked with solid rays).  The amplifiers are backed off \SI{9}{\decibel} from the one-dB-compression point.  The directions of the intermodulation products as predicted by $2\phi_1-\phi_2$ and $2\phi_2-\phi_1$ are marked with dotted rays.  The linear term has a null at the frequency of the intermodulation product.  Therefore, the linear term cannot be seen in (b), even though the linear term in Figure~\ref{fig:two_tone_tx_psd} has significant sidelobes around the frequency of the intermodulation product.}
		\label{fig:radpatterns_two_tone}
	\end{figure}

	The findings of Theorem~\ref{the:two_tone} can be seen in Figure~\ref{fig:two_tone_tx_psd} that shows the transmitted power spectral density and in Figure~\ref{fig:radpatterns_two_tone} that shows the radiation pattern at the frequency of pulse $\nu_2$ and the intermodulation product at $f = 2f_2 - f_1$.  It can be seen that the intermodulation product indeed is beamformed in the direction predicted by $2\phi_2 - \phi_1$.  
	
	\section{Measurement-Based Results}

\begin{figure}
		\centering
		\includegraphics{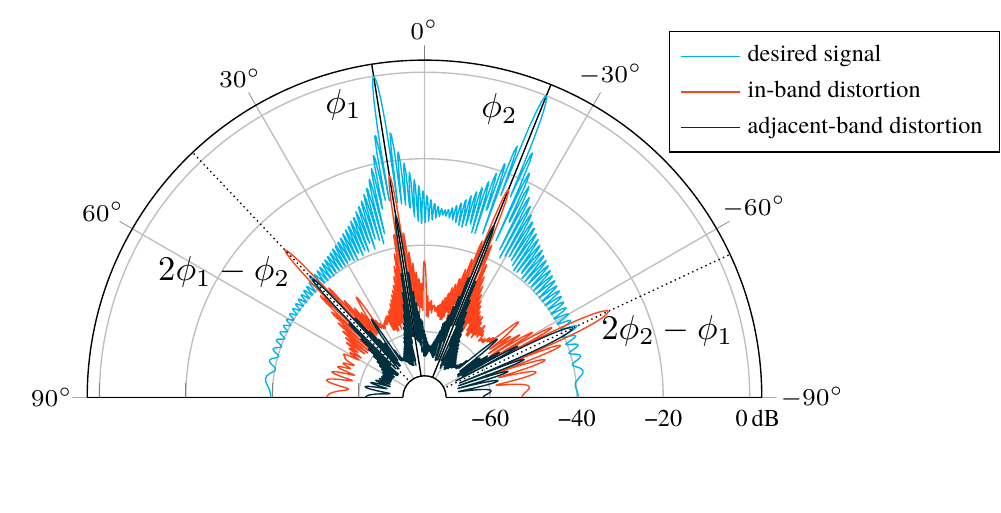}
		\vspace{-4ex}
		\caption{The radiation patterns of the desired signal and of the distortion, using a simulation of free-space propagation
		based on measured signals that have propagated through an actual amplifier in the laboratory.}
		\label{fig:measured_radpattern}
	\end{figure}

To illustrate and verify our theoretical results, we performed
measurements on a gallium-nitride (GaN), class~\textsc{ab} amplifier. The measurement were performed in the lab using the
on-line interface “web-lab” that is described in \cite{landin2015weblab}. Single-carrier
transmission with a root-raised cosine, roll-off 0.22, was considered.
Free-space (line-of-sight) propagation with a uniform linear array
(half-wavelength element spacing) was then simulated, assuming all
amplifiers were identical. Specifically, maximum-ratio precoding with
two directions was used to generate the transmit signals per
amplifier.  The amplified
signals were split up in desired signal and distortion, as
in \eqref{eq:partitioning192893881}, and the radiation patterns of
these two signal components were computed.  The amount of power
received in different directions was computed and the result is shown
in Figure~\ref{fig:measured_radpattern}.
	
	It can be seen in Figure~\ref{fig:measured_radpattern} that
	the desired signal is beamformed in the two desired
	directions.  Furthermore, both the in-band distortion and
	the out-of-band radiation are beamformed in the expected angles,
	which coincide with the angles derived in
	Section~\ref{sec:9885728}.  The amount of received in-band
	distortion in the direction of the users is
	approximately \SI{-22}{dB}.
	
	\section{Conclusion}
	
	\ifdraft
	\vspace{-1.5ex}
	\fi
	
	We have developed a framework for rigorous analysis of the
	spatial characteristics of nonlinear distortion from arrays.
	The theory can be used in system design to predict the effects
	of out-of-band radiation and to take distortion effects into
	account when, e.g., scheduling users in frequency and
	performing reciprocity calibration.  Our theory also
	characterizes the radiation pattern of the distortion and
	shows that the radiation pattern of the distortion resembles
	that of the desired signal, when there is a dominant user.  If
	there is no dominant user, the distortion is close to
	isotropic.

        The effect of the number of served users and the
	frequency-selectivity of the channel on the radiation pattern
	of the distortion was also studied and criteria for when the
	distortion can be viewed as isotropic are derived.  The
	effects of the distortion do not disappear, i.e.\ the
	received \SINR remains finite, as the number of antennas is
	increased.  The limit, however, is large even with low-end
	amplifiers and would therefore not constitute a significant
	impairment to a practical implementation.
	
	\ifdraft
	\vspace{-1.5ex}
	\enlargethispage{\baselineskip}
	\fi
		
	
	%
	
%
%
%
%
%

	\ifCLASSOPTIONcaptionsoff
	\newpage
	\fi

	
	
	\bibliographystyle{IEEEtran}
		
	\bibliography{bib_forkort_namn,bibliografi}

\begin{thebibliography}{10}
\providecommand{\url}[1]{#1}
\csname url@samestyle\endcsname
\providecommand{\newblock}{\relax}
\providecommand{\bibinfo}[2]{#2}
\providecommand{\BIBentrySTDinterwordspacing}{\spaceskip=0pt\relax}
\providecommand{\BIBentryALTinterwordstretchfactor}{4}
\providecommand{\BIBentryALTinterwordspacing}{\spaceskip=\fontdimen2\font plus
\BIBentryALTinterwordstretchfactor\fontdimen3\font minus
  \fontdimen4\font\relax}
\providecommand{\BIBforeignlanguage}[2]{{%
\expandafter\ifx\csname l@#1\endcsname\relax
\typeout{** WARNING: IEEEtran.bst: No hyphenation pattern has been}%
\typeout{** loaded for the language `#1'. Using the pattern for}%
\typeout{** the default language instead.}%
\else
\language=\csname l@#1\endcsname
\fi
#2}}
\providecommand{\BIBdecl}{\relax}
\BIBdecl

\bibitem{mollen2016OOB}
C.~Mollén, U.~Gustavsson, T.~Eriksson, and E.~G. Larsson, ``Out-of-band
  radiation measure for {MIMO} arrays with beamformed transmission,'' in
  \emph{Proc.\ {IEEE} Int.\ Conf.\ Commun.}, May 2016, pp. 1--6.

\bibitem{6690}
E.~G. Larsson, O.~Edfors, F.~Tufvesson, and T.~L. Marzetta, ``Massive {MIMO}
  for next generation wireless systems,'' \emph{{IEEE} Commun.\ Mag.}, vol.~52,
  no.~2, pp. 186--195, Feb. 2014.

\bibitem{bjornson2013massive}
E.~Björnson, J.~Hoydis, M.~Kountouris, and M.~Debbah, ``Massive {MIMO} systems
  with non-ideal hardware: {E}nergy efficiency, estimation, and capacity
  limits,'' \emph{{IEEE} Trans.\ Inf.\ Theory}, vol.~60, no.~11, pp.
  7112--7139, Nov. 2014.

\bibitem{blandino2017analysis}
S.~Blandino, C.~Desset, A.~Bourdoux, L.~V. der Perre, and S.~Pollin, ``Analysis
  of out-of-band interference from saturated power amplifiers in massive
  {MIMO},'' in \emph{Proc.\ Eur.\ Conf.\ Networks and Commun.}, Jun. 2017, pp.
  1--6.

\bibitem{UGUSGC14}
U.~Gustavsson, C.~{Sanchez Perez}, T.~Eriksson, F.~Athley, G.~Durisi, P.~N.
  Landin, K.~Hausmair, C.~Fager, and L.~Svensson, ``On the impact of hardware
  impairments on massive {MIMO},'' in \emph{Proc.\ {IEEE} Global Commun.\
  Conf.}, Dec. 2014.

\bibitem{zou2015impact}
Y.~Zou, O.~Raeesi, L.~Antilla, A.~Hakkarainen, J.~Vieira, F.~Tufvesson, Q.~Cui,
  and M.~Valkama, ``Impact of power amplifier nonlinearities in multi-user
  massive {MIMO} downlink,'' in \emph{{IEEE} Globecom Workshops}, Dec. 2015,
  pp. 1--7.

\bibitem{pedro2005comparative}
J.~C. Pedro and S.~A. Maas, ``A comparative overview of microwave and wireless
  power-amplifier behavioral modeling approaches,'' \emph{{IEEE} Trans.\
  Microw.\ Theory Tech.}, vol.~53, no.~4, pp. 1150--1163, Apr. 2005.

\bibitem{morgan2006generalized}
D.~R. Morgan, Z.~Ma, J.~Kim, M.~G. Zierdt, and J.~Pastalan, ``A generalized
  memory polynomial model for digital predistortion of {RF} power amplifiers,''
  \emph{{IEEE} Trans.\ Signal Process.}, vol.~54, no.~10, pp. 3852--3860, Sep.
  2006.

\bibitem{barrett1980formula}
J.~F. Barrett, ``Formula for output autocorrelation and spectrum of a
  {Volterra} system with stationary {Gaussian} input,'' in \emph{IEE
  Proceedings D - Control Theory and Applications}, vol. 127, no.~6.\hskip 1em
  plus 0.5em minus 0.4em\relax IET, Nov. 1980, pp. 286--289.

\bibitem{gard1999characterization}
K.~G. Gard, H.~M. Gutierrez, and M.~B. Steer, ``Characterization of spectral
  regrowth in microwave amplifiers based on the nonlinear transformation of a
  complex {Gaussian} process,'' \emph{{IEEE} Trans.\ Microw.\ Theory Tech.},
  vol.~47, no.~7, pp. 1059--1069, Jul. 1999.

\bibitem{raich2004orthogonal}
R.~Raich and G.~T. Zhou, ``Orthogonal polynomials for complex {Gaussian}
  processes,'' \emph{{IEEE} Trans.\ Signal Process.}, vol.~52, no.~10, pp.
  2788--2797, Oct. 2004.

\bibitem{raich2004orthogonal2}
R.~Raich, H.~Qian, and G.~T. Zhou, ``Orthogonal polynomials for power amplifier
  modeling and predistorter design,'' \emph{{IEEE} Trans.\ Veh.\ Technol.},
  vol.~53, no.~5, pp. 1468--1479, Sep. 2004.

\bibitem{sandrin1973spatial}
W.~Sandrin, ``Spatial distribution of intermodulation products in active phased
  array antennas,'' \emph{{IEEE} Trans.\ Antennas Propag.}, vol.~21, no.~6, pp.
  864--868, Nov. 1973.

\bibitem{hemmi2002pattern}
C.~Hemmi, ``Pattern characteristics of harmonic and intermodulation products in
  broadband active transmit arrays,'' \emph{{IEEE} Trans.\ Antennas Propag.},
  vol.~50, no.~6, pp. 858--865, Jun. 2002.

\bibitem{mollen2016outofbandArXiv}
\BIBentryALTinterwordspacing
C.~Moll{\'{e}}n, E.~G. Larsson, U.~Gustavsson, T.~Eriksson, and R.~W. Heath,
  Jr., ``Out-of-band radiation from large antenna arrays,'' \emph{ArXiv
  E-Print}, Nov. 2016, arxiv:1611.01359 [cs.IT]. [Online]. Available:
  \url{http://arxiv.org/abs/1611.01359}
\BIBentrySTDinterwordspacing

\bibitem{mollen2017nonlinearInThesis}
C.~Mollén and E.~G. Larsson, ``The {Hermite}-polynomial approach to the
  analysis of nonlinearities in signal processing systems,'' paper~E in
  “High-End Performance with Low-End Hardware: Analysis of Massive MIMO base
  Station Transceivers”, Ph.D. Dissertation, Linköping University, Sweden,
  2017.

\bibitem{proakis2002communication}
J.~G. Proakis and M.~Salehi, \emph{Communication Systems Engineering},
  2nd~ed.\hskip 1em plus 0.5em minus 0.4em\relax Prentice Hall, 2002.

\bibitem{stuber2001principles}
G.~L. St{\"u}ber, \emph{Principles of Mobile Communication}.\hskip 1em plus
  0.5em minus 0.4em\relax Springer, 2001, vol.~2.

\bibitem{faulkner2000effect}
M.~Faulkner, ``The effect of filtering on the performance of {OFDM} systems,''
  \emph{{IEEE} Trans.\ Veh.\ Technol.}, vol.~49, no.~5, pp. 1877--1884, Sep.
  2000.

\bibitem{cosovic2006subcarrier}
I.~Cosovic, S.~Brandes, and M.~Schnell, ``Subcarrier weighting: {A} method for
  sidelobe suppression in {OFDM} systems,'' \emph{{IEEE} Commun.\ Lett.},
  vol.~10, no.~6, pp. 444--446, Jun. 2006.

\bibitem{tom2013mask}
A.~Tom, A.~Sahin, and H.~Arslan, ``Mask compliant precoder for {OFDM} spectrum
  shaping,'' \emph{{IEEE} Commun.\ Lett.}, vol.~17, no.~3, pp. 447--450, Feb.
  2013.

\bibitem{tan2004reduced}
P.~Tan and N.~C. Beaulieu, ``Reduced {ICI} in {OFDM} systems using the ``better
  than" raised-cosine pulse,'' \emph{{IEEE} Commun.\ Lett.}, vol.~8, no.~3, pp.
  135--137, Mar. 2004.

\bibitem{falconer2002frequency}
D.~Falconer, S.~L. Ariyavisitakul, A.~Benyamin-Seeyar, and B.~Eidson,
  ``Frequency domain equalization for single-carrier broadband wireless
  systems,'' \emph{{IEEE} Commun.\ Mag.}, vol.~40, no.~4, pp. 58--66, Aug.
  2002.

\bibitem{li1997effects}
X.~Li and L.~J. Cimini, ``Effects of clipping and filtering on the performance
  of {OFDM},'' in \emph{Vehicular Technology Conference, 1997, IEEE 47th},
  vol.~3.\hskip 1em plus 0.5em minus 0.4em\relax IEEE, 1997, pp. 1634--1638.

\bibitem{tse2005fundamentals}
D.~Tse and P.~Viswanath, \emph{Fundamentals of Wireless Communication}.\hskip
  1em plus 0.5em minus 0.4em\relax Cambridge University Press, 2005.

\bibitem{marzetta2016fundamentals}
T.~L. Marzetta, E.~G. Larsson, H.~Yang, and H.~Q. Ngo, \emph{Fundamentals of
  Massive {MIMO}}.\hskip 1em plus 0.5em minus 0.4em\relax Cambridge University
  Press, 2016.

\bibitem{ghannouchi2009behavioral}
F.~M. Ghannouchi and O.~Hammi, ``Behavioral modeling and predistortion,''
  \emph{{IEEE} Microw.\ Mag.}, vol.~10, no.~7, pp. 52--64, Dec. 2009.

\bibitem{schetzen1980volterra}
M.~Schetzen, \emph{The {Volterra} and {Wiener} Theories of Nonlinear
  Systems}.\hskip 1em plus 0.5em minus 0.4em\relax Wiley, 1980.

\bibitem{mollen2016waveforms}
C.~Mollén, E.~G. Larsson, and T.~Eriksson, ``Waveforms for the massive {MIMO}
  downlink: {A}mplifier efficiency, distortion and performance,'' \emph{{IEEE}
  Trans.\ Commun.}, Apr. 2016.

\bibitem{ito1952complex}
K.~It{\^o}, ``Complex multiple {Wiener} integral,'' in \emph{Japanese journal
  of mathematics: transactions and abstracts}, vol.~22.\hskip 1em plus 0.5em
  minus 0.4em\relax The Mathematical Society of Japan, Dec. 1952, pp. 63--86.

\bibitem{ismail2015complex}
M.~Ismail and P.~Simeonov, ``Complex {Hermite} polynomials: their combinatorics
  and integral operators,'' \emph{Proc.\ Amer.\ Math.\ Soc.}, vol. 143, no.~4,
  pp. 1397--1410, Apr. 2015.

\bibitem{vieira2017reciprocity}
J.~Vieira, F.~Rusek, O.~Edfors, S.~Malkowsky, L.~Liu, and F.~Tufvesson,
  ``Reciprocity calibration for massive {MIMO}: {Proposal}, modeling, and
  validation,'' \emph{{IEEE} Trans.\ Wireless Commun.}, vol.~16, no.~5, pp.
  3042--3056, May 2017.

\bibitem{holloway2006use}
C.~L. Holloway, D.~Hill, J.~M. Ladbury, P.~F. Wilson, G.~Koepke, and J.~Coder,
  ``On the use of reverberation chambers to simulate a {R}ician radio
  environment for the testing of wireless devices,'' \emph{{IEEE} Trans.\
  Antennas Propag.}, vol.~54, no.~11, pp. 3167--3177, Nov. 2006.

\bibitem{federal2016fcc}
``{FCC} 16-89 report and order and further notice of proposed rulemaking,''
  \url{https://www.fcc.gov/document/spectrum-frontiers-ro-and-fnprm}, Federal
  Communications Commission, Tech. Rep., Jul. 2016, online: accessed
  2016-10-26.

\bibitem{landin2015weblab}
P.~N. Landin, S.~Gustafsson, C.~Fager, and T.~Eriksson, ``{WebLab}: {A}
  web-based setup for {PA} digital predistortion and characterization
  [application notes],'' \emph{{IEEE} Microw.\ Mag.}, vol.~16, no.~1, pp.
  138--140, Jan. 2015.

\end{thebibliography}
\end{document}